\documentclass[aps,twocolumn,floatfix,superscriptaddress,prx,showpacs]{revtex4-2}
\usepackage{graphicx}% Include figure files
\usepackage{epsfig}% Include figure files
\usepackage{amssymb,amsmath,amsfonts,hyperref}
\usepackage{wasysym}
\usepackage{latexsym}
\usepackage{verbatim}
\usepackage{float}
\usepackage[normalem]{ulem}
\usepackage{color}

\newcommand{\be}{\begin{eqnarray}}
\newcommand{\ee}{\end{eqnarray}}

\newcommand{\calR}{{\mathcal R}}
\newcommand{\calP}{{\mathcal P}}
\newcommand{\calI}{{\mathcal I}}

\begin{document}
%\preprint{APS/123-QED}
%draft
%\twocolumn[\hsize\textwidth\columnwidth\hsize\csname @twocolumnfalse\endcsname
\title{Dulmage-Mendelsohn percolation: Geometry of maximally-packed dimer models and topologically-protected zero modes on site-diluted bipartite lattices.}
\author{Ritesh Bhola}
\affiliation{\small{Department of Theoretical Physics, Tata Institute of Fundamental Research, Mumbai 400005, India}}
\author{Sounak Biswas}
\affiliation{\small{Rudolf Peierls Centre for Theoretical Physics, Oxford University, Oxford OX1 3PU, United Kingdom}}
\author{Md Mursalin Islam}
\affiliation{\small{Department of Theoretical Physics, Tata Institute of Fundamental Research, Mumbai 400005, India}}
\author{Kedar Damle}
\affiliation{\small{Department of Theoretical Physics, Tata Institute of Fundamental Research, Mumbai 400005, India}}
\begin{abstract}
The classic combinatorial construct of {\em maximum matchings} probes the random geometry of regions with local sublattice imbalance in a site-diluted bipartite lattice. We demonstrate that these regions, which host the monomers of any maximum matching of the lattice, control the localization properties of a zero-energy quantum particle hopping on this lattice. The structure theory of Dulmage and Mendelsohn provides us a way of identifying a complete and non-overlapping set of such regions. This motivates our large-scale computational study of the Dulmage-Mendelsohn decomposition of site-diluted bipartite lattices in two and three dimensions. Our computations uncover an interesting universality class of percolation associated with the end-to-end connectivity of such monomer-carrying regions with local sublattice imbalance, which we dub {\em Dulmage-Mendelsohn percolation}. Our results imply the existence of a monomer percolation transition in the classical statistical mechanics of the associated maximally-packed dimer model and the existence of a phase with area-law entanglement entropy of arbitrary many-body eigenstates of the corresponding quantum dimer model.  They also have striking implications for the nature of collective zero-energy Majorana fermion excitations of bipartite networks of Majorana modes localized on sites of diluted lattices, for the character of topologically-protected zero-energy wavefunctions of the bipartite random hopping problem on such lattices, and thence for the corresponding quantum percolation problem,  and for the nature of low-energy magnetic excitations in bipartite quantum antiferromagnets diluted by a small density of nonmagnetic impurities.

\end{abstract}


\maketitle

\section{Introduction}
\label{Introduction}

Static impurities are an important feature of many physical systems. This has long motivated studies of {\em quenched} disorder in the context of diffusion of fluids in disordered media, and electronic conduction and magnetism in disordered solids.

The random interconnectedness of a porous medium can affect the diffusion of a fluid through it in a strikingly crucial way if the end-to-end connectivity of a porous material is lost at a sharp porosity threshold. This key insight led Broadbent and Hammersley to initiate the theoretical study of percolation processes over sixty years ago as a paradigm for understanding such behaviour.  Universal features of such threshold behavior are expected to be shared by the corresponding  {\em percolation transition}  of diluted Euclidean lattices. This transition represents an elementary and intuitively compelling geometric example of universality and  scaling at a second-order critical point~\cite{Broadbent_Hammersley,Stauffer_Aharony_book,Christensen_Moloney_book}. Percolation theory has therefore come to occupy center-stage at the confluence of probability theory, geometry, and statistical physics in recent years. 

Unlike a classical fluid, quantum-mechanical particles do not simply diffuse through such a disordered medium; their wave-like character leads to interference phenomena that can render diffusion over large length scales impossible. This insight led to Anderson's nearly-concurrent development of the theory of localization for describing transport properties of the electron fluid in a dirty metal~\cite{Anderson}. This general theory addresses the effects of random potentials and random hopping amplitudes in models of electronic conduction~\cite{Abrahams_review}. 

The literature on {\em quantum percolation}, dating back nearly fifty years~\cite{Kirkpatrick_EggarterPRB1972}, constitutes the closest point of contact between studies of percolation on the one hand, and localization on the other. Quantum percolation is a special case of the localization problem. It models electronic conduction in disordered binary alloys by studying the behaviour of a quantum particle hopping with nearest-neighbor hopping amplitudes on a diluted lattice without any external random potentials~\cite{Shapir_Aharony_Harris_PRL82,Koslowski_Niessen_PRB90,Dillon_Nakanishi_EPJB14,
Daboul_Chang_Aharony_EPJB00,Meir_Aharony_Harris_EPL89,Chang_Lev_Harris_Adler_Aharony_PRL95,
Schubert_Fehske_PRB08,Schubert_Weisse_Fehske_PRB05}.

In the simplest versions of interest to us, the constituent which does not contribute to the conduction band of the alloy is represented by missing sites of a site-diluted {\em bipartite lattice} (which admits a decomposition into two sublattices, labeled ``$A$'' and ``$B$'' in our discussion below, with sites of one sublattice only having neighbors on the opposite sublattice). 
Can such a system of noninteracting electrons support ohmic conduction at some specified Fermi energy when the underlying lattice has end-to-end connectivity in the thermodynamic limit? Conversely, and more informally: Is there a regime of dilution in which a classical fluid percolates, but a quantum fluid does not? These are the key questions that motivate this body of work.

The random geometry of such diluted lattices is also probed by the classic combinatorial problem of {\em maximum matchings}~\cite{Lovasz_Plummer}. In a maximum matching of a lattice,  one attempts to cover the maximum number of sites of the lattice by hard-core dimers that each occupy a single link of the lattice. In this way, one attempts to {\em match} as many sites as possible with one of their neighbours. While no two dimers are allowed to touch at any site, some sites may remain unmatched, {\em i.e.}, host {\em monomers}, if the diluted lattice has no {\em perfect matchings} (equivalently, {\em fully-packed dimer covers}). This ensemble of maximum matchings, with some choice of positive weight for each maximum matching, defines a {\em maximally-packed dimer model} on the underlying lattice. 
%While such maximally-packed dimer models can therefore have a nonzero number of monomers  associated with it. However, additional monomers, over and above this minimum number, are forbidden. 

Like percolation and localization, the problem of maximum matchings also has a long and distinguished history, having seeded major developments in graph theory, combinatorics, and computer science. For instance, it was Edmonds's analysis of the computational complexity of his algorithm for finding maximum matchings~\cite{Edmonds} that led to the notion of the computational  complexity class P that plays a fundamental role in theoretical computer science.

For such maximally-packed dimer models on site-diluted bipartite lattices such as the square and honeycomb lattices in two dimensions, and the cubic lattice in three dimensions, our computational study establishes the presence of a nonzero monomer density in the thermodynamic limit for any nonzero dilution probability $n_{\rm vac} > 0$. As we will see below, monomers of {\em any} maximum matching live in well-demarcated regions of the lattice. We also study the dilution dependence of the end-to-end connectivity of these monomer-carrying regions. For a range of $n_{\rm vac}$ well within the geometrically-percolated phase of the diluted lattice, our computations reveal the striking presence of a localized phase of these monomers; this reflects the finite extent of each such monomer-carrying region.
\begin{figure}
	\includegraphics[width=\columnwidth]{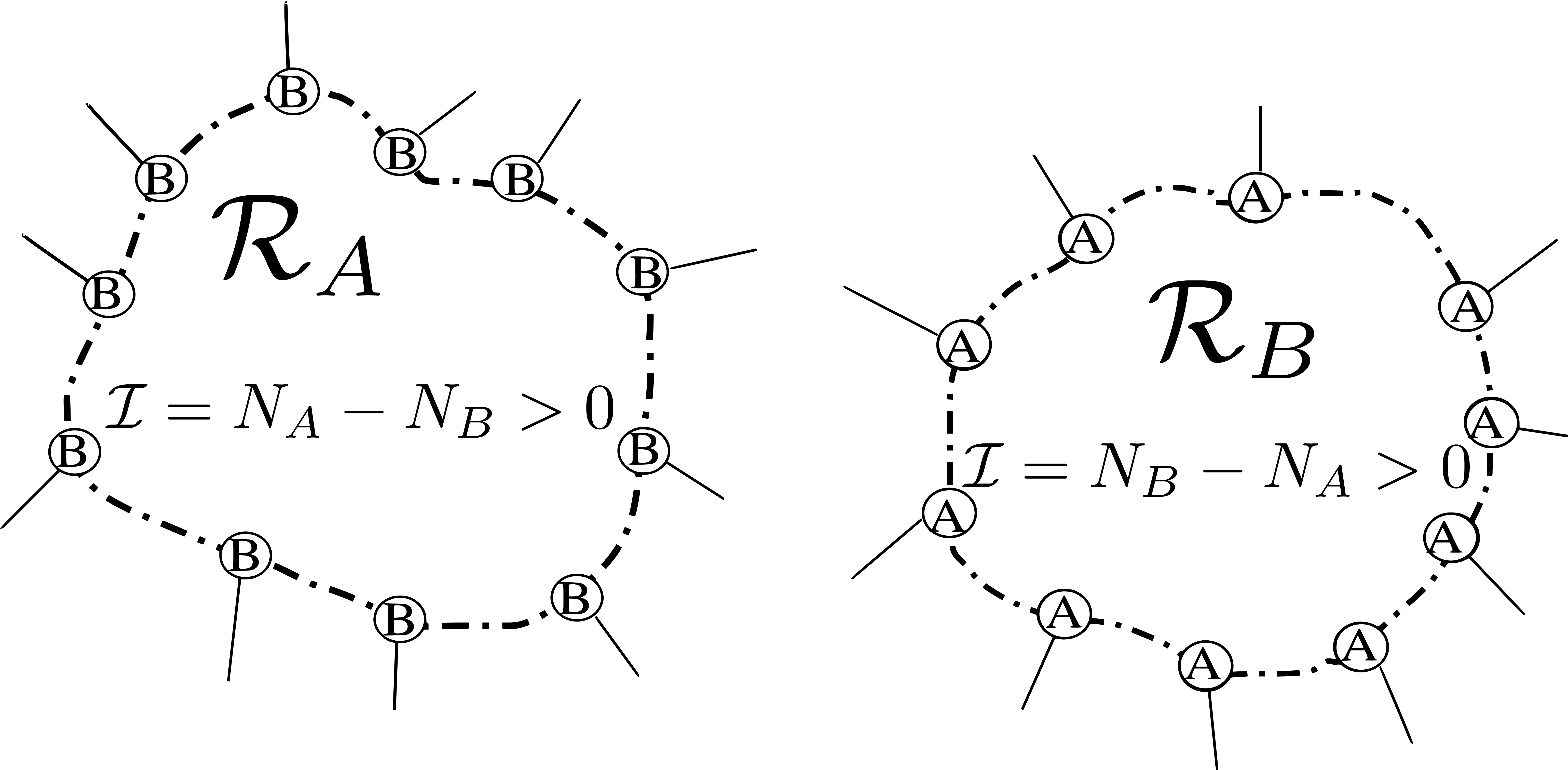}	
\caption{Topologically-protected zero modes of bipartite random hopping problems are expected to live in ${\mathcal R}$-type regions of the lattice, with local sublattice imbalance and a site boundary consisting entirely of sites belonging to the sublattice that is locally in the minority~{\protect{\cite{Sanyal_Damle_Motrunich_PRL}}}. See Sec.~\ref{Introduction}, Sec.~\ref{Executivesummary}, and Sec.~\ref{DM} for a detailed discussion.}
		\label{Rtyperegionsfigure}
\end{figure}

As $n_{\rm vac}$ is lowered further, our study uncovers interesting {\em percolation phenomena} displayed by the monomer-carrying regions. In the three dimensional case, we find a sharply-defined transition associated with the percolation of these connected monomer-carrying regions. This occurs at a nonzero $n_{\rm vac}^{\rm crit}$ that lies well within the geometrically-percolated phase of the underlying lattice. At an even lower dilution $n_{\rm vac}^{\rm low} < n_{\rm vac}^{\rm crit}$, we also observe a spontaneous sublattice symmetry-breaking transition of the monomer fluid, whereby the monomers break the statistical sublattice symmetry of the randomly-diluted bipartite lattice. 

In two-dimensions, we find that these monomer-carrying regions undergo an {\em incipient percolation} transition in the $n_{\rm vac} \to 0$ limit of vanishing site dilution. In this limit, large-scale properties of these regions exhibit universal behaviour controlled by a diverging length-scale that represents their typical size.  In three dimensions, large-scale properties in the vicinity of $n_{\rm vac}^{\rm crit}$ are again controlled by the divergent typical size of such regions. 
Our detailed computational study allows us to identify the correlation length exponent $\nu$ that controls the power-law divergence of this length scale in the vicinity of the critical point in three dimensions, and in the low-dilution limit in two dimensions:
${\nu_{\rm 2d}} = 5.1(9)$, $\nu_{\rm 3d} = 0.87(10)$. 

This universal critical behaviour represents an intrinsic property of the random geometry of the underlying site-diluted lattice in the following sense: Since every dimer in any maximum matching of a bipartite lattice pairs one $A$-sublattice site with an adjacent $B$-sublattice site, monomers in the corresponding maximally-packed dimer model are associated with regions of the diluted lattice with local sublattice imbalance. The critical exponent $\nu$ should thus be thought of as characterizing the random geometry of local sublattice imbalance in such site-diluted bipartite lattices.

Clearly, these results are of fundamental interest from the point of view of the statistical mechanics of such maximally-packed dimer models, and from the point of view of the graph theory of such random lattices, particularly percolation theory. However, at this point, it would be natural for any reader to wonder:  When the theoretical literature already abounds with several other interesting variations on the theme of geometric percolation~\cite{Kesten_review,Adler_review,DSouza_Nagler_review}, what motivated our interest in yet another variant, this one associated with monomers of maximum matchings? Indeed, what is the motivation for a computational study of maximum matchings of such lattices in the first place? And what does any of this have to do with the questions of Anderson localization and quantum percolation introduced at the very outset?

It is useful to answer these natural questions before proceeding further. In the remainder of this Introduction, we provide a colloquium-level overview that answers them and sketches a number of other interesting connections. Readers who prefer to first understand things at a technical level may wish at this point to go to the more detailed and technical overview in Sec.~\ref{Executivesummary}, and then circle back to the rest of this Introduction.

To proceed, we begin by noting that our original motivation came from two pieces of earlier work:
In one of these, Ref.~\onlinecite{Sanyal_Damle_Motrunich_PRL} noticed that the site-diluted honeycomb-lattice tight-binding model that describes low-energy carriers of undoped but diluted graphene had a nonzero density of single-particle states at exactly zero energy. These zero modes were argued to arise from so-called ``${\mathcal R}$-type'' regions of the lattice which have a {\em local} imbalance in the number of $A$-sublattice sites relative to $B$-sublattice sites (even when there is no such imbalance globally), and a boundary consisting entirely of sites belonging to the sublattice that is locally in the minority~\cite{Sanyal_Damle_Motrunich_PRL} (see Fig.~\ref{Rtyperegionsfigure}).

Crucially, this proposed mechanism also predicted that these zero modes are topologically-protected, in the sense that their {\em existence} is unaffected by changes in the actual values of the nonzero hopping amplitudes on each nearest-neighbor link, so long as the pattern of nearest-neighbor connectivity in the lattice does not change~\cite{Sanyal_Damle_Motrunich_PRL}. Associated with this robustness is a topologically-protected localization property of the corresponding wavefunction: it lives entirely within the ${\mathcal R}$-type region~\cite{Sanyal_Damle_Motrunich_PRL}.

In the other closely-related work, Ref.~\onlinecite{Sanyal_Damle_Chalker_Moessner} studied the effects of non-magnetic impurities in a SU(2)-symmetric version~\cite{Yao_Lee} of
Kitaev's model~\cite{Kitaev_anyon} for a Majorana spin liquid on the honeycomb lattice. Low-temperature properties of such systems are controlled by fermionic excitations, whose spectrum is obtained by solving a tight-binding model on the site-diluted honeycomb lattice with $\pi$ flux attached~\cite{Kitaev_anyon,Willans_Chalker_Moessner_PRB,Udagawa} to each vacancy. The topologically-protected zero-energy states identified earlier in the context of diluted graphene are expected to survive this flux attachment, albeit with modified wavefunctions~\cite{Sanyal_Damle_Motrunich_PRL,Sanyal_Damle_Chalker_Moessner}. 

In a striking corroboration of this prediction, the density of zero modes obtained numerically in Ref.~\onlinecite{Sanyal_Damle_Chalker_Moessner} in this problem with flux attachment matched the corresponding density in diluted graphene to within a few percent.
This strongly suggests the zero mode density is dominated by such topologically-protected zero modes, with fragile modes (that depend on a specific pattern of values for the nonzero hopping amplitudes) contributing an insignificant fraction of the total. 

If the vacancies are not permitted to get too close to each other, as was the case in the computations of Refs.~\onlinecite{Sanyal_Damle_Motrunich_PRL,Sanyal_Damle_Chalker_Moessner}, the smallest nontrivial example~\cite{Sanyal_Damle_Motrunich_PRL} of such topologically-protected zero modes on the site-diluted honeycomb lattice is a single mode associated with a ${\mathcal R}$-type region with six vacancies at specific positions near each other. The expected density of these ``${\mathcal R}_6$'' regions provides a simple but rigorous lower-bound on the density of topologically-protected zero modes~\cite{Sanyal_Damle_Motrunich_PRL}. However, the density of zero modes computed by multiprecision numerics far exceeds this bound in both cases, {\em i.e.}, with and without flux attachment~\cite{Sanyal_Damle_Motrunich_PRL,Sanyal_Damle_Chalker_Moessner}. This raises the question: If ${\mathcal R}_6$ regions are atypical, how does a generic ${\mathcal R}$-type region ``look'', and how many linearly-independent zero modes does it support? 

This is an interesting question, both in the context of such Kitaev-like models, and in the context of Hubbard-like models for electron-electron interactions in diluted undoped graphene. In the Kitaev systems, the density of zero modes controls the coefficient of a Curie-like term that dominates the low-temperature susceptibility~\cite{Sanyal_Damle_Chalker_Moessner}, while their wavefunction determines the spatial profile of the vacancy-induced magnetic moments that lead to this Curie term. Likewise, in the Hubbard model, Hartree-Fock mean field theory   suggests~\cite{Koga_Tsunetsugu} that the zero modes would be associated with local moment formation, with their wavefunctions again controlling the spatial profile of these local moments.

Since these topologically-protected modes are expected to live in ${\mathcal R}$-type regions with local sublattice imbalance, it is natural to explore the possibility that progress on this question can be made by thinking in terms of maximum matchings of the site-diluted lattice. After all, as we have already discussed, every dimer pairs one $A$-sublattice site with an adjacent $B$-sublattice site, while monomers of a maximally-packed dimer model are associated with local sublattice imbalance. 

Indeed, a similar line of reasoning led Longuet-Higgins~\cite{Longuet-Higgins} nearly sixty years ago to a pair of insightful exact results that partially anticipated related developments in the graph theory literature~\cite{Lovasz_Plummer}. Transcribed to the language used here, these results say the following: i) The number of monomers in any maximum matching of a bipartite lattice equals the number of linearly-independent topologically-protected zero-energy wavefunctions of a quantum-mechanical particle hopping on nearest-neighbor links of this lattice. ii) The spatial support of the topologically-protected zero-energy eigenspace of this hopping Hamiltonian is given by the set of all sites that can host a monomer in any maximum matching of the lattice. 

The first of Longuet-Higgins's results provides a simple way to independently confirm~\cite{Weik_etal} the relatively large density of topologically-protected zero modes~\cite{Sanyal_Damle_Motrunich_PRL}, far in excess of the simple lower bound of Ref.~\cite{Sanyal_Damle_Motrunich_PRL}. While this confirmation~\cite{Weik_etal} is reassuring, the initial impetus for our work came mainly from our attempts to go beyond Longuet-Higgins's second result in order to answer our earlier question: What is the spatial morphology of a ``typical'' ${\mathcal R}$-type region of the diluted lattice? 

This question can be restated more formally: Can we equip this eigenspace of zero modes with a ``natural'' basis that has a topologically-protected localization property?  Clearly, answering this requires us to go beyond Longuet-Higgins's second result, which only provides a {\em global} characterization of the spatial support of the topologically-protected null space of the hopping matrix as a whole.

Our approach to this question brings into play a graph-theoretical tool, the {\em Dulmage-Mendelsohn decomposition} of a bipartite graph, which provides a crucial structural characterization of such diluted lattices via the combinatorial problem of maximum matchings~\cite{Dulmage_Mendelsohn,Lovasz_Plummer,Pothen_Fan,Kavitha}. This allows us to use any one maximum matching of the diluted lattice to construct a set of non-overlapping connected regions of the lattice that  host  the monomers of {\em any} maximum matching. 

Our key insight is simply stated: These connected monomer-carrying regions identified using Dulmage and Mendelsohn's structure theory provide a natural construction of {\em a complete non-overlapping set} of ${\mathcal R}$-type regions of the lattice. 
Our construction shows that the number of linearly-independent topologically-protected zero modes localized within each such ${\mathcal R}$-type region is precisely equal to the number of monomers hosted by it in any maximum matching. 
\begin{figure*}
  \begin{tabular}{ccc}
	\includegraphics[width=0.66\columnwidth]{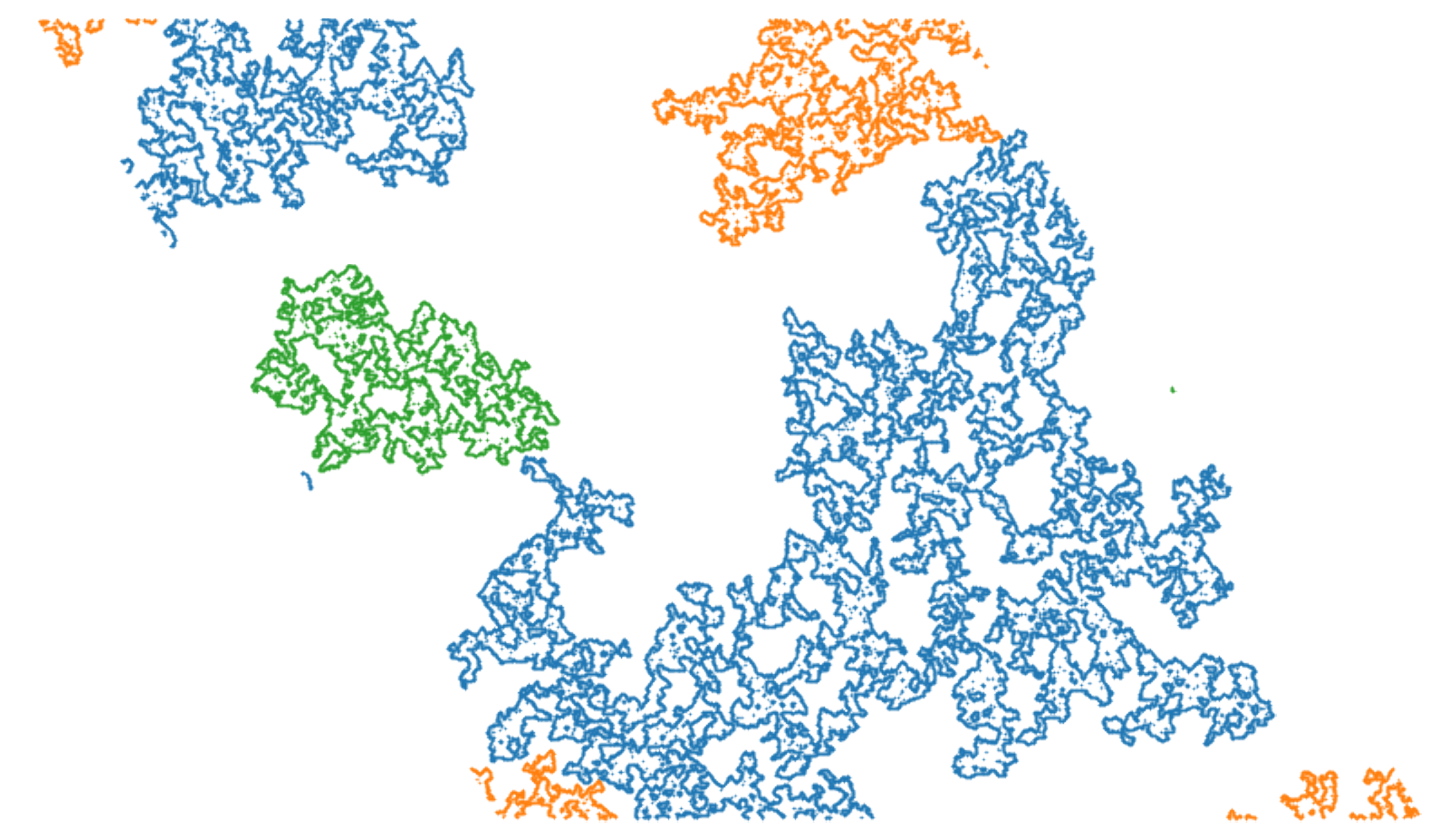} 	&\includegraphics[width=0.66\columnwidth]{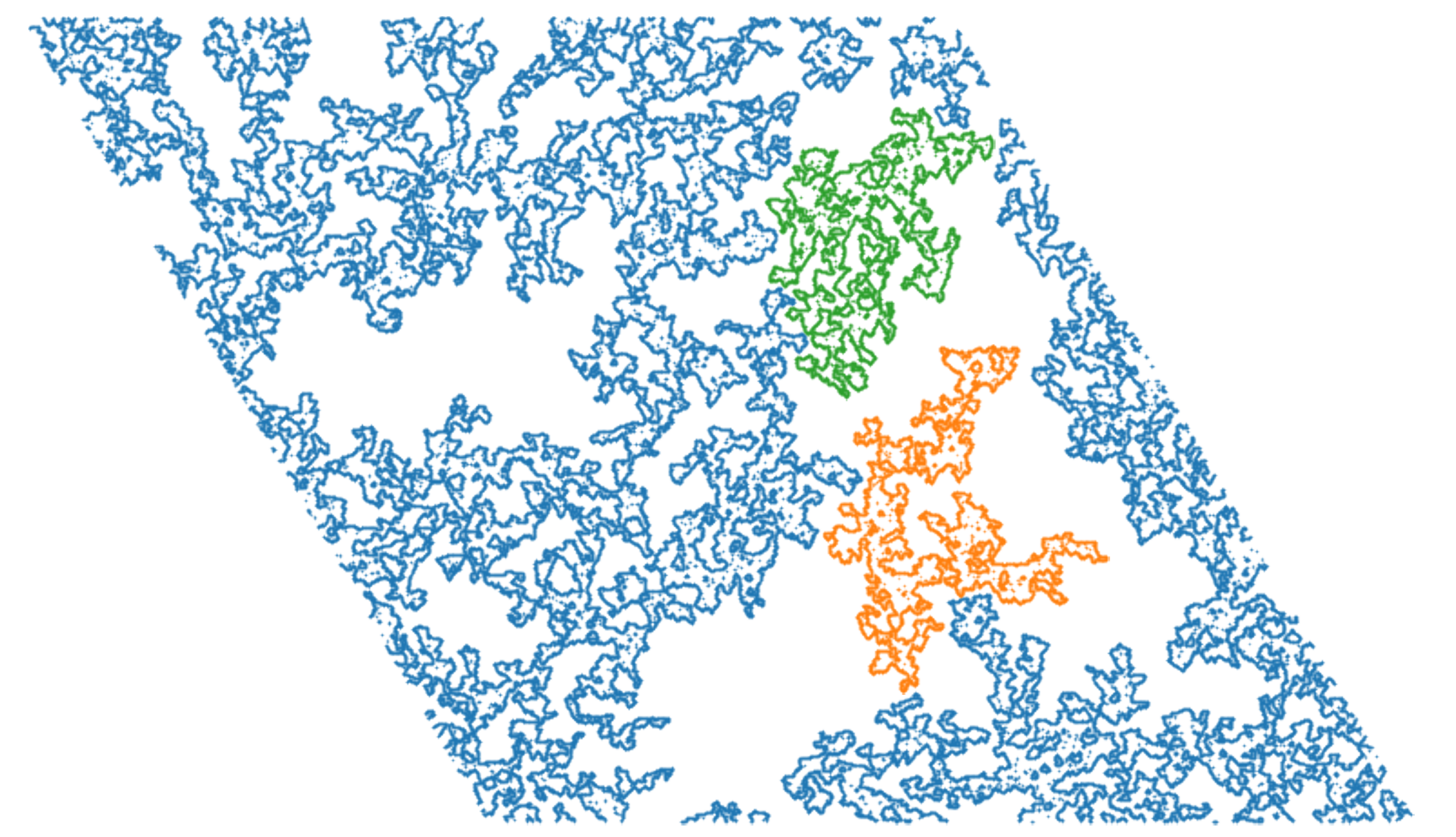}  &	\includegraphics[width=0.66\columnwidth]{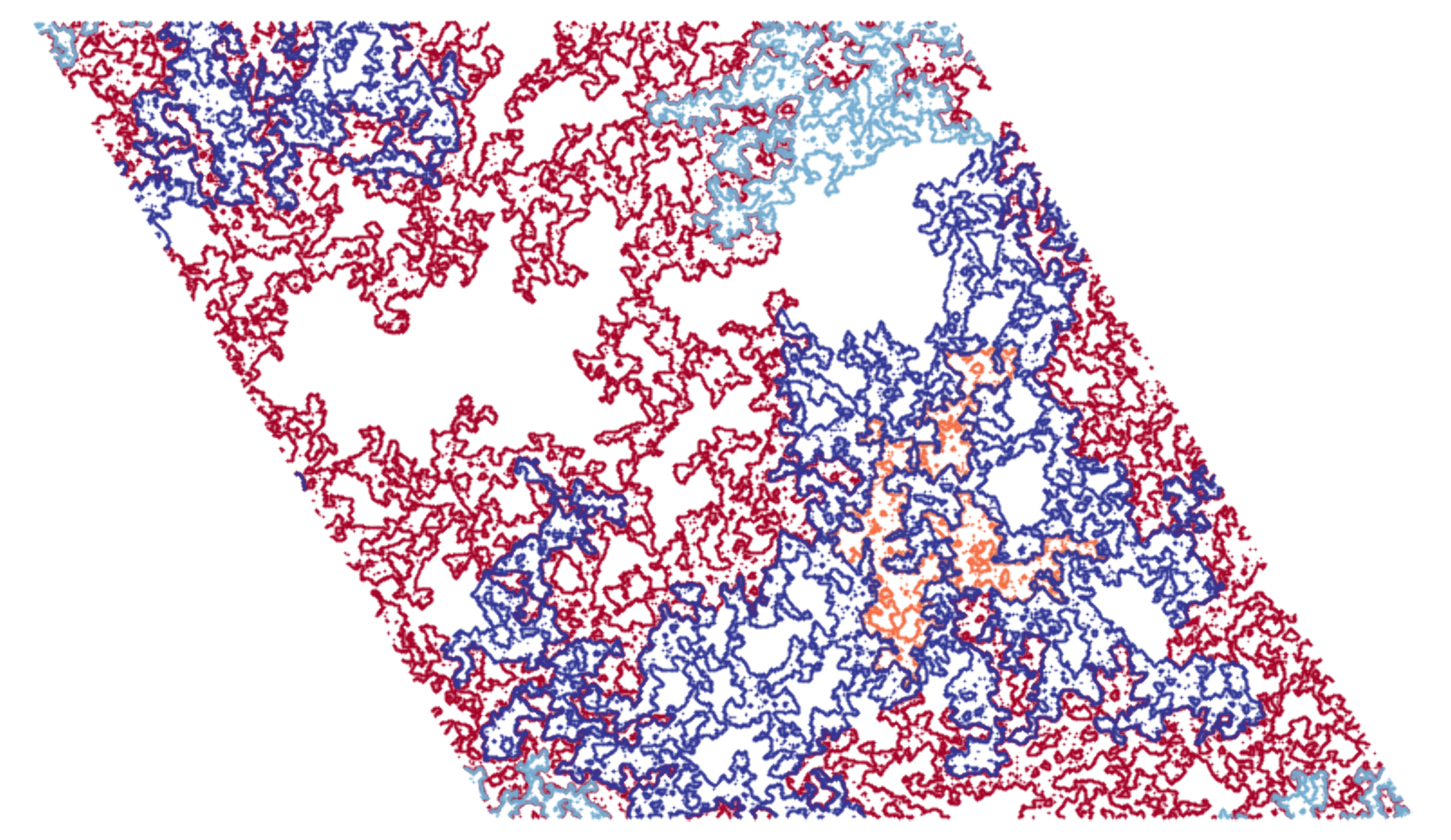} 		
  \end{tabular}
\caption{The left (middle) panel shows the boundary of the largest three ${\mathcal R}_A$-type (${\mathcal R}_B$-type) regions of
the Dulmage-Mendelsohn decomposition of a honeycomb lattice sample with $L=4000$ and periodic boundary conditions, with vacancy density $n_{\rm vac} = 0.08$. The right panel shows the largest two ${\mathcal R}_A$-type and ${\mathcal R}_B$-type regions of the same sample. The color-coding of the boundaries has been changed in the right panel (relative to the other two panels) for greater visibility. See Sec.~\ref{Introduction}, Sec.~\ref{Executivesummary} and Sec.~\ref{DM} for further details.}
		\label{Rtypepicture}
\end{figure*}

Thus, we obtain an alternate {\em local} proof of both of Longuet-Higgins's results using the structure theory of Dulmage and Mendelsohn. 
This has the advantage that it also provides a prescription for constructing an orthonormal basis of zero modes with a topologically-protected localization property.
The unusual monomer percolation phenomena advertised earlier correspond to the percolation transitions of these monomer-carrying components of the Dulmage-Mendelsohn decomposition.
Clearly, this has consequences for the corresponding quantum-mechanical wavefunctions.

Indeed, our identification of these {\em Dulmage-Mendelsohn percolation} phenomena has interesting implications for the zero-energy quantum percolation problem on such site-diluted bipartite lattices: In the two-dimensional case, a zero-energy quantum particle is localized for any nonzero dilution, albeit with a localization length that diverges as one approaches the $n_{\rm vac} \to 0$ limit of vanishing site dilution. There is thus no quantum percolation in this case, even deep within the classically percolated phase in which a classical fluid diffuses unhindered through the lattice.
This settles a question about which there has been some controversy in the literature~\cite{Shapir_Aharony_Harris_PRL82,Koslowski_Niessen_PRB90,Dillon_Nakanishi_EPJB14,
Daboul_Chang_Aharony_EPJB00,Meir_Aharony_Harris_EPL89}. 
 In the three-dimensional case, we conclude that the quantum percolation transition at zero energy is associated with the Dulmage-Mendelsohn percolation transition of the diluted lattice. A more detailed discussion of implications of our work for quantum percolation is in Sec.~\ref{QuantumPercolation&MajoranaPercolation} and Sec.~\ref{Outlook}.
 
The structure theory of Dulmage and Mendelsohn also provides a way of factorizing the partition function of such maximally-packed dimer models. Each factor is either associated with a monomer-carrying ${\mathcal R}$-type region, or a region of the lattice which is always perfectly matched in any maximum matching.  Remarkably, this factorized structure is valid not just for the simplest case of non-interacting dimer models (with Boltzmann probabilities determined entirely by weights of occupied links), but also in the presence of monomer and dimer interactions, so long as the dimer interaction terms only act on {\em flippable} (perfectly-matched) elementary plaquettes or on larger flippable loops, and the monomer interactions are sufficiently short-ranged. The corresponding quantum dimer models share these potential-energy terms with their classical analogs. Additionally, they have monomer-hopping terms,  as well as kinetic-energy terms that change the perfect matching of a flippable loop. For such quantum dimer models, this factorization property implies that any many-body eigenstate has a tensor product structure, with one factor associated with each of these regions.

Our computational study of two-dimensional site-diluted square and honeycomb lattices and the site-diluted cubic lattice in three dimensions reveals that the perfectly-matched regions remain finite and small in extent in the thermodynamic limit at any nonzero dilution. Thus, when there is no percolation of the monomer-carrying regions, both dimer and monomer correlations in the classical model remain strictly localized. For the analogous quantum dimer models in two and three dimensions, this also implies the existence of a phase with {\em area-law} entanglement entropy of eigenstates in the middle of the energy spectrum. This is discussed in more detail in Sec.~\ref{MBLprediction} and Sec.~\ref{Outlook}.

\begin{figure*}
  \begin{tabular}{cccc}
	\includegraphics[width=0.49\columnwidth]{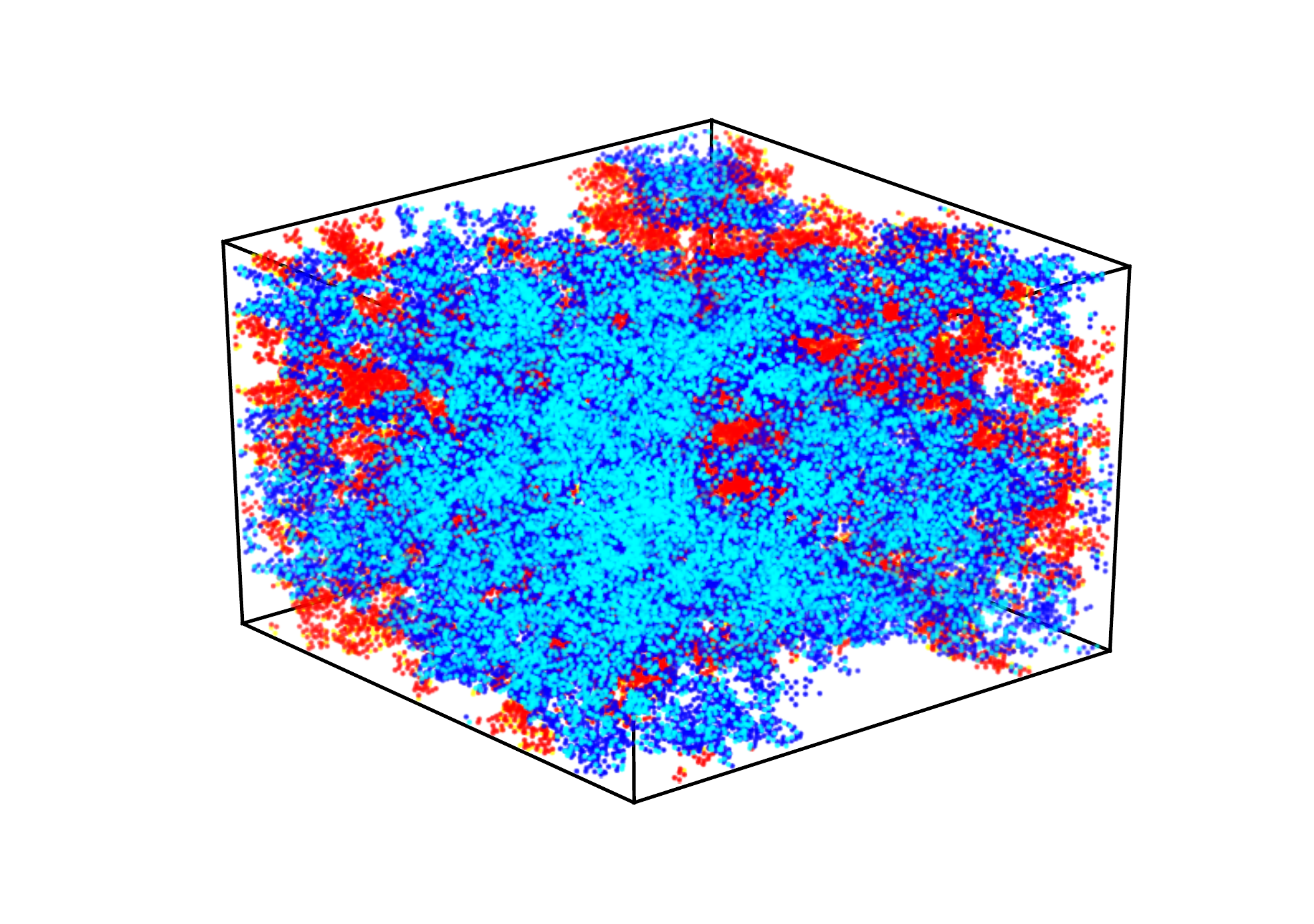} 	&\includegraphics[width=0.49\columnwidth]{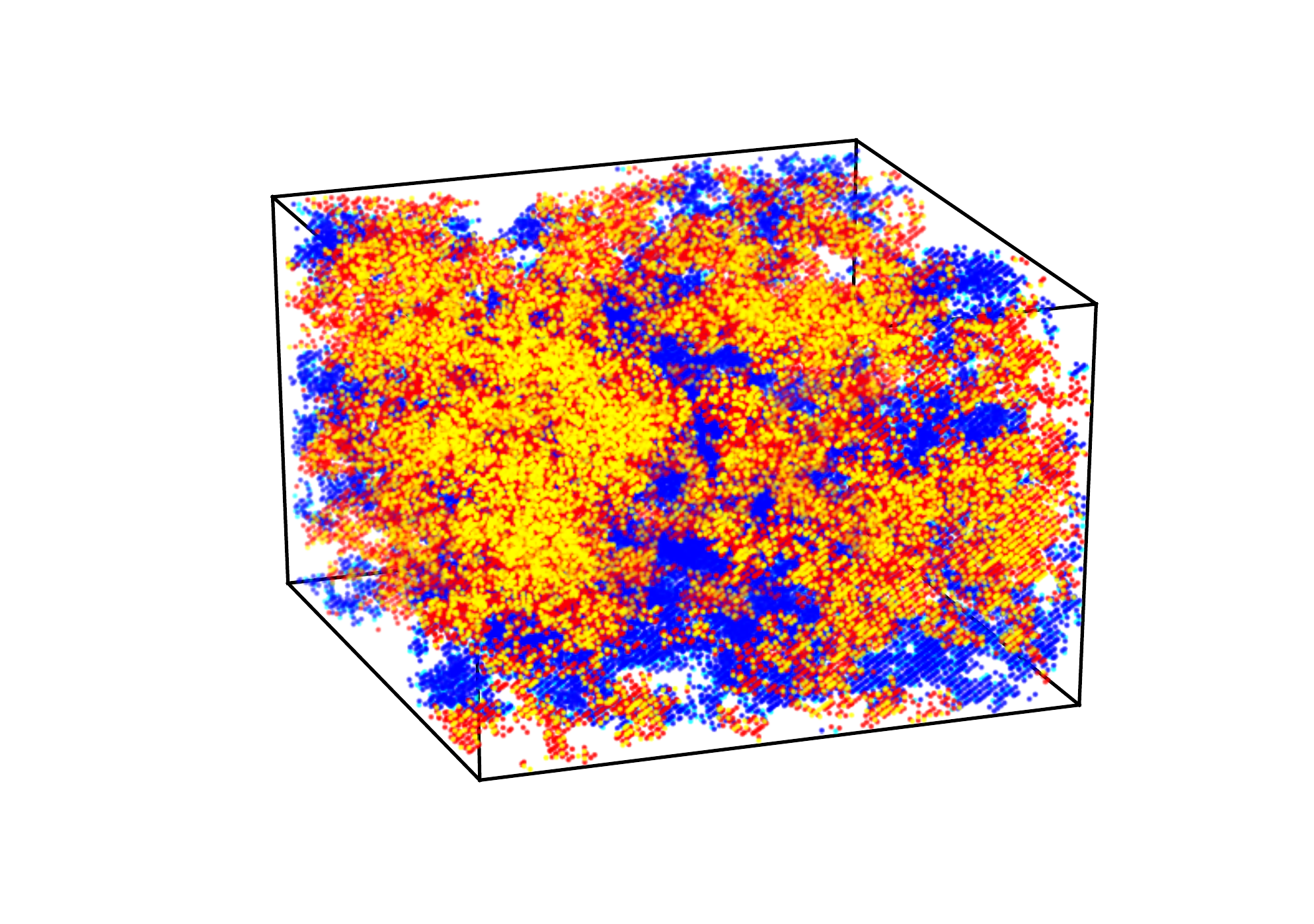}  &	\includegraphics[width=0.49\columnwidth]{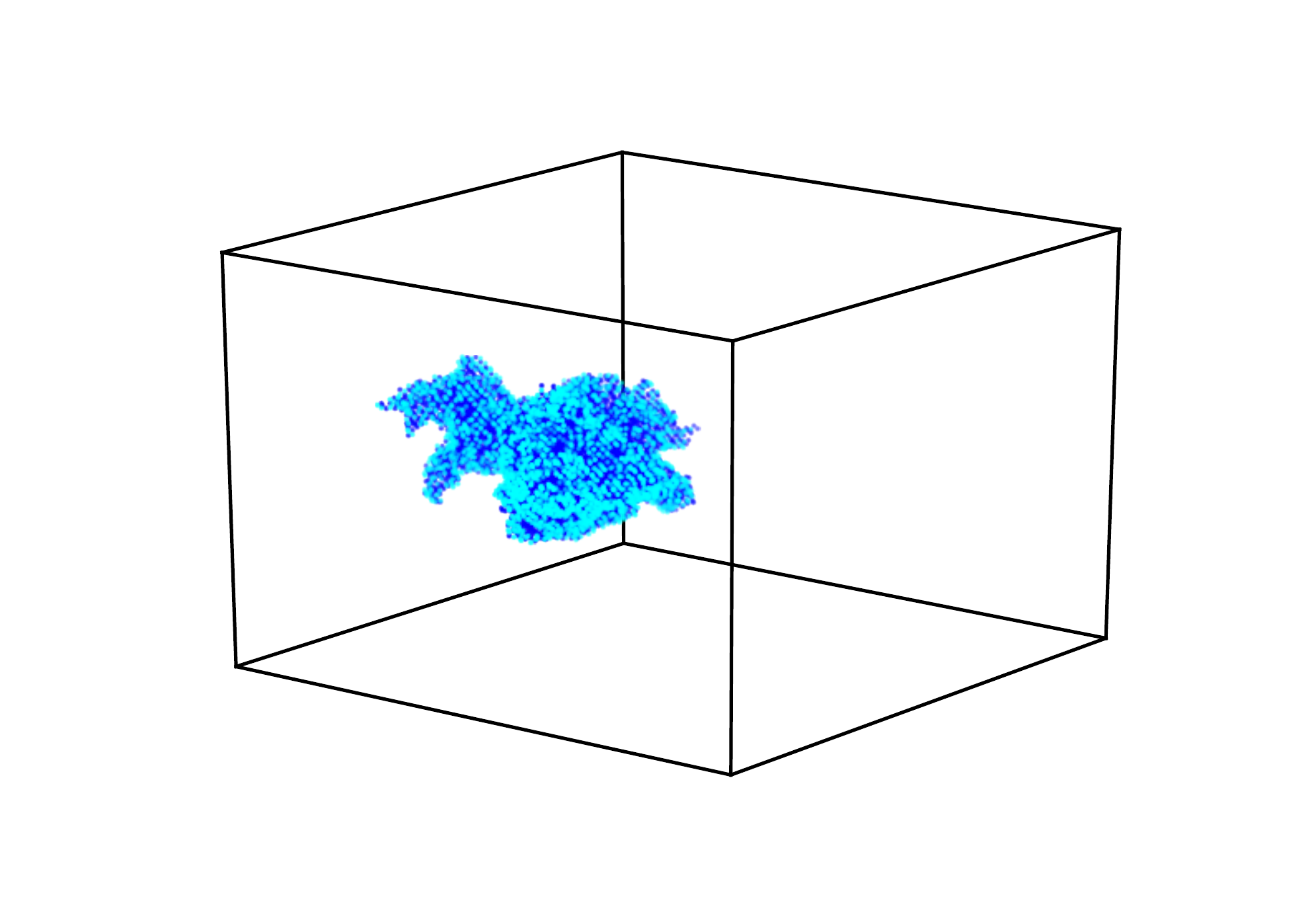}&
\includegraphics[width=0.49\columnwidth]{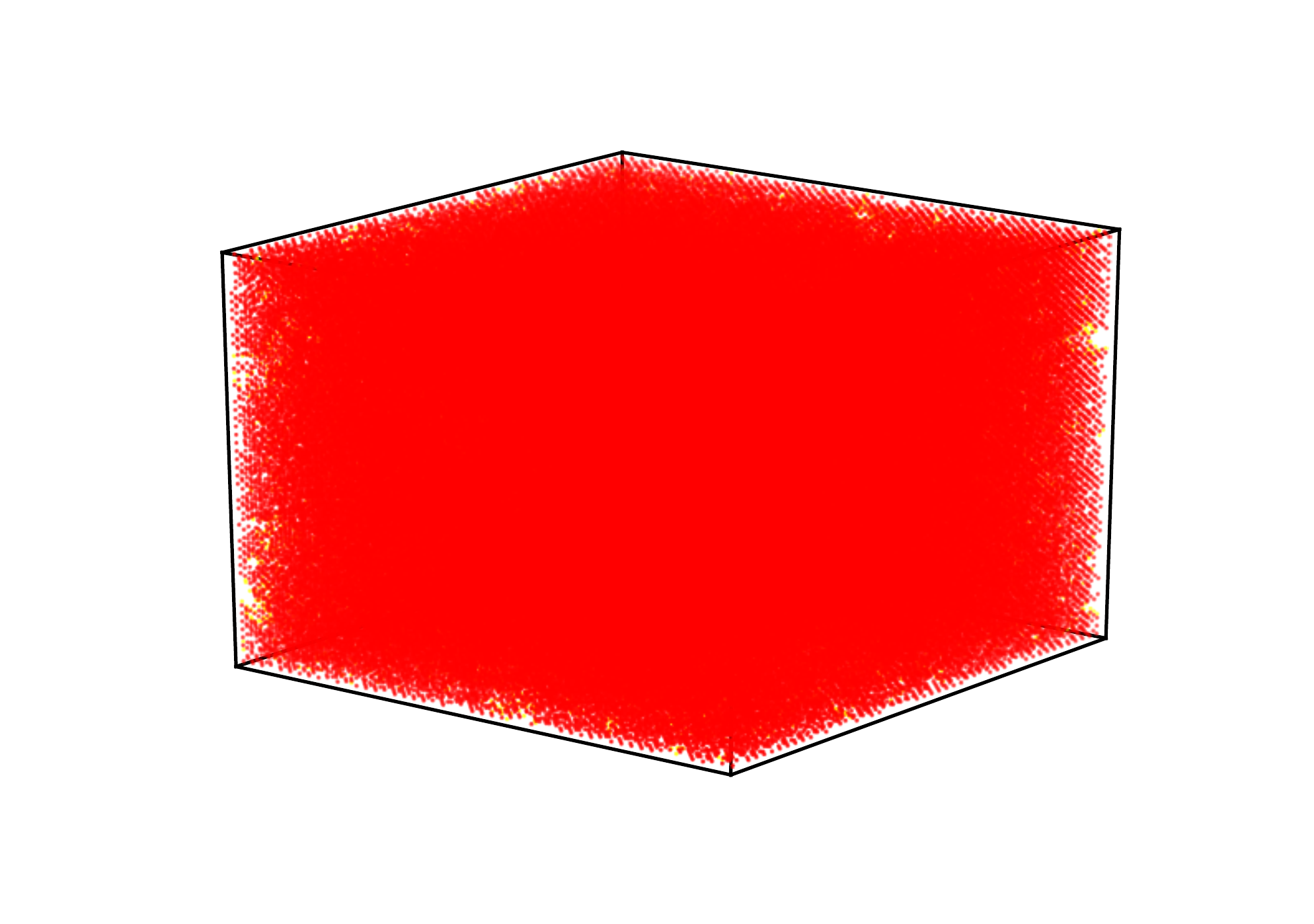}		 		
  \end{tabular}
\caption{Left two panels: At intermediate values of dilution $0.35 \lesssim n_{\rm vac} \lesssim 0.6$, the largest ${\mathcal R}_A$ (dark blue bulk with light blue surface) region and the largest ${\mathcal R}_B$-type region (red bulk with yellow surface) of the Dulmage-Mendelsohn decomposition of a site-diluted cubic lattice {\em both} percolate, as shown in this $L=100$ example from two different vantage points in these panels.  Right two panels: For smaller $n_{\rm vac}$, in each sample, {\em either} the largest ${\mathcal R}_A$-type region is small and the largest ${\mathcal R}_B$-type region percolates or vice-versa, as is evident from individual snapshots of the largest ${\mathcal R}_A$ (dark blue bulk with light blue surface) and the largest ${\mathcal R}_B$ region (red bulk with yellow surface) in the example shown. For such low dilution, each sample thus spontaneously breaks the sublattice symmetry that characterizes the ensemble of site-diluted lattices we consider. See Sec.~\ref{Introduction}, Sec.~\ref{Executivesummary} and Sec.~\ref{DM} for further details.} 
		\label{Rtypepicturecubic}
\end{figure*}

To reiterate the main message of all of the foregoing: Maximum matchings probe the random geometry of regions with local sublattice imbalance in a site-diluted bipartite lattice. These regions host the monomers of any maximum matching and control the localization properties of a zero-energy quantum particle hopping on this lattice. The structure theory of Dulmage and Mendelsohn provides us a way of identifying a complete and non-overlapping set of such regions, as well as regions of the lattice that are always perfectly matched in any maximum matching. This motivates our large-scale computational study of the Dulmage-Mendelsohn decomposition of site-diluted bipartite lattices in two and three dimensions. Our computations uncover an interesting universality class of Dulmage-Mendelsohn percolation phenomena associated with the end-to-end connectivity of such monomer-carrying regions (see Fig.~\ref{Rtypepicture} and Fig.~\ref{Rtypepicturecubic}), with striking implications for maximally-packed classical and quantum dimer models on such lattices, and for the quantum percolation problem. 

Having conveyed this main message, we now conclude this section by describing two other contexts in which our results yield interesting and significant conclusions. 
%The first concerns emergent (collective) Majorana fermion excitations of networks of localized Majorana modes. The second concerns the physics of  low-energy magnetic excitations of diluted antiferromagnets. 
The first of these refers to a network~\cite{Alicea_etal,Laumann_Ludwig_Huse_Trebst} whose nodes represent individual Majorana modes engineered to exist on some physical platforms (for instance, a topological superconductor device~\cite{Read_Green,Kitaev_chain}). The links of this network correspond to the dominant mixing amplitudes that couple these modes to each other. A missing node represents the absence of the corresponding localized Majorana mode, perhaps due to physical device being in the ``wrong'' regime of parameters due to disorder effects.

Although we expect these dominant mixing amplitudes to lead to most of the localized Majorana modes being lifted away from zero  energy, we may ask: Are there any collective zero-energy Majorana fermion excitations of the network as a whole?  If this network can be described by a site-diluted bipartite lattice in two or three dimensions, our results yield an interesting answer via a mapping to an equivalent problem of a quantum-mechanical particle hopping on this bipartite lattice: Every ${\mathcal R}$-type region with an {\em odd} number of linearly independent zero modes of this hopping problem  generically hosts a single such collective zero-energy Majorana fermion excitation which is perturbatively stable to the introduction of further-neighbor couplings in this network. The Dulmage-Mendelsohn percolation phenomena studied here thus have implications for the spatial structure of collective zero-energy Majorana fermion excitations of these networks, as discussed further in Sec.~\ref{QuantumPercolation&MajoranaPercolation} and Sec.~\ref{Outlook}.

The second of these contexts has to do with diluted quantum antiferromagnets.
In a series of papers largely motivated by experimental work on impurity effects in quantum antiferromagnets~\cite{Vajk_etal_Science2002}, Sandvik~\cite{Sandvik_PRB2002} and Wang~\cite{Wang_Sandvik_PRL2006,Wang_Sandvik_PRB2010} reported on a detailed quantum Monte Carlo (QMC) study of the $S=1/2$ Heisenberg antiferromagnet on the site-diluted square lattice. The basic conclusion was that long range antiferromagnetic order persists in the ground state all the way up to the classical percolation threshold of the diluted lattice. Indeed, it was argued that the critical percolating cluster at the geometric percolation transition has long range antiferromagnetic order in the ground state. As a result, the antiferromagnetic transition is driven by the underlying geometric transition, and occurs right at the geometric site-percolation threshold of the square lattice~\cite{Wang_Sandvik_PRL2006,Wang_Sandvik_PRB2010}.

These studies also explored the physics of the antiferromagnetically-ordered phase in the vicinity of this transition~\cite{Wang_Sandvik_PRL2006,Wang_Sandvik_PRB2010}.
An interesting finding in this regime was the presence of anomalously low-energy triplet excitations that dominated the low-energy behaviour of the ordered phase and had an effect on the quantum-critical scaling. These triplet excitations were argued to arise from regions of the lattice with local sublattice imbalance~\cite{Wang_Sandvik_PRB2010}, and were therefore modeled in terms of the spatial monomer distribution function of the corresponding dimer model. Motivated by this, Henley and collaborators  studied analogous phenomena on the Bethe lattice at percolation, and developed a Schwinger-boson mean field approach to this physics~\cite{Changlani_Ghosh_Pujari_Henley_PRL2013,Ghosh_Changlani_HenleyPRBB2015}.

Our work adds to this understanding in a direct and crucial way, since the ${\mathcal R}$-type regions constructed using the structure theory of Dulmage and Mendelsohn constitute a complete set of non-overlapping regions that host such low-energy triplet excitations of this diluted antiferromagnet. Our identification of incipient Dulmage-Mendelsohn percolation phenomena in the $n_{\rm vac} \to 0$ limit, deep in the geometrically-percolated phase of the site-diluted square lattice, then implies that these triplet excitations of the diluted $S=1/2$ antiferromagnet have their own  hitherto-unsuspected  critical behaviour deep inside the antiferromagnetic phase; this is unconnected with the geometric percolation driven antiferromagnetic transition itself. It also opens up interesting avenues for further work, which are discussed in Sec.~\ref{dilutedAFM} and Sec.~\ref{Outlook}.

As mentioned earlier, this introductory discussion sets the stage for a more technical overview in Sec.~\ref{Executivesummary} of our principal ideas and results. This more technical overview can be read in conjunction with  Sec.~\ref{Discussion}, which provides additional theoretical perspective and a more detailed discussion of the various implications of our results, and Sec.~\ref{Outlook}, which discusses interesting questions thrown up by our work. The middle third of our article describes the precise theoretical arguments, the computational methods, and the detailed analyses of numerical data that lead to these results. A road map through this part of the paper, and through the Appendix which augments this material, is provided at the end of Sec.~\ref{Executivesummary}.

\section{Technical overview: Models, approach, and results}
\label{Executivesummary}

\begin{figure}
a)~\includegraphics[width=\columnwidth]{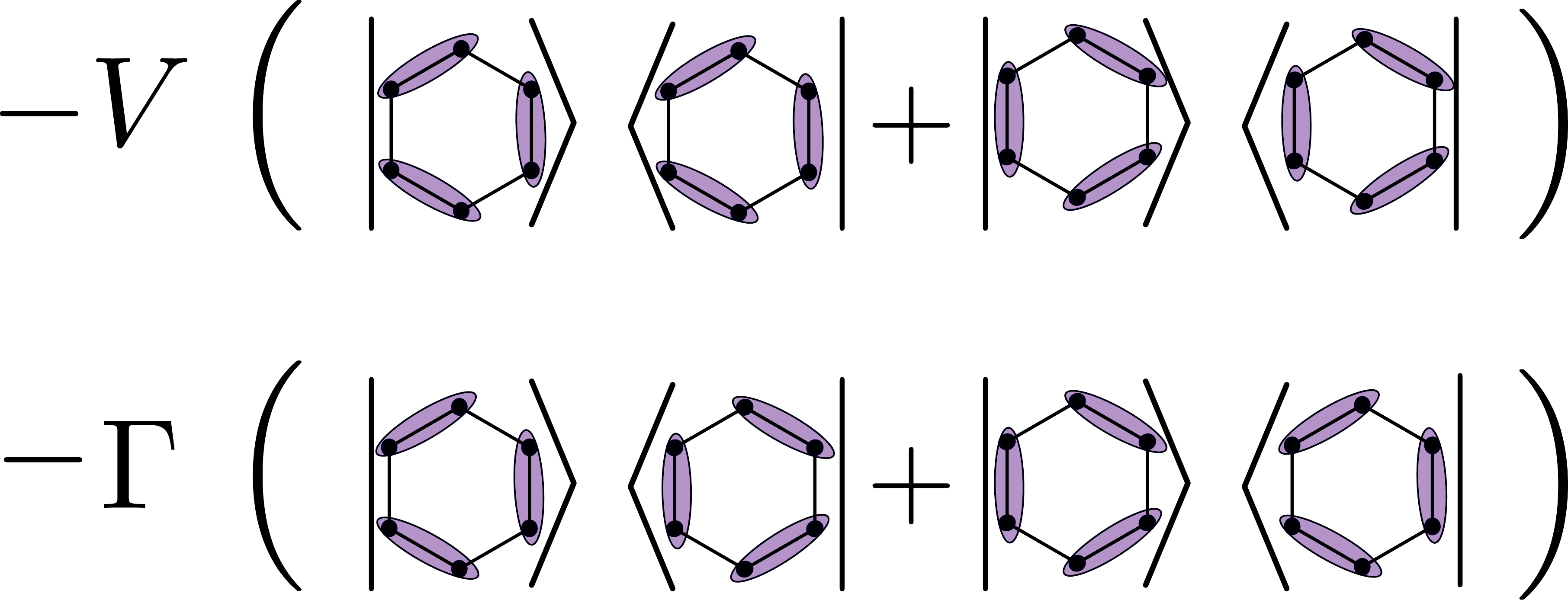}\\
\vspace{0.5cm}
b)~\includegraphics[width=\columnwidth]{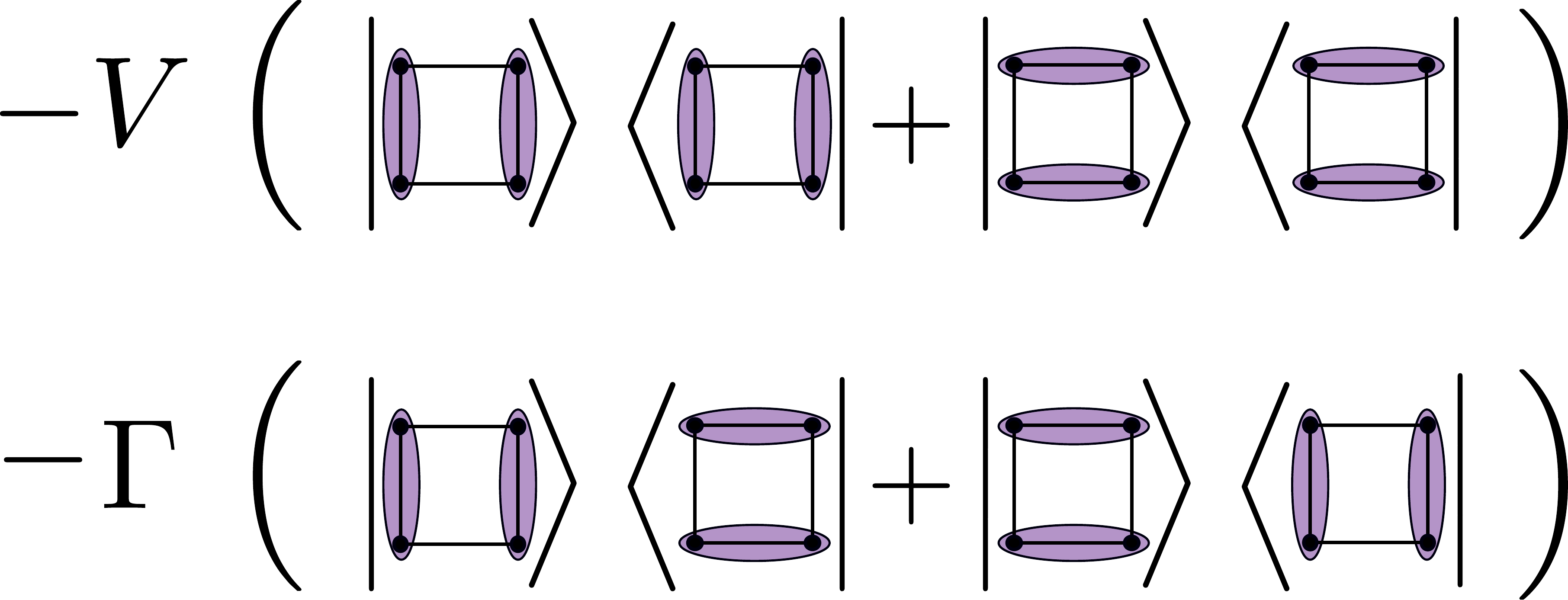}\\	
\vspace{0.5cm}
c)~\includegraphics[width=\columnwidth]{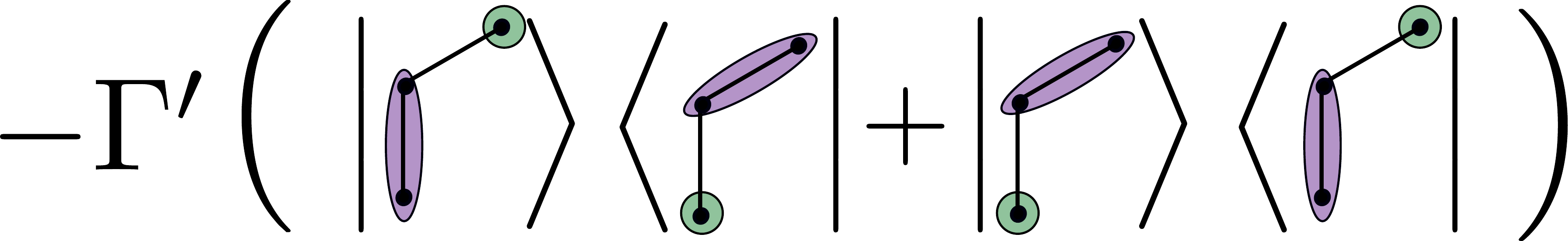}\\	
\caption{a) Potential energy ($V$) and ring exchange ($\Gamma$) terms on flippable elementary plaquettes in the maximally-packed honeycomb-lattice quantum dimer models of interest to us. b) Potential energy ($V$) and ring exchange ($\Gamma$) terms on flippable elementary plaquettes in such models on square and cubic lattices. c) Monomer hopping term ($\Gamma^{'}$) in these maximally-packed quantum dimer models. See Sec.~\ref{Executivesummary} and Sec.~\ref{QMD} for detailed discussion.}
		\label{Quantumdimermodel}
\end{figure}

We consider site-diluted bipartite lattices such as the square and honeycomb lattice in two dimensions, and the cubic lattice in three dimensions, with compensated dilution $n_{\rm vac}$, {\em i.e.,} exactly equal numbers of surviving sites on the two sublattices. On the one hand, we are interested in the classical statistical mechanics of maximally-packed dimer models, {\em i.e.}, of the ensemble of maximum matchings of such lattices, with partition function
\begin{eqnarray}
Z &=& \sum_{{\mathcal C}} \exp(-S({\mathcal C}))\nonumber \\
S &=& -\sum_{\langle r r' \rangle} w_{r r'} n_{r r'}({\mathcal C}) + \dots \; ,
\label{Z_monomerdimer}
\end{eqnarray}
where $n_{\langle r r' \rangle}({\mathcal C})$ is the dimer occupation number of the link $\langle r r'\rangle$ in maximally-packed configuration ${\mathcal C}$, and $\exp(w_{r r'})$ (with real $w_{r r'} > 0 $) is the associated bond weight. In contrast to the undiluted case, this ensemble of maximally-packed dimer configurations may have a nonzero number of monomers associated with it if the disordered lattice has no perfect matchings. The ellipses refer to monomer and dimer interaction terms. Monomer interactions are restricted to be short-ranged in nature, only coupling two monomers if they are at next-nearest neighbor sites (there can be no additional nearest-neighbour monomer interactions since monomers of a maximum matching already obey a nearest-neighbor exclusion constraint). The dimer interaction terms are defined on {\em flippable} (perfectly-matched) elementary plaquettes or larger flippable loops on the lattice (for the simplest such terms, see the precise definitions given in Fig.~\ref{Quantumdimermodel}). 

Additionally, we are interested in the many-body spectrum of the corresponding maximally-packed quantum dimer models. The simplest such quantum dimer model inherits its ``potential energy'' terms from the corresponding classical dimer model defined above. In addition, it has two kinds of ``kinetic energy'' terms: ring-exchange terms that flip the state of flippable elementary plaquettes with amplitude $\Gamma$, and monomer kinetic-energy terms that hop a monomer to a next-nearest-neighbor location with amplitude $\Gamma^{'}$ (see Fig.~\ref{Quantumdimermodel} for the precise form of the Hamiltonian in the simplest case). Additional kinetic-energy terms that change the perfect matching of larger flippable loops are also permitted, as are terms that represent longer-range hopping of monomers (along an alternating sequence of unmatched and matched links).

On the other hand, we are also interested in the zero-energy wavefunctions of a quantum-mechanical particle hopping with possibly random hopping amplitudes on links of such a lattice. The corresponding Hamiltonian for a gas of non-interacting fermions at zero chemical potential $\mu=0$ reads: 
\begin{eqnarray}
H_{F} &=& -\sum_{\langle r r' \rangle} t_{r r'} c^{\dagger}_r c_{r'} + {\rm h.c.} \; ,
\label{H_fermion}
\end{eqnarray}
where $c^{\dagger}_r$ creates a canonical fermion at site $r$ of this site-diluted lattice and $t_{r r'}$  is the possibly random hopping amplitude on nearest-neighbor link $\langle r r'\rangle$ of this lattice.

Additionally, we are interested in characterizing topologically-robust zero-energy Majorana fermion excitations of a bipartite network of localized Majorana modes $\eta_r$ described by the Majorana Hamiltonian
\begin{eqnarray}
  H_{\rm Majorana} &=& i\sum_{\langle r r' \rangle} a_{r r'} \eta_{r}\eta_{r'}\; ,
\label{H_Majorana}
\end{eqnarray}
where $a_{r r'}$ is a real antisymmetric matrix that describes the mixing between Majorana modes located on neighbouring sites of a site-diluted bipartite lattice.
% Networks described by $H_{\rm Majorana}$~\cite{Alicea_etal,Laumann_Ludwig_Huse_Trebst} can potentially be realized for instance by exploiting the localized Majorana fermion excitations of topological superconductors~\cite{Read_Green,Kitaev_chain}, adding to the motivation for this question. 

In all these settings, we focus on topologically-protected aspects which depend only on the connectivity of the underlying lattice while being insensitive to the actual values of the corresponding quantum amplitudes or weights or interaction energies. To achieve this, we rely crucially on the {\em Dulmage-Mendelsohn decomposition}~\cite{Lovasz_Plummer,Dulmage_Mendelsohn,Pothen_Fan,Kavitha}  of a bipartite graph. Using this structure theory, which provides us a classification of sites of a bipartite lattice into three types, we construct two classes of non-overlapping connected regions ${\mathcal R}_{i}$ ($i=1\dots N_{\mathcal R}$) and ${\mathcal P}_j$ ($j=1 \dots N_{\mathcal P}$) that together cover the diluted lattice. These have an intuitively appealing characterization from the point of view of a maximally-packed dimer model defined on the diluted lattice: monomers of this maximally-packed dimer model are all confined to live in these ${\mathcal R}$-type regions, which are of two types: ${\mathcal R}_A$-type regions in which the monomers only live on $A$-sublattice sites, and ${\mathcal R}_B$-type regions in which the monomers only live on $B$-sublattice sites. In any maximum matching, a given ${\mathcal R}$-type region $\calR_i$ hosts the same fixed nonzero number $\calI_i$ of monomers. On the other hand, ${\mathcal P}$-type regions are parts of the lattice that are always
perfectly matched in any maximum matching.

For the maximally-packed classical dimer models defined in Eq.~\ref{Z_monomerdimer}, we establish a complete factorization of the partition function, with each ${\mathcal R}$-type and ${\mathcal P}$-type region independently contributing one factor to the partition function. As a result, correlations between monomers of the maximally-packed dimer model are strictly localized within individual ${\mathcal R}$-type regions, and dimer correlations are likewise localized within individual ${\mathcal R}$-type and ${\mathcal P}$-type regions. For the corresponding quantum dimer models, we demonstrate that an arbitrary many-body eigenstate can be written as a tensor product of eigenstates of individual ${\mathcal R}$-type and ${\mathcal P}$-type regions. As a result, the entanglement entropy across a cut that partitions the system into two halves is controlled by the typical size of the ${\mathcal R}$-type and ${\mathcal P}$-type regions that such a cut passes through. This implies that many-body eigenstates even in the middle of the spectrum have area-law entanglement entropy whenever ${\mathcal R}$-type and ${\mathcal P}$-type regions remain finite in the thermodynamic limit.

For the bipartite random hopping problem, we provide an alternative ``local'' proof of the graph-theoretic identity~\cite{Longuet-Higgins,Lovasz_Plummer}  between the number of topologically-protected zero-energy states and the number of monomers in any maximum matching of the underlying lattice. This local proof establishes that the wavefunctions of such topologically-protected zero-energy states can be chosen to live entirely within individual ${\mathcal R}$-type regions. $\calI_i$ such linearly-independent zero modes coexist on the $A$-sublattice ($B$-sublattice) sites of each ${\mathcal R}_A$-type (${\mathcal R}_B$-type) region $\calR_i$.
  
Our construction of this orthonormal basis implies a topologically-protected localization property of the basis-independent zero-energy on-shell Green function $\Delta G(r,r')$: $\Delta G(r,r')$ is nonzero if and only if $r$ and $r'$ belong to the same ${\mathcal R}$-type region. This has interesting consequences for the zero-temperature, zero-magnetic-field conductivity tensor of $H_{F}$. For the Majorana Hamiltonian $H_{\rm Majorana}$, we show that each ${\mathcal R}$-type region with an {\em odd} value for ${\mathcal I}$ generically supports a {\em single} topologically-protected zero-energy Majorana fermion excitation that is perturbatively stable to additional mixing amplitudes between further neighbors.

All of this motivates our detailed computational study of the random geometry of these ${\mathcal R}$-type and ${\mathcal P}$-type regions of site-diluted bipartite lattices in two and three dimensions. The basic picture that emerges from our results is common to the two-dimensional and three-dimensional cases, and can be summarized thus: The ${\mathcal P}$-type regions remain finite in extent in the thermodynamic limit over the entire range of dilution studied. At any nonzero $n_{\rm vac}$ in the thermodynamic limit, we find a nonzero number density $n_{\mathcal R}$ ($n_{\mathcal P}$)  of ${\mathcal R}$-type (${\mathcal P}$-type) regions in the thermodynamic limit (Fig.~\ref{NRNPmw_combined}), as well as a nonzero number density of ``odd'' ${\mathcal R}$-type regions with odd imbalance ${\mathcal I}$ (this implies a nonzero density of collective topologically-protected Majorana fermion excitations of Eq.~\ref{H_Majorana}). These ${\mathcal R}$-type (${\mathcal P}$-type) regions contain a nonzero fraction $m_{\rm tot}$ ($m_{\mathcal P}$) of sites of the undiluted lattice in the thermodynamic limit (Fig.~\ref{NRNPmw_combined}), with the ${\mathcal R}$-type regions hosting a nonzero monomer density $w$ in the maximally-packed dimer model of Eq.~\ref{Z_monomerdimer} (equivalently, a density $w$ of topologically-protected zero modes in the bipartite hopping problem of Eq.~\ref{H_fermion}). 

In the small-$n_{\rm vac}$ limit,  we find that most of the sites of the diluted lattice belong to ${\mathcal R}$-type regions, with $m_{\rm tot} \to 1-n_{\rm vac}$,  although $w \to 0$ quite rapidly as $n_{\rm vac} \to 0$ (Fig.~\ref{NRNPmw_combined}). In the range of system sizes ($L$) accessible to us, the typical size of ${\mathcal R}$-type regions is much larger than the mean spacing between zero modes (Fig.~\ref{length-scalebasic}) as $n_{\rm vac} \to 0$; in fact it appears to be only limited by the finite size $L$ (Fig.~\ref{Rgandxi}). The dominant contribution to both $m_{\rm tot}$ and $w$ in this regime comes from such large ${\mathcal R}$-type regions (Fig.~\ref{smallfraction_m} and Fig.~\ref{maxfraction_mwcubic}). Moreover, in this regime, the dominant contribution to $m_{\rm tot}^{\rm odd}$, the fraction of sites of the undiluted lattice belonging to odd ${\mathcal R}$-type regions, also comes from the large system-size-limited odd ${\mathcal R}$-type regions (see Figs.~\ref{oddsmallfraction_m},~\ref{oddmaxfraction_cubic} in Appendix). This basic picture is equally valid in both two and three dimensions.

Going beyond this basic picture, we find interesting differences between the two-dimensional and three-dimensional cases: 
In two dimensions, both ${\mathcal R}$-type and ${\mathcal P}$-type regions remain localized at any nonzero $n_{\rm vac}$ accessible to our numerical study. However, the $n_{\rm vac} \to 0$ limit exhibits {\em incipient} percolation of ${\mathcal R}$-type regions. As a result, large-scale aspects of the random geometry of ${\mathcal R}$-type regions are universal in the small-$n_{\rm vac}$ limit. We obtain a numerical estimate of ${\nu_{\rm 2d}} = 5.1 \pm 0.9$ for the correlation length exponent that characterizes this universality class, and place a bound $\eta_{\rm 2d} \lesssim 0.06$ on the value of the corresponding anomalous exponent $\eta$. A particularly interesting aspect of this incipient {\em Dulmage-Mendelsohn percolation} phenomenon is that this nontrivial behaviour is exhibited in the limit of vanishing vacancy density, raising the intriguing possibility of being accessible to a rigorous analysis in the $n_{\rm vac} \to 0^{+}$ limit (although existing results in the mathematical literature are considerably weaker in nature~\cite{Aldred_Anstee_Locke,Anstee_Blackman_Yang}).

In contrast, on the cubic lattice in three dimensions, we find that there is a nonzero dilution threshold $n_{\rm vac}^{\rm crit}=0.5956(5)$ below which (above which) ${\mathcal R}$-type regions percolate through the lattice (remain bounded in extent). In the vicinity of this {\em Dulmage-Mendelsohn percolation} transition, the random geometry of ${\mathcal R}$-type regions exhibits critical scaling behaviour. We obtain a numerical estimate of $\nu_{\rm 3d} =0.87 \pm 0.1$ for the correlation length exponent of this critical point, and place an upper bound $\eta_{\rm 3d} \lesssim 0.03$ on the corresponding anomalous exponent. Immediately below $n_{\rm vac}^{\rm crit}$ in the thermodynamic limit, there are {\em two} infinite (percolating) clusters, one ${\mathcal R}_A$-type region {\em and} another ${\mathcal R}_B$-type region. This behaviour persists throughout this intermediate-$n_{\rm vac}$ phase, until $n_{\rm vac}$ is lowered below $n_{\rm vac}^{\rm low} = 0.35(1)$, which represents the upper-limit of a small-$n_{\rm vac}$ phase with {\em spontaneous sublattice symmetry breaking}. In this low-$n_{\rm vac}$ phase, each random sample has exactly one percolating ${\mathcal R}$-type region, which can be {\em either} a ${\mathcal R}_A$-type {\em or}  a ${\mathcal R}_B$-type region.

%These Dulmage-Mendelsohn percolation phenomena in two and three dimensions clearly have interesting and robust implications for all three sets of questions: equilibrium correlation functions of the maximally-packed dimer model, zero-energy wavefunctions of the bipartite random hopping problem, particle hopping on the lattice, and the surviving Majorana excitations of bipartite Majorana networks.

Importantly, all these interesting phenomena occur well inside the geometrically percolated phase of the underlying site-diluted lattice.  Clearly, this Dulmage-Mendelsohn percolation, incipient or otherwise, is a monomer-percolation phenomenon exhibited by the ensemble of maximum matchings of such lattices. In addition, for the classical statistical mechanics of the maximally-packed dimer models of Eq.~\ref{Z_monomerdimer}, our two-dimensional results, in conjunction with the factorization property of $Z$, imply that monomer and dimer correlation functions remain {\em strictly localized} for arbitrarily small but nonzero dilution (for caveats and details, see Sec.~\ref{MD} and Sec.~\ref{subsec:analysisoverview}). This is, at least at first sight, a rather remarkable property of such two-dimensional dimer models.  Moreover, our results suggest that the corresponding monomer correlation functions could in principle exhibit critical scaling in the $n_{\rm vac} \to 0$ limit, possibly with an additional set of independent exponents that characterize the long-distance behaviour of these functions in this limit.

Similarly, in three dimensions, our results imply a  monomer-percolation transition at $n_{\rm vac}^{\rm crit}$. This separates an intermediate-dilution percolated phase from a high-dilution phase in which monomers are strictly localized. Additionally, we predict a sublattice symmetry-breaking transition of the monomer gas at $n_{\rm vac}^{\rm low}$; this transition separates a low-dilution percolated phase with spontaneous sublattice symmetry breaking from the intermediate dilution phase of the monomer gas. 

For the corresponding maximally-packed quantum dimer models, our results immediately imply that many-body eigenstates even in the middle of the spectrum have area-law entanglement entropy for any nonzero $n_{\rm vac}$ in the two-dimensional case (for caveats and details, see Sec.~\ref{QMD}, Sec.~\ref{subsec:analysisoverview} and Sec.~\ref{MBLprediction}). For the three-dimensional cubic-lattice quantum dimer model, our results imply the existence of a phase with such area-law behavior for $n_{\rm vac} > n_{\rm vac}^{\rm crit}$. In this three-dimensional case, the Dulmage-Mendelsohn percolation transition at $n_{\rm vac}^{\rm crit}$ is thus expected to correspond to an entanglement phase transition separating a $n_{\rm vac} < n_{\rm vac}^{\rm crit}$ phase with more conventional volume-law entanglement entropy from a $n_{\rm vac} > n_{\rm vac}^{\rm crit}$ area-law phase (for details and a caveat, see Sec.~\ref{QMD} and Sec.~\ref{MBLprediction}).

For systems described by $H_F$, the understanding developed here sheds considerable light on some long-standing questions in the literature on quantum percolation. Since $\Delta G(r,r')$ is only nonzero when $r$ and $r'$ both belong to the same ${\mathcal R}$-type region, our results in two dimensions imply that there is no delocalized phase and therefore no quantum percolation transition for any nonzero $n_{\rm vac}$ when the chemical potential is set to the particle-hole symmetric value $\mu = 0$. Instead, our results point to the existence of a different universality class of scaling behaviour associated with incipient wavefunction percolation in the $n_{\rm vac} \to 0$ limit (for caveats and details, see Sec.~\ref{GF}, Sec.~\ref{subsec:analysisoverview}, and Sec.~\ref{QuantumPercolation&MajoranaPercolation}). 

On the cubic lattice with chemical potential $\mu=0$, our results establish the existence of a localized phase for $n_{\rm vac} > n_{\rm vac}^{\rm crit}$. Thus, for $n_{\rm vac} \in (n_{\rm vac}^{\rm crit}, n_{\rm vac}^{\rm geom})$ (where $n_{\rm vac}^{\rm geom}$ is the threshold for ordinary site percolation on the cubic lattice), quantum percolation is forbidden although a classical fluid percolates. In conjunction with the earlier literature, our results also imply that there is a delocalization transition precisely at the Dulmage-Mendelsohn percolation threshold $n_{\rm vac}^{\rm crit}$ (for details and a caveat, see Sec.~\ref{GF} and Sec.~\ref{QuantumPercolation&MajoranaPercolation}). This provides an interesting perspective on the nature of the quantum percolation transition in three dimensions. Further, our results strongly suggest that this  delocalized phase has a sublattice symmetry-breaking transition at $n_{\rm vac}^{\rm low}$.

Our results also have interesting consequences for bipartite Majorana networks described by $H_{\rm Majorana}$. In two dimensions, our results establish the existence of a nonzero density of topologically-robust zero-energy Majorana fermion excitations which are localized for any nonzero $n_{\rm vac}$. Additionally, they suggest the existence of an incipient Majorana percolation phenomenon, whereby  these collective zero-energy Majorana fermion excitations undergo a delocalization transition in the $n_{\rm vac} \to 0$ limit (for caveats and details, see Sec.~\ref{GF}, Sec.~\ref{subsec:analysisoverview}, and Sec.~\ref{QuantumPercolation&MajoranaPercolation}). In the three-dimensional case, our results establish the presence of a nonzero density of strictly-localized zero-energy Majorana fermion excitations for $n_{\rm vac} > n_{\rm vac}^{\rm crit}$, and strongly suggest the existence of two different delocalized phases below $n_{\rm vac}^{\rm crit}$, separated from each other by the sublattice symmetry-breaking transition at $n_{\rm vac}^{\rm low}$ (for details and a caveat, see Sec.~\ref{GF} and Sec.~\ref{QuantumPercolation&MajoranaPercolation}).

The precise arguments and detailed analyses that lead to these results are presented in the next few sections:
In Sec.~\ref{DM}, we introduce the Dulmage-Mendelsohn decomposition of a bipartite graph. In Sec.~\ref{Consequences}, we  use this graph-theoretical tool to construct the connected components ${\mathcal R}_i$ and ${\mathcal P}_j$ of interest to us, and establish key properties of these ${\mathcal R}$-type and ${\mathcal P}$-type regions. In Sec.~\ref{MD}, we use these to derive consequences for monomer and dimer correlations in maximally-packed classical dimer models. In Sec.~\ref{QMD}, we discuss consequences for the entanglement entropy of arbitrary many-body eigenstates of the corresponding quantum dimer models. In Sec.~\ref{ZM}, we use these graph-theoretical ideas to provide an alternate proof of the correspondence, alluded to earlier,  between the number of zero modes and the number of monomers. In Sec.~\ref{GF} we establish a topologically-protected localization property of the on-shell zero-energy Green function of the hopping problem, and discuss consequences for the conductivity tensor of the free-fermion Hamiltonian $H_F$ (Eq.~\ref{H_fermion}) in the limit of vanishing temperature and external magnetic field. This is followed in Sec.~\ref{Majorana} by an analysis of implications for the perturbative stability of collective Majorana fermion excitations of bipartite Majorana networks. 

In Sec.~\ref{Computational}, we describe our
   computational methods and define the various geometric
    quantities of interest to us. This is followed in
     Sec.~\ref{Geometry:Basic} by our results for the basic properties of the random geometry of ${\mathcal R}$-type and ${\mathcal P}$-type regions in the thermodynamic limit. Next, we provide a detailed account of our finite-size scaling analysis of Dulmage-Mendelsohn percolation in Sec.~\ref{percolation&incipientpercolation}. This includes a precise characterization of the spontaneous breaking of sublattice symmetry in three dimensions. As noted earlier, the last two sections provide additional theoretical perspective and a more detailed discussion of the implications of our results, as well as some suggestions for promising lines of further enquiry. Finally, the Appendix is devoted to a description of the morphology of the largest ${\mathcal R}$-type regions that dominate the low dilution limit, and to a separate study of scaling properties of ${\mathcal R}$-type regions with odd monomer number ${\mathcal I}$. This separate study is motivated by the fact that such regions control the spatial structure of perturbatively-stable zero-energy Majorana fermion excitations of the Majorana networks defined in Eq.~\ref{H_Majorana}.

\section{The Dulmage-Mendelsohn decomposition}
\label{DM}
  The structure theory of bipartite graphs developed by Dulmage and Mendelsohn~\cite{Lovasz_Plummer,Dulmage_Mendelsohn,Pothen_Fan,Kavitha} is simple and elegant, but perhaps not well-known to physicists. Here, we attempt to remedy this with a self-contained account, which also sets up our notation. Our description below follows the treatment of Ref.~\cite{Kavitha}.

  Consider the maximally-packed dimer model on a bipartite graph, {\em i.e.}, with the proviso that the number of monomers is restricted to the minimum possible value. Dulmage and Mendelsohn showed that this ensemble of maximum matchings defines a  unique structural decomposition of the underlying graph, independent of the maximum matching one starts with.

  To establish this, start with any one maximum matching, which has monomers at lattice sites $h_{k}$, with $k=1 \dots W$. Here, $W$ is the number of unmatched sites in any maximum matching. Now, consider {\em alternating paths} starting from unmatched vertices that host monomers: These are paths that begin at any unmatched vertex $h_k$, traverse any one of the unmatched (unoccupied by dimers) links emanating from it and subsequently go along an alternating sequence of matched and unmatched links of the lattice without visiting any site more than once or traversing any link more than once. Maximum matchings are characterized by the absence of {\em augmenting paths}, {\em i.e.}, the absence of alternating paths of odd length which start and end at unmatched sites~\cite{Lovasz_Plummer}.   Indeed, if there was such a path, one could have added one more dimer to the system by switching the occupied and unoccupied links of this path, and the original matching would not have maximum cardinality.

Using this, one can classify sites into three groups, odd (o-type), even (e-type) and unreachable (u-type): Unreachable sites cannot be reached by such an alternating path starting from any monomer. o-type sites can be reached by an alternating path of odd length starting from some monomer. And even sites are those that can be reached by an alternating path of even length starting from some monomer; this class also includes the unmatched sites themselves. These three groups of sites are disjoint. To see this, we first note that the definitions themselves imply that the set of u-type sites is disjoint from the other two sets. Additionally, in a bipartite lattice, the same site cannot be reached both by even-length and odd-length alternating paths. For if this were possible, one would either have an augmenting path connecting two monomers, or have an odd-length alternating cycle (simple loop). Neither can exist since we are working with a {\em maximum} matching of a {\em bipartite} graph. 

Further, it is easy to check that this decomposition is unique, no matter which maximum matching one starts with. To see this, one notes that the overlap of two maximum matchings consists of edges that are matched (covered by dimers) by both maximum matchings, vertices that are left unmatched by both maximum matchings, overlap loops, and overlap paths. Here, overlap loops are cycles whose links are alternately covered by dimers belonging to the two maximum matchings that have been superposed. Similarly, overlap paths are paths that start at an unmatched site of one maximum matching and end at an unmatched site of the other maximum matching; this is guaranteed by the fact that both matchings are maximum. These paths also have links alternately covered by dimers belonging to the two maximum matchings.

Sites left unmatched by both maximum matchings, and sites at the ends of links covered by dimers in both maximum matchings clearly have the same type label (e-type, o-type, or u-type) starting with either maximum matching. Sites on the overlap loops also have the same type label starting from either maximum matching. Again, this is because the lattice is bipartite and all such cycles have even length. For the paths, we have already noted that they must start at an unmatched site of one maximum matching and end at an unmatched site of the other maximum matching. They must therefore have an equal number of matched edges (dimers) from each maximum matching and be even in length. Therefore, sites on the paths also have the same type label starting from either maximum matching.

Thus, this classification into odd, even and unreachable represents a fundamental structural property of the lattice itself. Additionally, it is clear from the definitions that every o-type site must be paired with some e-type site by a dimer in any maximum matching.
Likewise, in any maximum matching, every u-type site must always be matched to another u-type site by a dimer, while an e-type site can either be unmatched or matched with some o-type site by a dimer. 

Next we note that two e-type sites cannot be connected by a nearest-neighbor link of the lattice. For if such a link were present, it would either imply the existence of an odd cycle in the lattice, or the existence of an augmenting path starting from a monomer in our matching; since we are dealing with a maximum matching of a bipartite lattice, neither is possible. 

Additionally, an e-type site cannot be a neighbor of a u-type site either. To see this one first recalls from the above that any link between a u-type site and e-type site cannot have a dimer on it, since the e-type site is reached from a monomer of a maximum matching by an alternating path that ends in a matched link (covered by a dimer), and is therefore matched with an o-type site. Given this, it is clear that the existence of such a link, which must necessarily be unmatched, would allow for this alternating path to be extended to reach the u-type site via this unmatched link, contradicting the fact that it is unreachable.

  \section{Implications}
  \label{Consequences}
  We now present five arguments that use this structural decomposition to provide information on i) monomer and dimer correlation functions of the maximally-packed classical dimer models defined in Eq.~\ref{Z_monomerdimer}, ii) entanglement entropy of eigenstates of the corresponding maximally-packed quantum dimer models, iii)
  topologically-protected zero mode wavefunctions in the free-fermion problem Eq.~\ref{H_fermion}, iv) the corresponding on-shell zero-energy Green function
  $\Delta G$ and the conductivity tensor for the free-fermion Hamiltonian Eq.~\ref{H_fermion},  and v) the perturbative stability of zero-energy Majorana fermion excitations in the corresponding bipartite Majorana networks with Hamiltonian Eq.~\ref{H_Majorana}.
  
All our arguments rely on using the Dulmage-Mendelsohn classification of sites into o-type, e-type and u-type sites to define a further decomposition of the lattice into non-overlapping connected regions ${\mathcal R}_{i}$ ($i=1\dots N_{\mathcal R}$) and ${\mathcal P}_j$ ($j=1 \dots N_{\mathcal P}$). We define these as follows: Color all o-type and e-type sites red. Also color red all links that connect any o-type site to any e-type site. Color all u-type sites blue. Also color blue all links between two u-type sites. And delete all links between u-type sites and o-type sites (e-type sites are never neighbors of u-type sites), as well as all links between two o-type sites (two e-type sites never have a link between them).
  Our lattice now decomposes into $N_{\mathcal R} +N_{\mathcal P}$ connected components (due to the deletion of links described above):  Red components ${\mathcal R}_i$ ($i=1 \dots N_{\mathcal R}$) consisting of connected ${\mathcal R}$-type regions, and blue components ${\mathcal P}_j$ ($j=1 \dots N_{\mathcal P}$) consisting of connected ${\mathcal P}$-type regions. From the definitions, it is straightforward to see that no monomers live in the ${\mathcal P}$-type regions, which thus represent parts of the lattice that are always perfectly matched by any maximum matching. It is also clear that monomers of any maximum matching always live on e-type sites inside a ${\mathcal R}$-type region.

Additionally, we note that the boundary sites of any ${\mathcal R}$-type region, {\em i.e.,} those sites which have neighbors belonging to other ${\mathcal R}$-type or ${\mathcal P}$-type regions,  are always of o-type. Further, since all o-type sites of a ${\mathcal R}$-type region are matched to an e-type site of the same region, and monomers live only on e-type sites, this implies that the number of e-type sites in any ${\mathcal R}$-type region exceeds the number of o-type sites {\em in spite of all boundary sites being o-type}. Additionally, we note that these ${\mathcal R}$-type regions come in two ``flavours'' ${\mathcal R}_A$ and ${\mathcal R}_B$: ${\mathcal R}_A$-type  ( ${\mathcal R}_B$-type) regions have all their e-type sites on the $A$ ($B$) sublattice and all their o-type sites on the $B$ ($A$) sublattice.
  
Finally, we note that each ${\mathcal P}$-type region ${\mathcal P}_j$ itself can be uniquely decomposed into subregions defined to ensure that each overlap loop (in the ensemble of overlap loops obtained by superimposing any two perfect matchings of ${\mathcal P}_j$) lives entirely within a single subregion of ${\mathcal P}_j$. This is related to the so-called ``fine decomposition'' of Dulmage and Mendelsohn used by Pothen and Fan in their algorithm for block triangularization of matrices~\cite{Dulmage_Mendelsohn,Pothen_Fan}.
 As we will see in our subsequent discussion, this additional ``fine structure'' of ${\mathcal P}$-type regions can play a potentially crucial role in determining the dimer correlation length of the maximally-packed dimer model, especially  if ${\mathcal P}$-type regions are large in size.

\subsection{Maximally-packed classical dimer models: Monomer and dimer correlations}
\label{MD}

We now consider the statistical mechanics of the maximally-packed dimer model, Eq.~\ref{Z_monomerdimer}. Our discussion relies on three crucial observations: i) Any alternating path starting from an unmatched site of a maximum matching lies entirely within a single ${\mathcal R}$-type region. In other words, the number of monomers inside any given ${\mathcal R}$-type region remains the same in all maximum matchings. ii) The links deleted during our construction (of connected ${\mathcal R}$-type and ${\mathcal P}$-type regions) never host a dimer in any maximum matching of the full lattice. To establish this, we simply note that the boundary sites of any ${\mathcal R}$-type region are o-type sites, which must be matched to e-type sites within that region itself, and all sites of any ${\mathcal P}$-type region are matched among themselves within that region. iii) Since monomers only live on e-type sites of a ${\mathcal R}$-type region, and the boundary of ${\mathcal R}$-type regions is made up entirely of o-type sites, two monomers on next-nearest neighbor sites must lie in the same ${\mathcal R}$-type region. 

The last observation implies that each monomer interaction term in the maximally-packed dimer models defined in Eq.~\ref{Z_monomerdimer} always acts entirely within any one ${\mathcal R}$-type region.  On the other hand, the second observation implies that each {\em flippable} (perfectly matched)  elementary plaquette or larger flippable loop must lie entirely within a single ${\mathcal R}$-type or ${\mathcal P}$-type region. As a result, each dimer interaction term in Eq.~\ref{Z_monomerdimer} also acts entirely within a single ${\mathcal R}$-type or ${\mathcal P}$-type region.
Therefore, the ensemble of maximum matchings can be generated from any one maximum matching by making all possible rearrangements of dimers (keeping their numbers fixed)
{\em independently} {\em within} each connected component ${\mathcal R}_i$ and each connected component ${\mathcal P}_j$. 

Indeed, the classical statistical mechanics of the corresponding maximally-packed dimer models defined in Eq.~\ref{Z_monomerdimer} can be studied by separately considering each connected component ${\mathcal R}_i$ and ${\mathcal P}_j$. Each ${\mathcal P}_j$ is guaranteed to independently host a fully-packed dimer model, with its sites always being perfectly matched amongst themselves in any maximum matching, and inter-dimer interactions acting entirely within the region. On the other hand, since each ${\mathcal R}$-type region hosts the same fixed nonzero number of monomers in any maximum matching, one has an independent maximally-packed dimer model defined on each ${\mathcal R}$-type region, again with dimer and monomer interaction terms acting entirely within each region.  

The classical partition function $Z$ of such maximally-packed dimer models thus factorizes: 
\begin{eqnarray}
Z&=&\left(\prod_i Z_{{\mathcal P}_i} \right ) \times \left(\prod_j Z_{{\mathcal R}_j} \right)\; .
\label{factorization}
\end{eqnarray}
 This factorization is topological in nature, in the sense that it is independent of numerical values of the nonzero bond fugacities assigned to each link and only depends on the pattern of nearest-neighbor links of the graph. It immediately implies that monomer and dimer correlations do not extend beyond individual connected components. 
The foregoing implies that the typical size of a ${\mathcal R}$-type region places an upper bound on the length scale over which correlations can propagate in the gas of monomers.

For dimer correlations, the situation is more complicated, for  the average two-point function of dimers has two contributions that could be very different from each other: One arising from dimer correlations within each ${\mathcal R}$-type region (which hosts a maximally-packed dimer model) and the other arising from perfectly-matched ${\mathcal P}$-type regions (each of which hosts a fully-packed dimer model). In the site-diluted lattices  we study, it is not {\em a priori}  obvious which of these dominates the long-distance behaviour of the average two-point dimer correlator. The issue is the following: Although the ${\mathcal P}$-type regions turn out to be typically much smaller than ${\mathcal R}$-type regions in the regimes of interest to us, there is no {\em a priori} guarantee
that their contribution is negligible compared to that of ${\mathcal R}$-type regions. This is because the dimer correlation
length in the ${\mathcal R}$-type regions can potentially be much smaller than the typical size of such a region, due to
the presence of a coexisting gas of monomers. 

Clearly, a detailed numerical study of this question would depend on lattice-level details and is somewhat removed from the focus of the present work. Here, we confine ourselves to noting that this question becomes tricky if the dimer correlation length in ${\mathcal R}$-type regions is of the same order of magnitude as the typical size of ${\mathcal P}$-type regions. In this case, the fine decomposition of ${\mathcal P}$-type regions~\cite{Dulmage_Mendelsohn,Pothen_Fan} comes into play: As we have noted in the previous section, each ${\mathcal P}$-type region itself can be decomposed into subregions that  are defined to ensure that overlap loops between two perfect matchings of ${\mathcal P}$ live entirely within a single subregion. This implies a further factorization of each $Z_{{\mathcal P}_i}$ in Eq.~\ref{factorization}, and means that dimer correlations cannot extend beyond these individual subregions identified by the fine decomposition. In this case, it is the typical size of these subregions that
potentially plays the key role in determining the dimer correlation length.

\subsection{Maximally-packed quantum dimer models: Entanglement entropy of arbitrary eigenstates}
\label{QMD}

This analysis also leads to an interesting conclusion about entanglement properties of eigenstates of a natural quantum version of the maximally-packed dimer model on such diluted lattices, whose Hamiltonian has been displayed in Fig.~\ref{Quantumdimermodel} and defined in the accompanying discussion in Sec.~\ref{Executivesummary}. 
%The Hamiltonian of such a 
%maximally-packed quantum dimer model has two kinds of off-diagonal ``kinetic energy'' terms: usual {\em ring-exchange} terms that act on flippable elementary plaquettes of the lattice and flip the state of such plaquettes with amplitude $\Gamma$, and monomer-hopping terms that move a monomer to a next-nearest-neighbor site along some alternating path with amplitude $\Gamma^{'}$. In addition, there are the usual diagonal ``potential energy'' terms that act on these flippable elementary plaquettes in a maximum matching and assign an energy $V$ to such plaquettes.

From the properties of ${\mathcal R}$-type and ${\mathcal P}$-type regions established in  Sec.~\ref{DM}, and the analysis of the classical case presented in the preceding section, it is clear that all the terms in the Hamiltonian of such a quantum dimer model also act entirely within a single ${\mathcal R}$-type or ${\mathcal P}$-type region. This includes dimer potential-energy and kinetic-energy terms defined on arbitrary flippable loops, next-nearest neighbor monomer interactions, and any monomer kinetic-energy terms that hop a monomer by two or more units along an alternating path starting at that unmatched site. An immediate consequence of this is that every eigenstate of such quantum dimer models can be written as a tensor product of some eigenstates of the individual ${\mathcal P}$-type and ${\mathcal R}$-type regions:
\begin{eqnarray}
|\Psi \rangle &=& \otimes_{i} |\psi_{{\mathcal P}_i}\rangle \otimes_{j} |\psi_{{\mathcal R}_j}\rangle  \; .
\label{productstate}
\end{eqnarray}
If these regions remain finite in extent in the thermodynamic limit, this immediately implies that $|\Psi\rangle$ cannot have volume-law entanglement entropy even if it is in the middle of the many-body spectrum. Indeed, in this case, the entanglement entropy of $|\Psi\rangle$ across a cut that partitions the system into two halves must have area-law scaling due to its tensor product structure. The prefactor of this area-law scaling will clearly be determined by the typical sizes of the ${\mathcal R}$-type and ${\mathcal P}$-type regions through which this cut passes.
This conclusion holds independent of the values of $\Gamma$, $\Gamma^{'}$ and $V$, including in the presence of quenched randomness in their values. 

We emphasize again that this tensor product structure of arbitrary many-body eigenstates is not entirely fragile: Any ring-exchange or dimer potential-energy terms that act on any flippable loops preserve this structure, as do next-nearest-neighbor interactions between the monomers, and monomer kinetic-energy terms that move a monomer along an alternating path. As noted in the preceding section, this is because any such flippable loop or alternating paths must lie entirely within a single ${\mathcal P}$-type or ${\mathcal R}$-type region, and two monomers on next-nearest-neighbor sites must also be in the same ${\mathcal R}$-type region. However, an interaction between monomers on next-next-nearest neighbor sites can couple a ${\mathcal R}_A$-type region to a neighboring ${\mathcal R}_B$-type region. Therefore, in presence of such extended inter-monomer interactions, many-body eigenstates are no longer guaranteed to have this tensor-product structure.

  \subsection{Bipartite random-hopping models: Topologically-protected zero modes}
  \label{ZM}

  We now discuss topologically-protected zero modes of the hopping problem. By topologically-protected zero modes, we mean zero modes whose existence is robust to changes in the actual values of the hopping amplitudes, and depends only on the connectivity of the graph, {\em i.e.}, on whether a given hopping amplitude is zero or nonzero. Such modes are robust to bond disorder, since their existence is unaffected by randomness or modulations in the hopping amplitudes.

In Ref.~\cite{Sanyal_Damle_Motrunich_PRL}, the nonzero density of zero modes of the tight-binding model for diluted graphene was argued to arise from the presence of so-called ${\mathcal R}_{A}$-type ( ${\mathcal R}_{B}$-type) regions of the site-diluted sample, having more $A$ ($B$) sites than $B$ ($A$) sites, but a boundary consisting of only $B$ ($A$) sites.
  From the foregoing discussion, it is now clear that the connected components ${\mathcal R}_i$ defined here provide a consistent and convenient (not to mention elegant) construction of these ${\mathcal R}$-type regions of 
  Ref.~\cite{Sanyal_Damle_Motrunich_PRL}, thus justifying the commonality of nomenclature.

  This allows us to now make a precise argument for the number and structure of such topologically-protected zero modes:
  Let the number of o-type sites in any one particular ${\mathcal R}$-type region be $N_{\rm o}$ and the number of e-type sites be $N_{\rm e}$, with their difference ${\mathcal I}= N_{\rm e} -N_{\rm o}$ being a positive number equal to the number of monomers hosted by this region in any maximum matching. We now establish that there are ${\mathcal I}$ linearly-independent topologically-protected solutions of the zero-energy Schr\"{o}dinger equation with the property that the corresponding wavefunctions are only nonzero on e-type sites within this one region. To establish this, we write down the system of Schr\"{o}dinger equations that must be satisfied by any zero-energy wavefunction of this form, and show that this is a rectangular system with $N_{\rm e}$ variables and $N_{\rm o}$ equations. 
  
To see this, first note that the Schr\"{o}dinger equation on all e-type sites of this ${\mathcal R}$-type region is trivially satisfied since the wavefunction is zero on all the neighbours of each of these e-type sites. This is true because e-type sites always have o-type neighbours. This is guaranteed by the fact that there are no e-type sites on the boundary of a ${\mathcal R}$-type region (as we have seen earlier, the boundary consists entirely of o-type sites). Further, all e-type neighbours of each o-type site belonging to this region, including e-type neighbours of o-type sites on the boundary of this region, lie {\em within} this ${\mathcal R}$-type region. The zero energy Schr\"{o}ndinger equations to be satisfied by our wavefunction therefore reduce to $N_{\rm o}$ constraints (one on
each of the o-type sites of this region) that must be satisfied by $N_{\rm e}$ variables (corresponding to the wavefunction amplitudes on the $N_{\rm e}$ e-type sites of this region).
  
  The minimum number of linearly independent solutions of this system of equations equals $N_{\rm e} -r_{\rm max}$, where
  $r_{\rm max}$ is the maximum rank of the corresponding rectangular matrix with $N_{\rm e}$ columns and $N_{\rm o}$ rows. Since $N_{\rm o} < N_{\rm e}$, $r_{\rm max} \leq N_{\rm o}$, implying the existence at least $N_{\rm e} -N_{\rm o} \equiv {\mathcal I}$ linearly independent solutions to these equations, independent of the precise values of the nonzero hopping amplitudes. Thus, we have established that each ${\mathcal R}$-type region ${\mathcal R}_i$ constructed using the structure theory of Dulmage and Mendelsohn contributes exactly ${\mathcal I}_i$ topologically-protected zero modes in the quantum mechanics of a particle hopping on the lattice. The total number of topologically-protected zero modes is thus
  $W=\sum_i {\mathcal I}_i$, {\em i.e.}, exactly equal to the number of monomers in any maximum matching of the lattice.

 This argument provides an alternate proof of the well-known graph-theoretic identity between the number of topologically-protected zero modes  $W$ and the number of monomers in any maximum matching~\cite{Longuet-Higgins,Lovasz_Plummer}. In contrast to the standard ``global'' approaches~\cite{Longuet-Higgins,Lovasz_Plummer} that makes use of determinants to prove this, our ``local'' argument uses the structure of the connected components ${\mathcal R}_i$ to provide a constructive proof. In the process, we uncover a crucial aspect of the topologically-protected zero-energy eigenspace, namely, the fact that it is possible to choose a basis such that each zero-energy wavefunction of this basis lives entirely within one ${\mathcal R}$-type region, with ${\mathcal I}_i$ such basis functions co-existing in region ${\mathcal R}_i$ (for $i=1 \dots N_{\mathcal R}$).

  \subsection{Quantum percolation: On-shell zero-energy Green function and conductivity tensor}
  \label{GF}
Next, we consider the on-shell zero-energy Green function: $\Delta G(r,r') \equiv \sum_{\alpha} \psi_{\alpha}(r) \psi_{\alpha}^{*}(r') + h.c.$, where the sum on $\alpha$ is over $W$ orthonormal zero-energy eigenfunctions that make up any choice of basis for the zero-energy eigenspace, and $r$ and $r'$ represent position coordinates of points on the lattice. By construction, $\Delta G$ defined in this manner is independent of this choice of basis. It is therefore particularly convenient to evaluate it using the basis described above, {\em i.e.},  consisting of ${\mathcal I}_i$ orthonormal wavefunctions that live entirely within each ${\mathcal R}$-type region ${\mathcal R}_i$ for $i= 1 \dots N_{\mathcal R}$. There is of course considerable residual freedom in choosing an orthonormal set to span this ${\mathcal I}_i$ dimensional subspace of wavefunctions that live within a single region ${\mathcal R}_i$. But again, $\Delta G$ is of course independent of these choices. 

Employing such a basis, we see that $\Delta G(r,r')$ in only nonzero if $r$ and $r'$ both belong to the same ${\mathcal R}$-type region. 
Thus, our argument identifies a {\em generic} maximally-localized choice of zero-energy eigenbasis whose localization property is topological in nature and implies a corresponding topologically-protected localization property of the on-shell zero-energy Green function. While the behaviour of $\Delta G$  within any given ${\mathcal R}_i$ will depend on the strength of the disorder in the hopping amplitudes in that region, our argument shows that $\xi_{\rm G}$, the localization length of $\langle |\Delta G|^2 \rangle(r-r')$ (where the angular brackets represent averaging over quenched disorder) must be bounded above by the typical size $\xi$ of ${\mathcal R}$-type regions: $\xi_{\rm G} < \xi$.

To appreciate the consequences of this for transport properties of Eq.~\ref{H_fermion} at the particle-hole symmetric chemical potential $\mu =0$, we recall the careful analysis in Ref.~\cite{Baranger_Stone} of the conductivity tensor ${\underline{\sigma}}(r,r')$ of free-fermion systems. In the limit of vanishing temperature and magnetic field, ${\underline{\sigma}}(r,r')$ has an expression given entirely in terms of $\Delta G(r,r')$. As a result, the localization length $\xi_{\rm loc}$ that governs the spatial dependence of $\langle {\underline{\sigma}} \rangle (r-r')$ (where the angular brackets again denote averaging over quenched disorder) must also be bounded above by the typical size of ${\mathcal R}$-type regions: $\xi_{\rm loc} < \xi$.

Thus, if ${\mathcal R}$-type regions remain bounded in size, this implies that the corresponding fermionic system is guaranteed to be an insulator at $\mu = 0$ in the zero temperature limit. On the other hand, if the typical size of ${\mathcal R}$-type
regions diverges, $\xi_{\rm loc}$ and $\xi_{\rm G}$ could still be finite due to the localized character of the zero mode wavefunctions inside the largest ${\mathcal R}$-type regions in a sample. Although this cannot be ruled out, one expects~\cite{Evers_Mirlin,Motrunich_Damle_Huse_GadeWegnerPRB} that zero energy wavefunctions of bipartite random hopping problems are not exponentially localized, suggesting that a phase with delocalized ${\mathcal R}$-type regions likely corresponds to a delocalization transition in the transport properties of the free fermions as well.

\subsection{Bipartite Majorana networks: Zero-energy Majorana fermion excitations}
\label{Majorana}
Our understanding of the spatial structure of topologically-protected zero mode wavefunctions also
leads us to an interesting conclusion regarding the nature of collective zero-energy Majorana fermion excitations of bipartite Majorana networks~\cite{Alicea_etal,Laumann_Ludwig_Huse_Trebst} modeled by the Hamiltonian:
\begin{eqnarray}
  H_{\rm Majorana} &=& i\sum_{\langle r r' \rangle} a_{r r'} \eta_{r} \eta_{r'}\; ,
  \label{H_Majorana_repeat}
\end{eqnarray}
where $r$, $r'$ represent the locations of individual Majorana modes, $a_{r r'}$ is a real antisymmetric matrix that represents the mixing of nearest-neighbor Majorana modes in this bipartite network, and we have neglected quartic interaction terms whose effects~\cite{Affleck_Rahmani_Pikulin,Li_Franz} are assumed unimportant in our analysis below. 
Specifically, we demonstrate that ``odd'' ${\mathcal R}$-type regions (with odd values of ${\mathcal I}$) are expected to generically host a single topologically-protected  zero-energy Majorana fermion excitation of the bipartite Majorana network described by Eq.~\ref{H_Majorana_repeat}. This mode survives the leading effects of additional next-nearest-neighbor couplings. In contrast, ${\mathcal R}$-type regions with even ${\mathcal I}$ do not generically host such robust zero-energy Majorana fermion excitations. 

To see this, we first use the bipartite structure of the network to define a new basis, whose $B$-sublattice components have a phase $\exp(i\pi/2)$ relative to the corresponding components of the original basis in which $a_{rr'}$ is written. The $A$-sublattice components of the new basis are identical to those of the original basis. In this new basis, the matrix of unperturbed mixing amplitudes is now a real symmetric matrix $t^{(0)}_{rr'}$ that defines a bipartite random hopping problem. The collective zero-energy Majorana excitations of the bipartite network are now obtained from the zero modes of this real symmetric matrix, which has matrix elements $t^{(0)}_{r_A r_B} = t^{(0)}_{r_B r_A} = a_{r_A r_B}$. Thus, for bipartite networks whose geometry is that of a diluted bipartite lattice, this maps to a special case of the bipartite random hopping problem on such lattices (with {\em real} hopping amplitudes)~\cite{Motrunich_Damle_Huse_PRB1}.

Next, we consider the perturbative effect of weak additional mixing amplitudes that couple nearby modes on the same sublattice. In other words, we write $i\tilde{a}_{r r'} = ia_{r r'} + i\delta a_{r r'}$, where the real antisymmetric matrix $\delta a$ represents weak mixing between modes on nearby sites of the same sublattice of this network. In the new basis, we have a random hopping problem defined by the hermitean hopping amplitudes $t_{rr'} = t^{0}_{rr'} + \delta t_{r r'}$ with
\begin{eqnarray}
&&t^{(0)}_{r_A r^{'}_A} = t^{(0)}_{r_B r^{'}_B}  = 0 \; ,\nonumber \\
&&t^{(0)}_{r_A r_B}  = t^{(0)}_{r_B r_A} = a_{r_A r_B} \; ,\nonumber \\
&& \delta t_{r_A r_B} = 0 \; , \nonumber \\
&&\delta t_{r_A r^{'}_A} =i \delta a_{r_A r^{'}_A} \; ,\nonumber \\
&&\delta t_{r_B r^{'}_B} = i \delta a_{r_B r^{'}_B} \; .
\label{Majorana_perturbedinnewbasis}
\end{eqnarray}

To understand the effects of this perturbation on the collective zero-energy Majorana fermion excitations to leading order in perturbation theory,
we study the effect of $\delta t$ on the zero modes of the unperturbed random hopping problem. In order to do this, we must project the perturbation $\delta t$ into the zero-energy subspace of $t^{(0)}_{r r'}$ and diagonalize the resulting matrix.
This is greatly facilitated by thinking in terms of the maximally-localized choice of basis described in the previous section, with each unperturbed zero-energy basis state living entirely within any one ${\mathcal R}$-type region, and ${\mathcal I}_i$ such basis states coexisting in region ${\mathcal R}_i$ for $i=1 \dots N_{\mathcal R}$. Further, since $t^{(0)}$ is a real symmetric matrix, these basis vectors are real in this particular case.

Consider first a perturbation $\delta t$ which mixes next-nearest neighbor sites on the same sublattice, but does not have any other nonzero matrix elements. Observe that all e-type sites (in the language of Sec.~\ref{DM}) of a ${\mathcal R}_A$ (${\mathcal R}_B$) region belong to the $A$ ($B$) sublattice, and boundary sites of ${\mathcal R}_A$ (${\mathcal R}_B$) regions are all o-type, and belong to the $B$ ($A$)
sublattice. Additionally, recall that the basis states have nonzero amplitudes only on e-type sites. As a result of this structure, the projection of $\delta t_{r r'}$ into the unperturbed zero-energy subspace has a block-diagonal form in this basis if $\delta t_{r r'}$ is zero beyond next-nearest neighbor sites (in the sense of connectivity, not geometric distance). Indeed, when $\delta t_{r r'}$ does not extend beyond next-nearest neighbor sites, this projection decomposes into $N_{\mathcal R}$ independent blocks corresponding to the ${\mathcal R}$-type regions of the lattice, with
each region ${\mathcal R}_i$ contributing a block of
size ${\mathcal I}_i \times {\mathcal I}_i$ for $i= 1 \dots N_{\mathcal R}$.

Further analysis depends crucially on the fact that each of these ${\mathcal I}_i \times {\mathcal I}_i$ blocks inherits the ``chiral'' symmetry of the original problem~\cite{Motrunich_Damle_Huse_PRB1}, which guarantees that a positive eigenvalue $+\lambda$ always has a partner at $-\lambda$. To see this is true, note that $\delta t$ is pure imaginary and antisymmetric, and all states in our chosen basis of unperturbed zero modes are real, implying that each of these blocks is a pure imaginary antisymmetric matrix.  As a result of this, one can immediately conclude that every region
${\mathcal R}_i$ with odd ${\mathcal I}_i$ will host {\em at least one} collective Majorana zero mode that survives the leading-order
effects of any perturbation $\delta t_{r r'}$ that does not extend beyond next-nearest neighbors. This lends additional significance to ${\mathcal R}$-type regions with odd imbalance ${\mathcal I}$.
These robust zero-energy Majorana excitations hosted by such ${\mathcal R}$-type regions with odd ${\mathcal I}$ are also potentially stable to the leading effects of additional longer-range but weak mixing terms. However, this depends on the detailed morphology of the
${\mathcal R}$-type regions and the range of the perturbation.

Finally, we note that the argument sketched here is somewhat analogous to earlier results in the one-dimensional case of Majorana wires (as exemplified by paradigmatic Kitaev chain~\cite{Kitaev_chain}). In the thermodynamic limit of the original Kitaev chain, there is one Majorana mode at each end of the chain in the topological phase. Together, these two form a single complex fermion whose wavefunction is split between the two ends of the chain, and whose energy goes to zero exponentially in the thermodynamic limit. In multichannel generalizations of this situation, one finds that Majorana modes at one end of the wire can generically mix due to additional local mixing terms in the Hamiltonian, leading to a situation in which each end of the wire has either a single Majorana mode or none, depending on whether the number of original Majorana modes at each end was odd or even~\cite{Potter_Lee}.

\section{Computational methods and observables}
\label{Computational}

 The arguments presented in Sec.~\ref{Consequences} motivate our computational study of the random geometry of the Dulmage-Mendelsohn decomposition of site-diluted bipartite lattices. We focus on random dilution of $L\times L$ honeycomb (square) lattices consisting of $N_{\rm sites} = 2L^2$ ($N_{\rm sites} = L^2$) sites in two dimensions, and cubic lattices with $N_{\rm sites} = L^3$ sites in three dimensions, with periodic boundary conditions and  even $L$. In all our studies, we restrict ourselves to samples with compensated dilution, so that each random sample in our ensemble has an equal number of $A$ and $B$ sublattice sites. This ensures that our dataset is not ``contaminated'' by spurious effects associated with a global imbalance in the number of $A$ and $B$-sublattice sites.

We consider two disorder ensembles in the two-dimensional case: In one of these,  we have chosen to impose a nearest and next-nearest neighbour exclusion constraint on the position of the vacancies to largely eliminate the possibility of subregions of the lattice disconnecting completely from the rest of the graph at small values of the dilution $n_{\rm vac}$. In the other ensemble, each site can independently be replaced by a vacancy, with dilution probability $n_{\rm vac}$. With the global compensation constraint in place in both ensembles, we find that the universal aspects we focus on here are independent of the exclusion constraints. Therefore, in the two-dimensional case, we only present the results for the ensemble with exclusion constraints. Likewise, we confine ourselves to a study of the independently-diluted case in three dimensions, again with global compensation.

%the system typically has multiple disconnected components. Some fraction of these disconnected components, especially the smaller disconnected components, are of course imbalanced, and have additional trivial zero modes associated with this imbalance. However, the small values of dilution $n_{\rm vac}$ considered here place the sample deep within the percolated phase from the point of view of geometric percolation. As a result, the effects we focus on are dominated by the contribution of the largest geometric component, and are largely insensitive to this difference in correlations of the vacancy ensemble. The  exception is  some non-critical quantities (such as $N_{\mathcal R}$, the total number of ${\mathcal R}$-type regions) that depend sensitively on the lack of exclusion constraints amongst vacancies. 

Our tests of the efficiency of various maximum-matching algorithms described in Ref.~\cite{Duff_Kaya_Ucar} suggest that the Breadth-First-Search (BFS) algorithm with pruning of search branches outperforms the others (including the Hopcroft-Karp algorithm which has the theoretical advantage in terms of worst-case complexity) in the two-dimensional case at small dilution. On the other hand, on the diluted cubic lattice, we find that the Pothen-Fan algorithm outperforms both the Hopcroft-Karp and the BFS algorithm for vacancy densities smaller than $\approx 0.2$, with the BFS algorithm being the worst of the three in this regime. However, the performance of the Pothen-Fan algorithm deteriorates very rapidly and dramatically when the vacancy density is increased beyond $\approx 0.2$. For these higher vacancy densities, we have to rely on either the Hopcroft-Karp algorithm or the BFS algorithm. For vacancy concentrations up to  $\approx 0.35$, the Hopcroft-Karp algorithm outperforms the BFS algorithm, while the BFS algorithm outperforms the Hopcroft-Karp algorithm by a small margin for vacancy densities above $\approx 0.35$. 

To increase the computational efficiency of our matching code, we choose to use the maximum matching at a lower (higher) vacancy concentration to obtain an initial condition for the matching algorithm at the next higher (lower) concentration, and work out way up (down) a grid of concentrations; the two-dimensional data with exclusion constraints is obtained by working up along an ascending sequence of $n_{\rm vac}$, while most of the data without exclusion constraints is obtained by going down along a descending sequence of $n_{\rm vac}$. This gives us one random sample at each concentration. This process is then repeated  many times to generate our ensemble. For most of the data shown, we use an ensemble consisting of at least $3000$ samples at the largest sizes, with smaller size data being obtained by averaging over a substantially larger number of samples. In both two and three dimensions, we are limited by the time needed to find a maximum matching to sizes that correspond to $N_{\rm sites} \lesssim 10^9$.

Once we have constructed a maximum matching of the diluted lattice corresponding to a given sample, we can construct alternating paths starting from the unmatched sites to obtain the connected ${\mathcal R}$-type regions ${\mathcal R}_i$ ($i=1 \dots N_{\mathcal R}$) that contain the monomers of any maximum matching. In practice, we find it convenient to first use these paths to label each site odd (o-type), even (e-type) or unreachable (u-type) to obtain the Dulmage-Mendelsohn decomposition defined in Sec.~\ref{DM}. Based on this labeling, we then use a simple and efficient burning algorithm to construct the connected ${\mathcal R}$-type and ${\mathcal P}$-type regions defined in Sec.~\ref{Consequences}. With this in hand, we measure a number of statistical properties of the random geometry of these regions.
\begin{figure*}
  \begin{tabular}{cc}
	a) \includegraphics[width=\columnwidth]{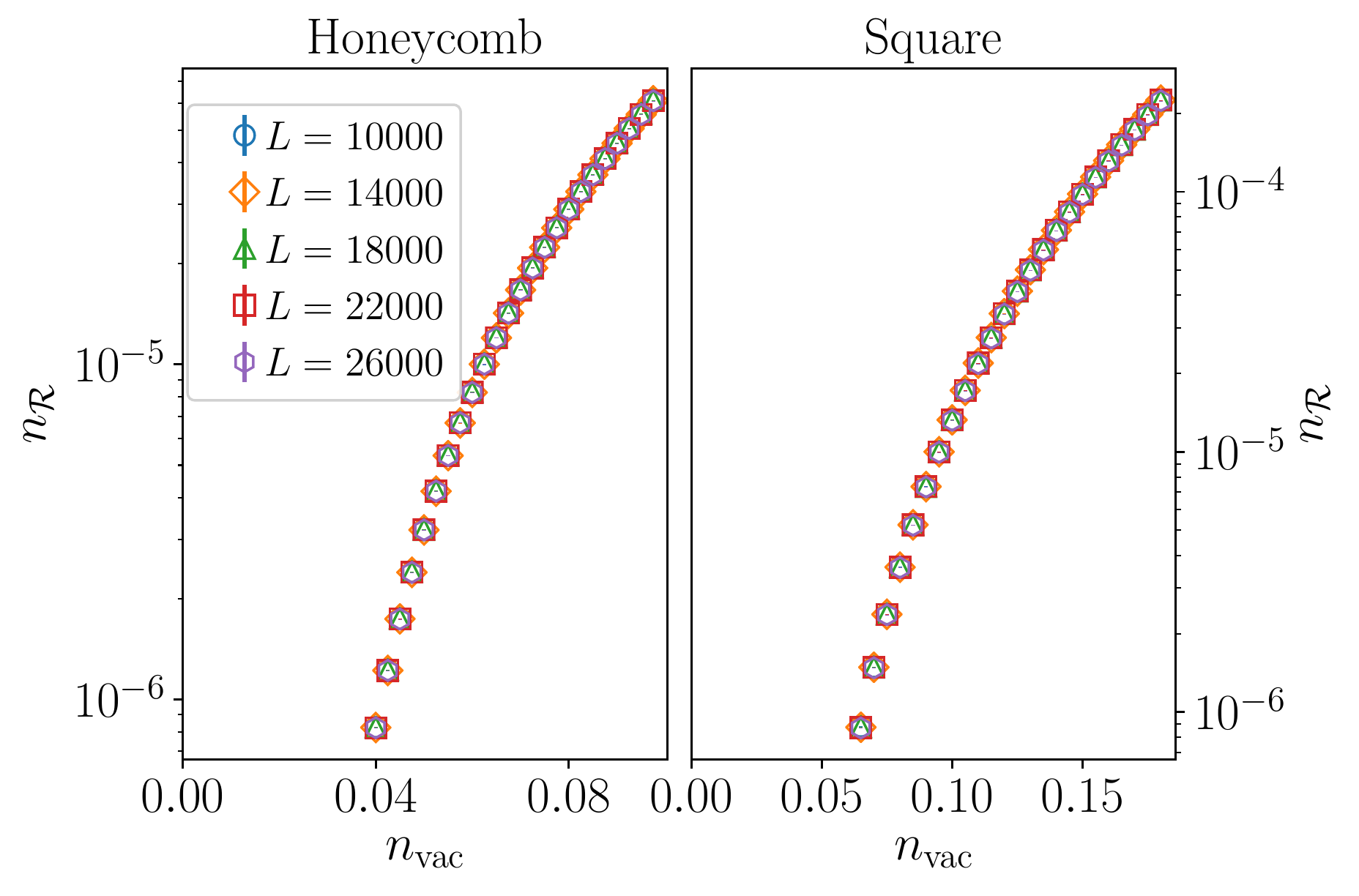} &
    b) \includegraphics[width=\columnwidth]{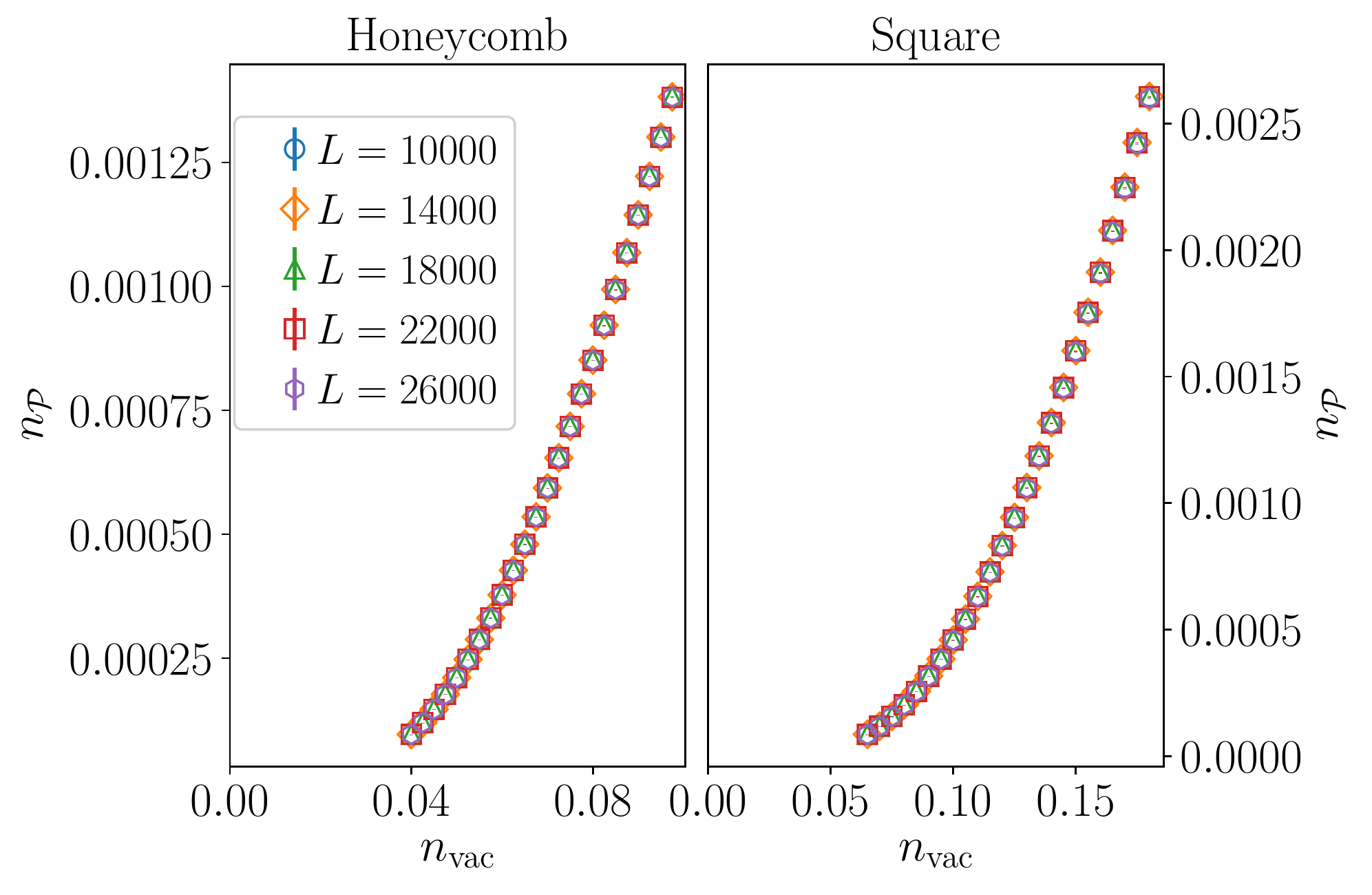} \\
	c) \includegraphics[width=\columnwidth]{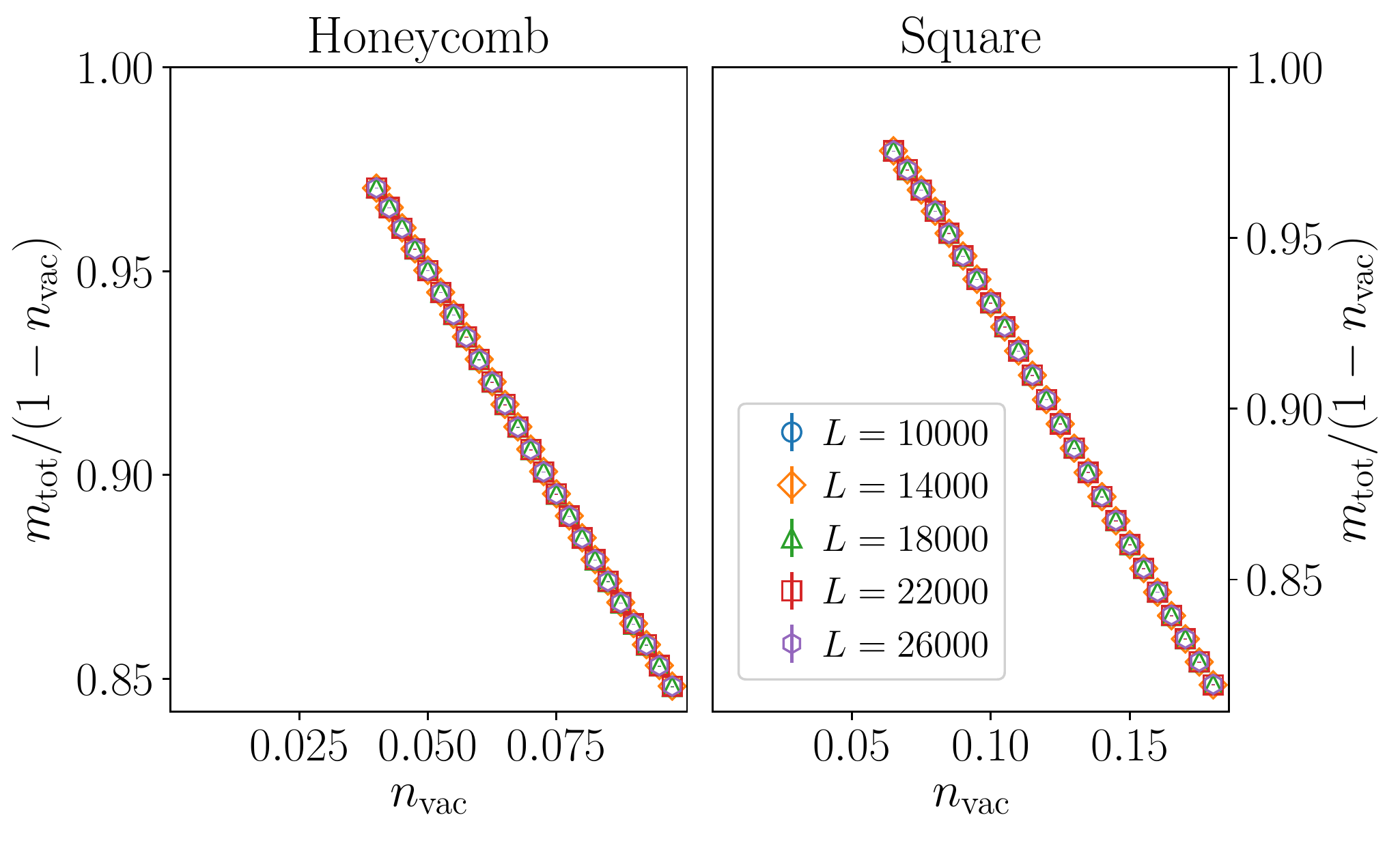} &
    d) \includegraphics[width=\columnwidth]{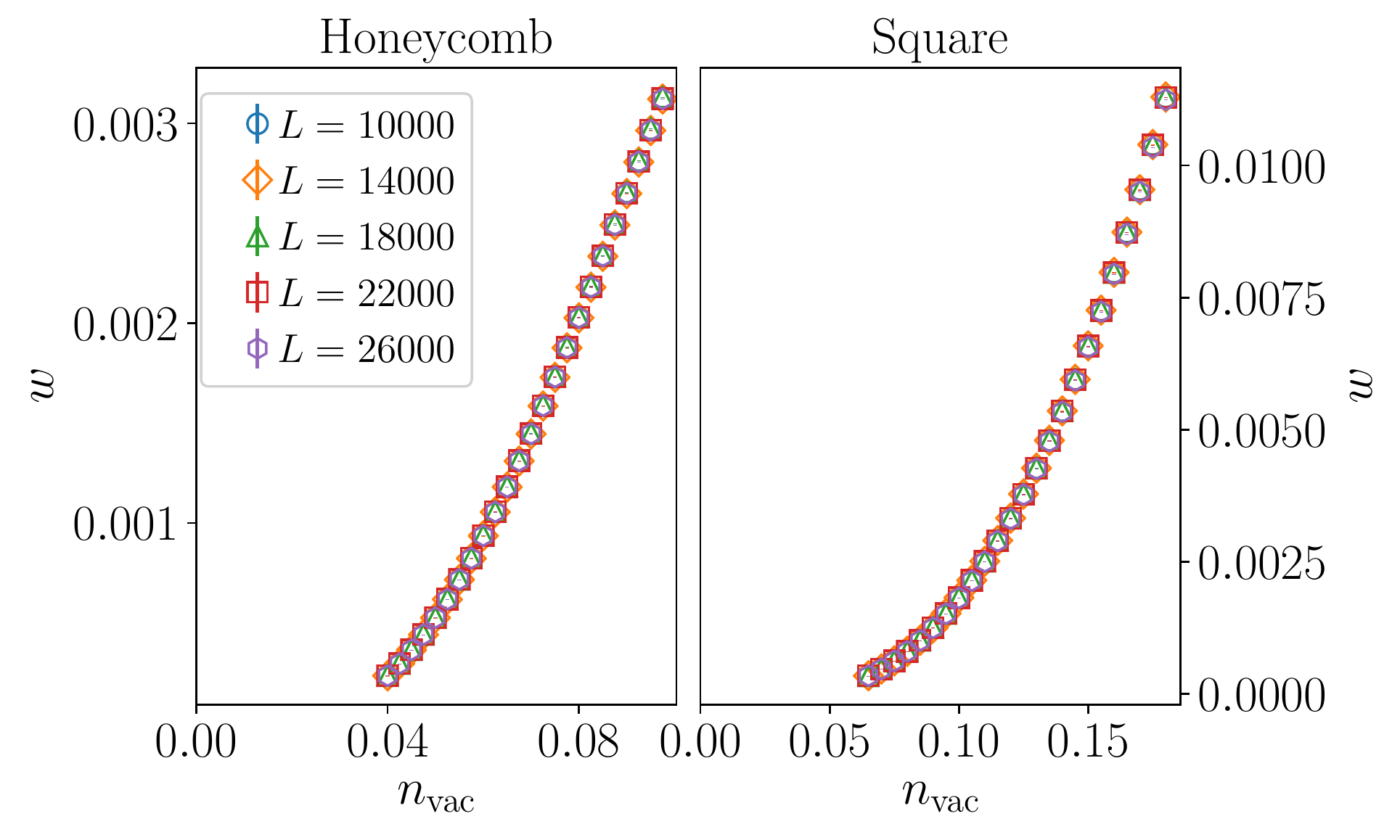}\\
    e) \includegraphics[width=\columnwidth]{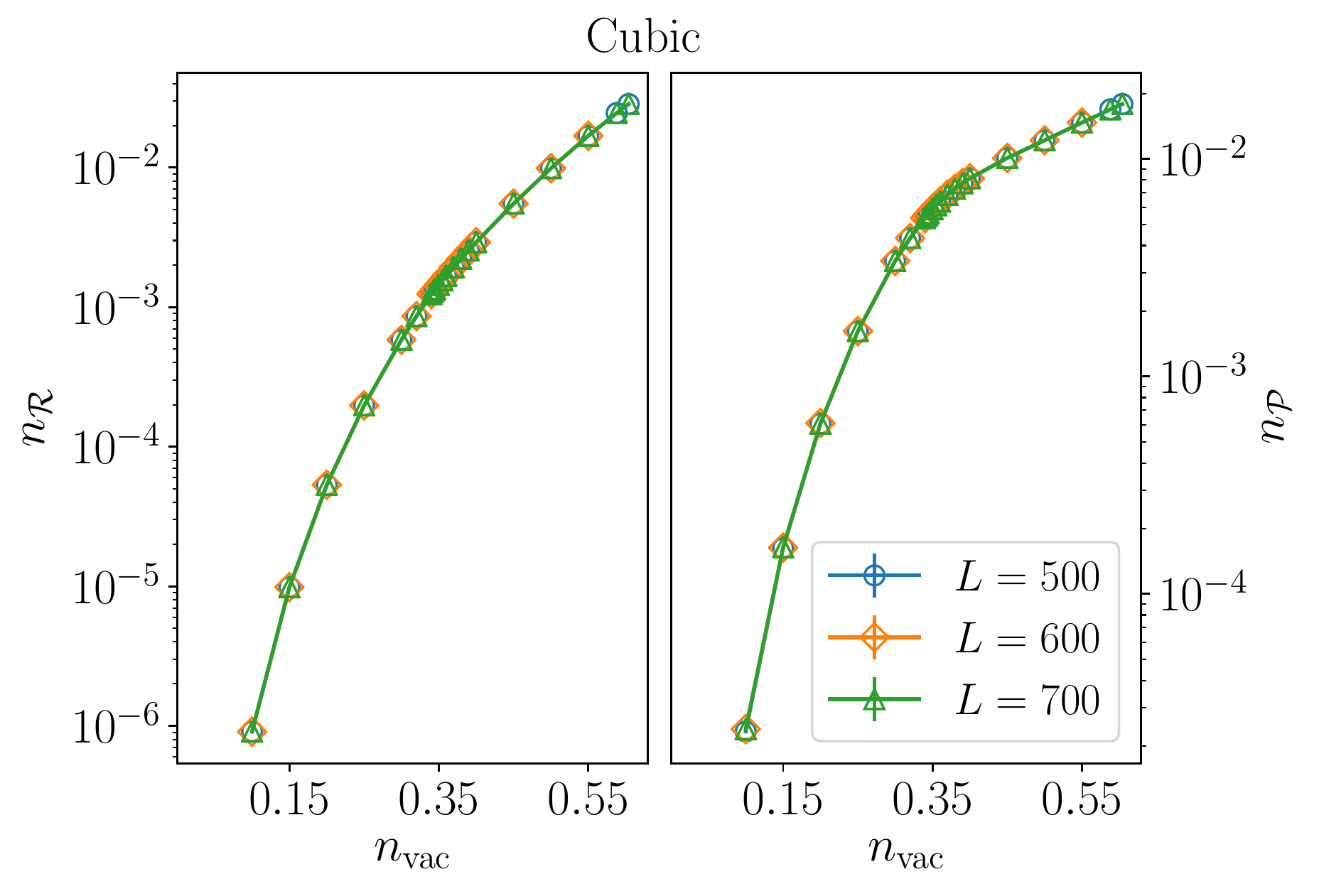} &
    f) \includegraphics[width=\columnwidth]{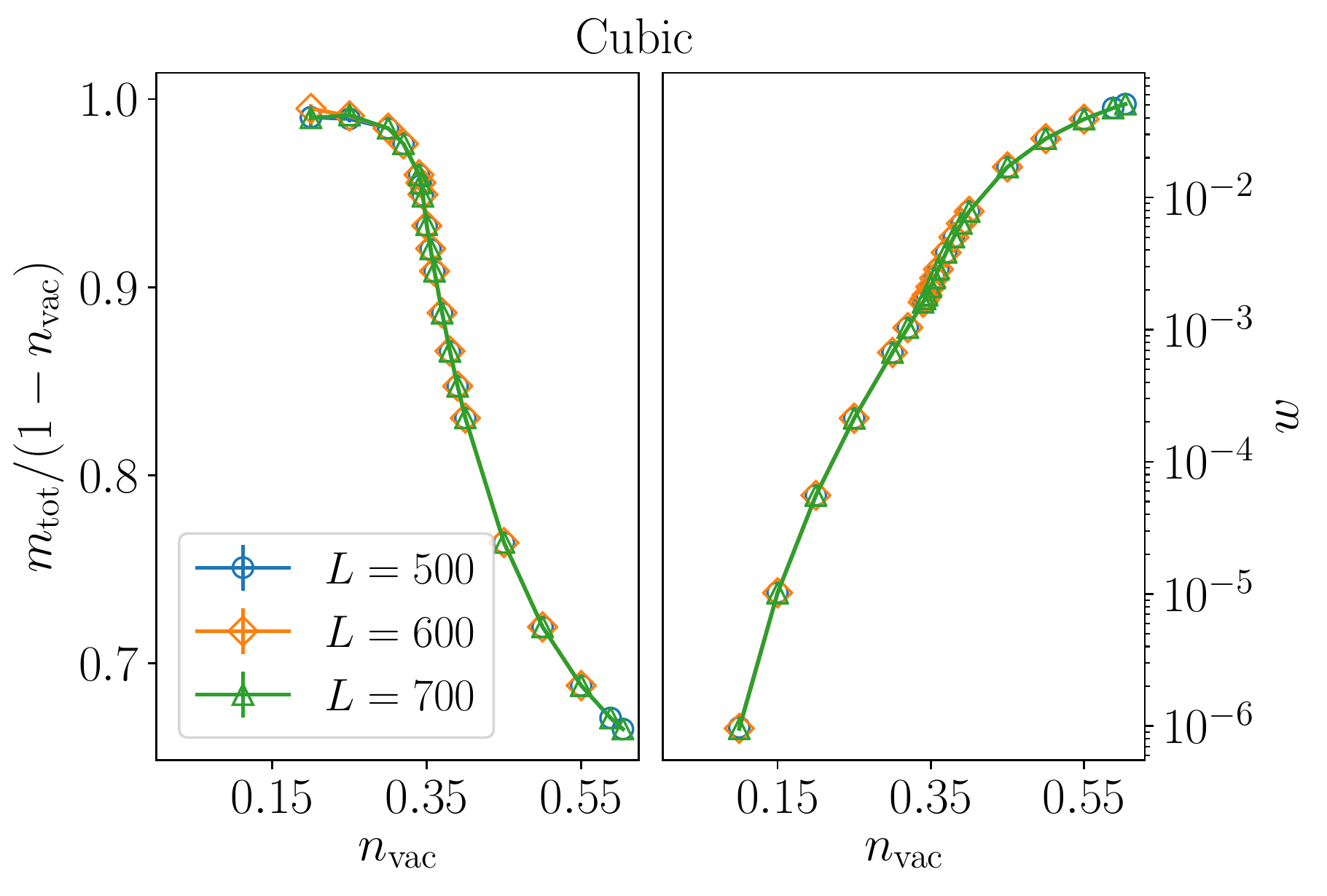}     
  \end{tabular}
\caption{a), b), e): The mean number densities $n_{\mathcal R}$ and $n_{\mathcal P}$ of ${\calR}$-type and ${\calP}$-type regions of the Dulmage-Mendelsohn decomposition of site-diluted honeycomb, square and cubic lattices are nonzero in the thermodynamic limit and decrease rapidly as we approach the limit of vanishing dilution $n_{\rm vac} \to 0$ in all three cases.  c), d), f): The sample-averaged total mass density of ${\mathcal R}$-type regions, $m_{\rm tot}$, is nonzero in the thermodynamic limit and appears to tend towards $1-n_{\rm vac}$ in the limit of small dilution $n_{\rm vac} \to 0$ on the honeycomb, square and cubic lattices. This density is normalized to the number of sites in the undiluted sample, so that $n_{\rm vac} + m_{\rm tot}+ m_{\mathcal P} = 1$ in all three cases (here $m_{\mathcal P}$ is the corresponding sample-averaged total mass density of ${\mathcal P}$-type regions). The sample-averaged density of zero modes, $w$, is also nonzero in the thermodynamic limit and goes rapidly to zero in the small-$n_{\rm vac}$ limit. Sec.~{\protect{\ref{Geometry:Basic}}} for details.}
\label{NRNPmw_combined}
\end{figure*}

The most basic of these are $\langle M_{\rm tot} \rangle=  \langle \sum_i m_i \rangle $, the
ensemble-averaged sum of the mass (number of sites) $m_i$ of ${\mathcal R}$-type regions ${\mathcal R}_i$, the ensemble-averaged number $\langle N_{\mathcal R}\rangle $ ($\langle N_{\mathcal P} \rangle $) of
${\mathcal R}$-type (${\mathcal P}$-type) regions in a sample, and the ensemble-averaged total
 number of zero modes $\langle W \rangle= \langle \sum_i {\mathcal I}_i \rangle$ in a sample. Anticipating our results, we also
  define the corresponding densities in the $N_{\rm sites} \to \infty$ thermodynamic limit: 
  $n_{\mathcal R} \equiv \langle N_{\mathcal R} \rangle/N_{\rm sites}$,
   $n_{\mathcal P} \equiv \langle N_{\mathcal P} \rangle /N_{\rm sites}$,
    $m_{\rm tot} \equiv \langle M_{\rm tot} \rangle /N_{\rm sites}$ and $w \equiv \langle W \rangle /N_{\rm sites}$. As expected, these densities are found to be self-averaging for large enough system size $L$, with smaller $n_{\rm vac}$ requiring a larger size $L$ for the self-averaging property to hold.

A nonzero $w$ defines a length scale $l_w \equiv 1/w^{1/d}$ 
where $d$ is the spatial dimensionality. This length scale $l_w$
 is a measure of the ``typical distance'' between zero modes if
  one thinks of them as localized objects. Note however that this interpretation of $l_w$ cannot be taken too literally if the wavefunctions of individual modes are spread out on a scale comparable to $l_w$, or, more precisely, if the length-scale $\xi_{\rm G}$ that controls the spatial dependence of $\langle \Delta G^2 \rangle (r-r')$, is comparable to $l_w$. Independent of this caveat, we will find it
   illuminating in our subsequent discussion to compare our results for $l_w$ to another length
    scale $l_{\rm vac}$, which represents the typical distance between vacancies:
     $l_{\rm vac} = 1/n_{\rm vac}^{1/d}$.

We also find it useful to separately keep track of the mass $m^A_{\rm max}$ ($m^B_{\rm max}$ ) of the largest (by mass) ${\mathcal R}_A$-type (${\mathcal R}_B$-type) region in each sample. For the larger (by mass) of these two regions, we also keep track of the imbalance ${\mathcal I}_{\rm max}$, and the size ${\mathcal B}_{\rm max}$ of the boundary. This size ${\mathcal B}_{\rm max}$ is defined in terms of the number of surviving links to the rest of the lattice from o-type sites that make up the boundary of the largest ${\mathcal R}$-type region in each sample. 

In addition, we also keep track of the number of deleted links ${\mathcal D}^{\rm w}_{\rm max}$ (${\mathcal D}^{\rm nw}_{\rm max}$) that would have connected a e-type site (o-type site) of this region to a neighboring site that now hosts a vacancy.
The superscripts ``${\rm w}$'' and ``${\rm nw}$'' stand for the labels ``wavefunction'' and
``nonwavefunction''. This refers to the fact that the zero mode wavefunctions of a ${\mathcal R}$-type region have amplitude only on the e-type sites of the ${\mathcal R}$-type region. The underlying intuition that motivates this measurement is this: Roughly speaking, since ${\mathcal I} = N_{\rm e} - N_{\rm o}$ (see Sec.~\ref{ZM}), we expect vacancies adjacent to e-type sites to ``seed'' zero modes in this  ${\mathcal R}$-type region, while vacancies adjacent to o-type sites ``eliminate'' zero modes.

It is also useful to measure a suitably-defined length scale associated with the size of large ${\mathcal R}$-type regions. Two different but related length scales come to mind. The first is $R_{\rm max}$, the sample-averaged radius of gyration of the largest (by mass) ${\mathcal R}$-type region in each sample.
The second length scale $\xi$ admits a natural interpretation as a correlation length associated with a sample-averaged geometric correlation function $C(r-r')$ which gets a contribution of $+1$ from a sample if both $r$ and $r'$ are in the same ${\mathcal R}$-type region in that sample, and zero otherwise~\cite{Stauffer_Aharony_book,Christensen_Moloney_book}. In terms of $C$, we define
\begin{eqnarray}
2\xi^2 &=& \frac{\sum_{r,r'} C(r-r') |r-r'|^2}{\sum_{r,r'} C(r-r') }
\label{xidefn}
\end{eqnarray} 
In classical bond percolation, this  correlation function and the associated correlation length map to corresponding properties of the $q$-state Potts model in the $q \to 1$ limit~\cite{Stauffer_Aharony_book,Christensen_Moloney_book}. 
 \begin{figure*}
	\begin{tabular}{cc}
	  \includegraphics[width=\columnwidth]{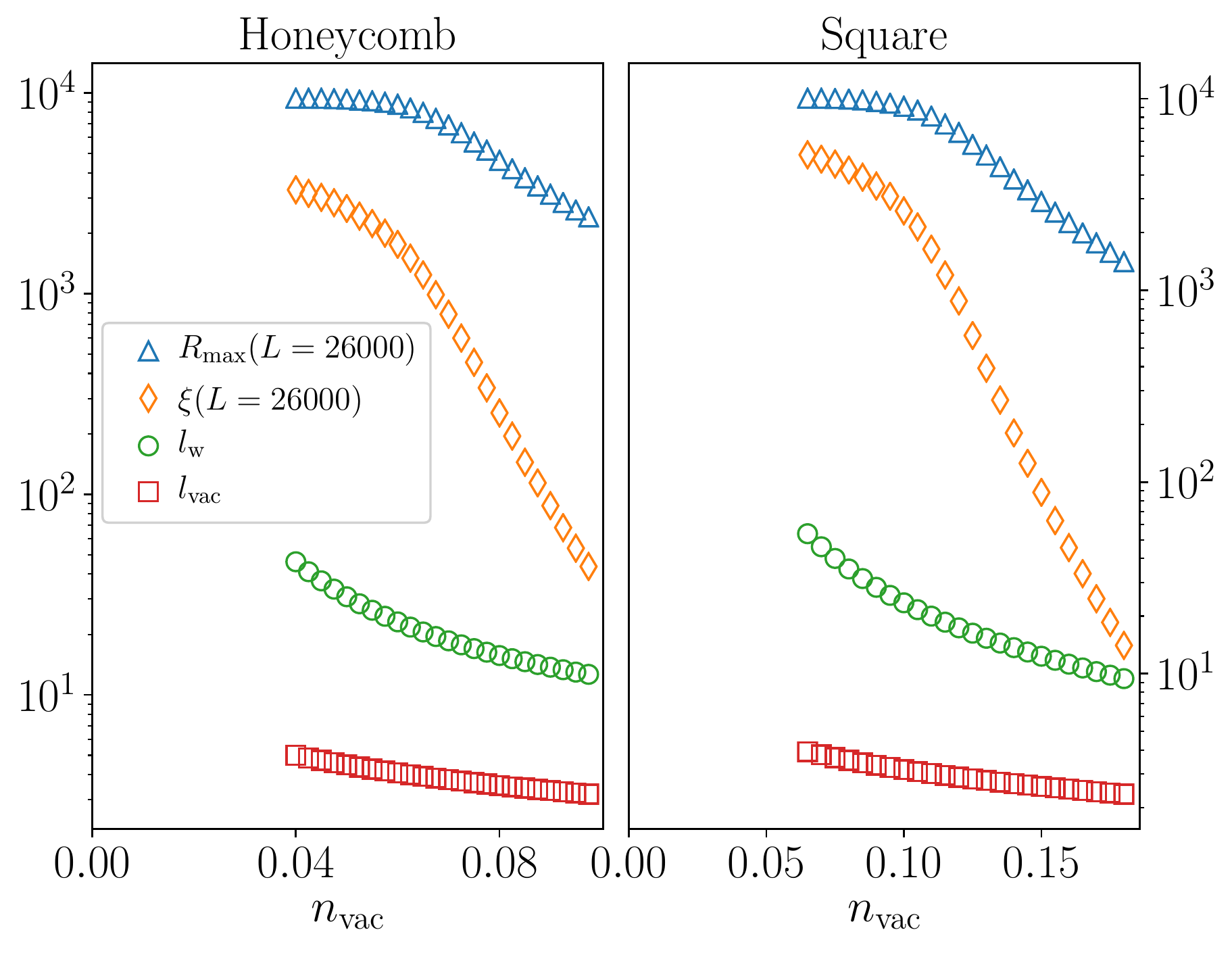} &
	  \includegraphics[width=\columnwidth]{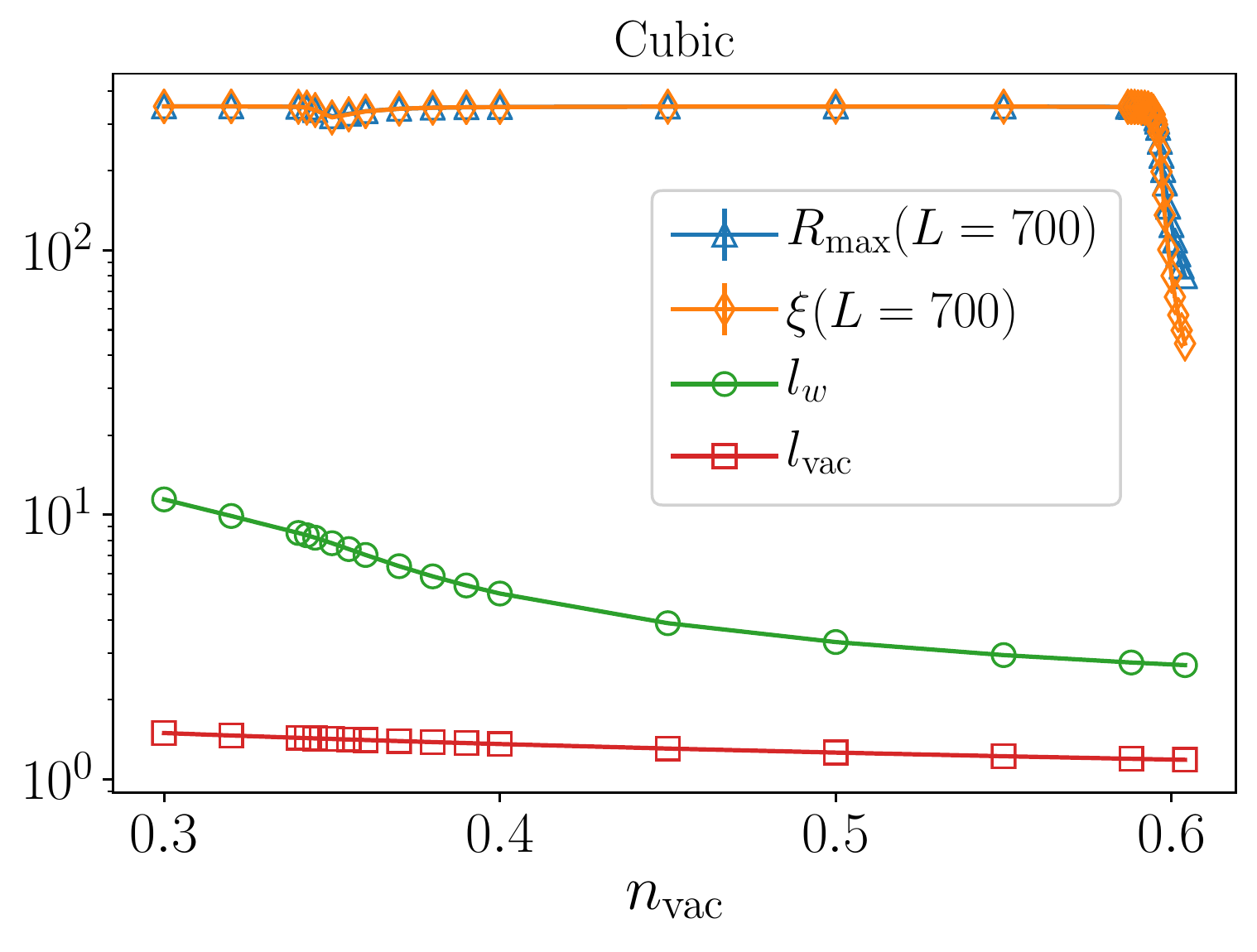} 
          \end{tabular}
		\caption{The mean separation between vacancies, $l_{\rm vac} = 1/n_{\rm vac}^{1/d}$, the mean distance between zero modes,
		$l_w = 1/w^{1/d}$ (where $d$ is the spatial dimension), the correlation length $\xi$ associated with the sample-averaged correlation function $C(r-r')$, and $R_{\rm max}$, the sample-averaged radius of gyration of the largest ${\mathcal R}$-type region in a sample, plotted as functions of the vacancy density $n_{\rm vac}$. Note that $\xi$ and $R_{\rm max}$ track each other and both are much larger than the other two length scales in the small-$n_{\rm vac}$ limit, being limited only by the system size $L$ in this regime in the range of sizes accessible to our numerical work. See Sec.~{\protect{\ref{Geometry:Basic}}} for details. }
		\label{length-scalebasic}
\end{figure*}

By thinking in terms of contributions of each ${\mathcal R}$-type region to the double summation in Eq.~\ref{xidefn}, we see that
$\xi^2$ is also the mean-square radius of gyration of ${\mathcal R}$-type regions, with the average taken to be weighted by
 $m^2$, where $m$ is the mass of a ${\mathcal R}$-type region. Thus, we have the expression
 \begin{eqnarray}
   \xi^2 &=& \frac{\langle \sum_{i=1}^{N_{\mathcal R}} m_i^2 R_i^2 \rangle }{\langle \sum_{i=1}^{N_{\mathcal R}} m_i^2 \rangle} \;, \nonumber \\
&=& \frac{\sum_{m} m^2R^2_mN_m}{\sum_{m} m^2N_m}  \;,
   \label{Definition_xi}
   \end{eqnarray}
   where the angular brackets in the first expression indicate averaging over the ensemble of diluted samples, and $N_m$ and $R^2_m$ in the second expression are defined respectively to be the mean number of ${\mathcal R}$-type regions of mass $m$ and the mean square radius of gyration of such regions in this ensemble. 

The toroidal geometry of our samples introduces a subtlety in our actual measurements of these length-scales and causes the strict correspondence between the two definitions of $\xi$ in Eq.~\ref{xidefn} and Eq.~\ref{Definition_xi} to break down.
This is because a definition of $|r-r'|$ (in Eq.~\ref{xidefn}) that takes this geometry into account would involve computing the {\em shortest distance} on the torus between $r$ and $r'$ for each pair of points $r$ and $r'$ that belong to a particular ${\mathcal R}$-type region. The corresponding computational cost of using Eq.~\ref{xidefn} to obtain $\xi$ scales as $L^{2d}$ if there is even one such region that contains ${\mathcal O}(L^d)$ sites; this introduces a significant inefficiency in the computation.

Fortunately, there is a simple work-around. The idea is to use Eq.~\ref{Definition_xi} instead of Eq.~\ref{xidefn} and compute the radius of gyration of each cluster using a method borrowed from the image-processing literature~\cite{Bai_Breen}: This involves making two passes through each cluster. The first pass computes a particular definition of center of mass which is well-adapted to the toroidal geometry and provides a sensible location of the center of mass of the cluster. The second pass then computes the mean-square shortest distance of each point of the cluster from this center of mass. In all our work, we adopt this definition of the radius of gyration of each cluster, and use Eq.~\ref{Definition_xi} to obtain $\xi^2$. Although the strict equivalence with Eq.~\ref{xidefn} is now lost, we have checked that this computationally convenient redefinition shares all the qualitative features of the length-scale originally defined in Eq.~\ref{xidefn}.
 \begin{figure*}
\begin{tabular}{cc}
		a) \includegraphics[width=\columnwidth]{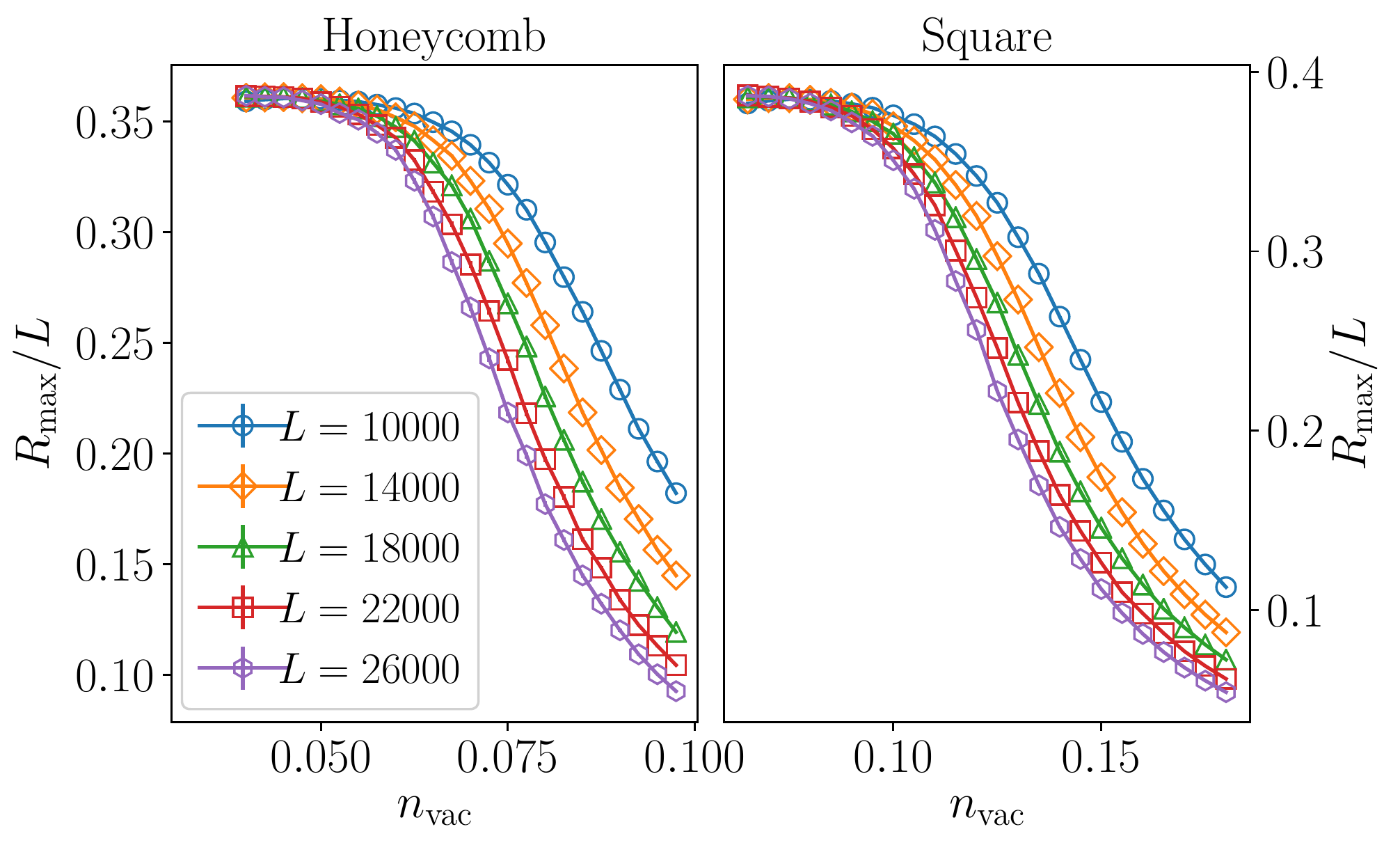} &
		b) \includegraphics[width=\columnwidth]{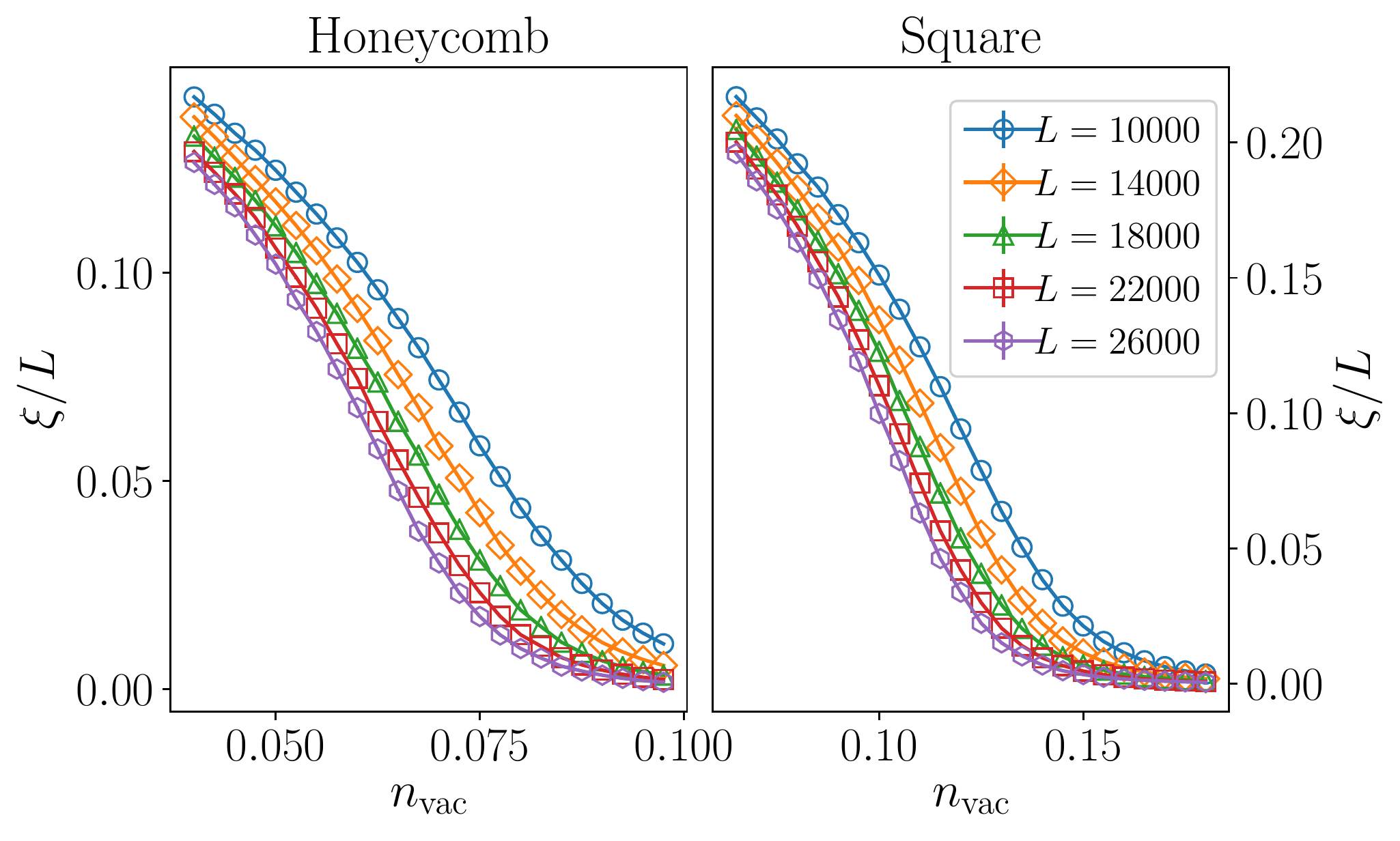}\\
        c)  \includegraphics[width=\columnwidth]{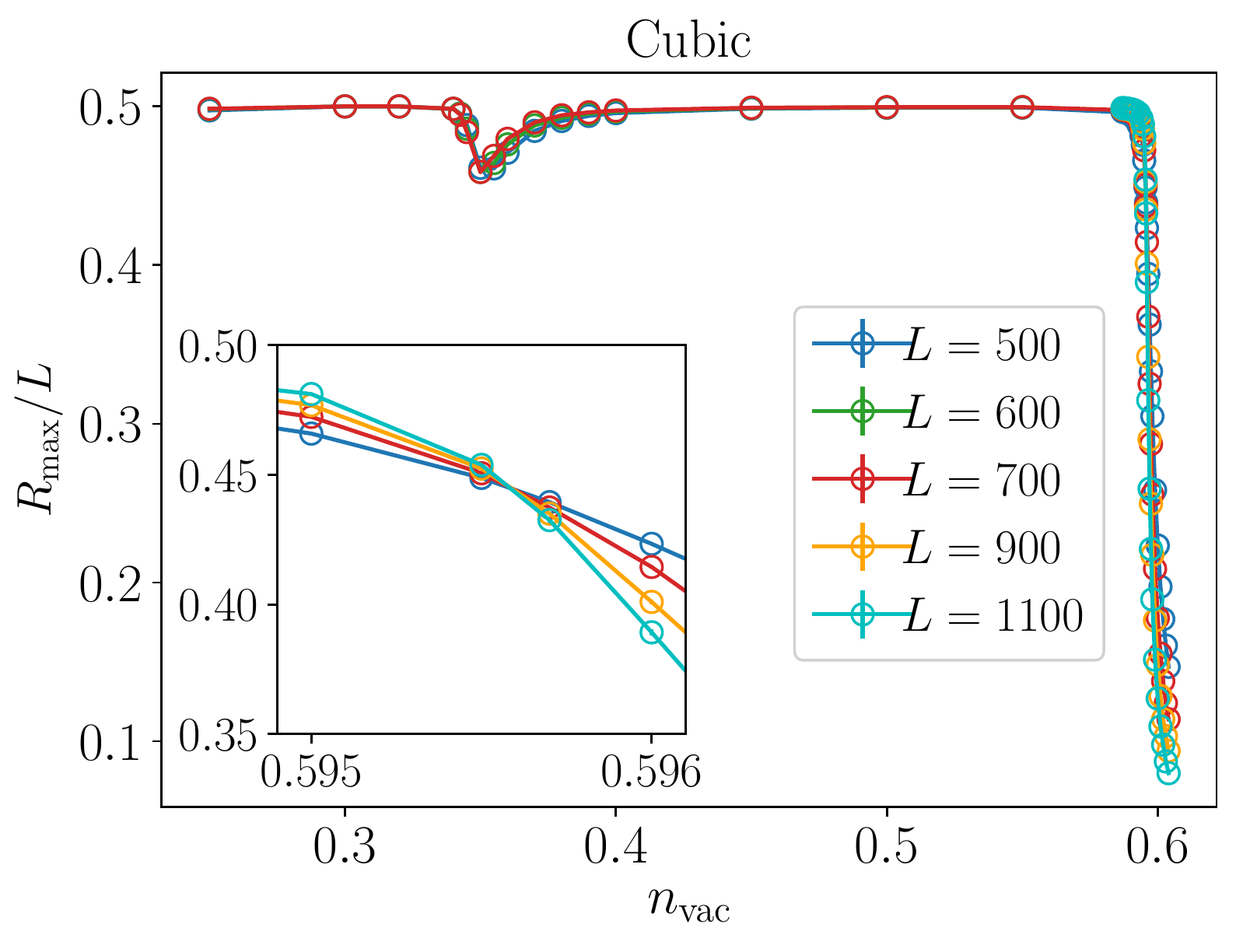} &    
         d) \includegraphics[width=\columnwidth]{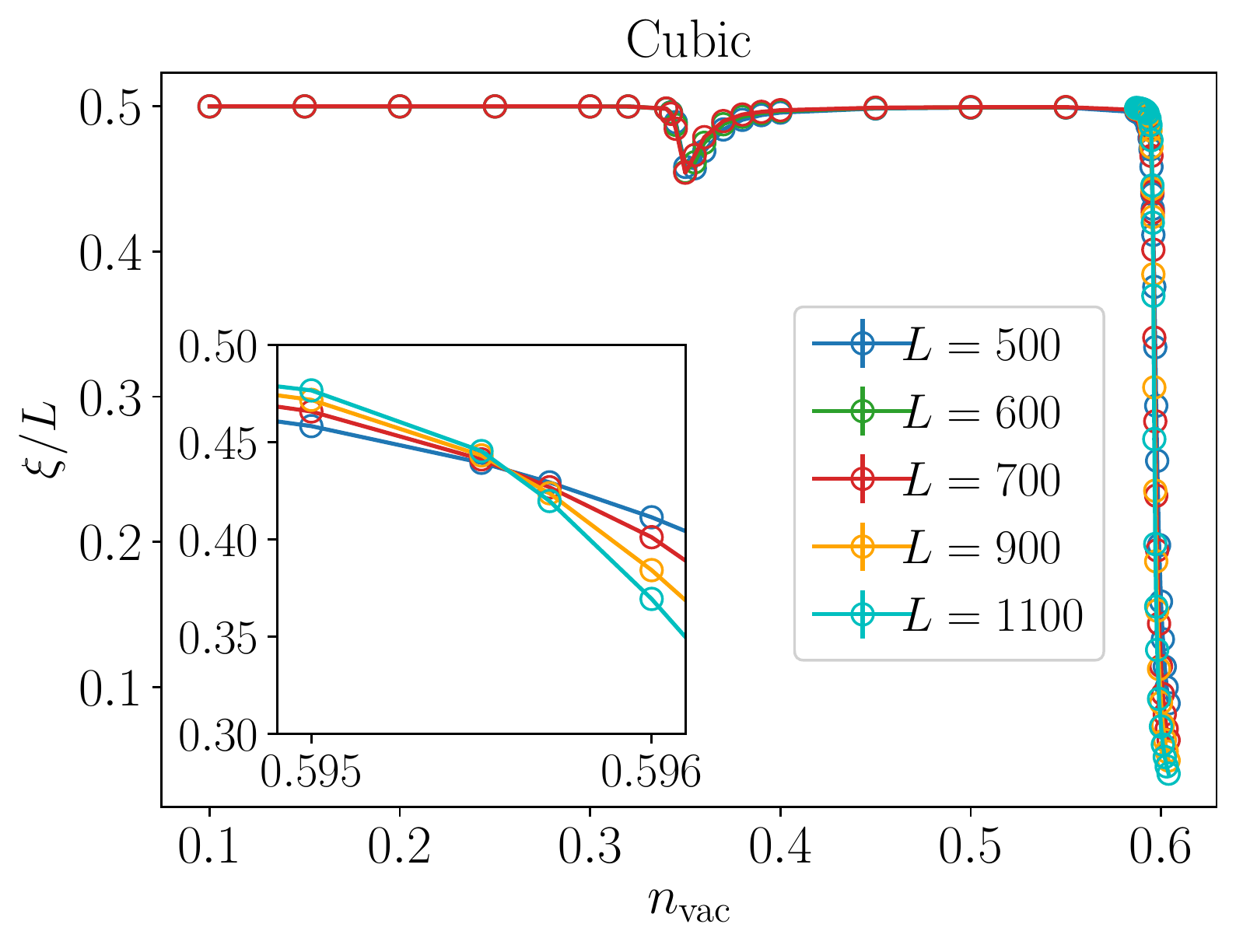}        
\end{tabular}
		\caption{The sample-averaged radius of gyration, $R_{\rm max}$, of the largest ${\mathcal R}$-type region in a sample, as well as the correlation length $\xi$ associated with the sample-averaged geometric correlation function both increase rapidly in the small-$n_{\rm vac}$ limit to values that are set by the system size $L$ for accessible system sizes in two and three dimensions. In two dimensions, curves of the dimensionless ratios $R_{\rm max}/L$ and $\xi/L$ corresponding to different $L$ do not cross over the range of $n_{\rm vac}$ accessible to our numerics. However, on the cubic lattice, there is a clear crossing in the vicinity of $n_{\rm vac} \approx 0.6$ (insets of c), d)). See Sec.~{\protect{\ref{Geometry:Basic}}} for details.}
\label{Rgandxi}
\end{figure*}

In addition, we also probe the statistics of the mass $m$ of ${\mathcal R}$-type regions by measuring the susceptibility $\chi$, which is defined in terms of the sample-averaged geometric correlation function
$C(r-r')$ in the following natural way:
\begin{eqnarray}
\chi &=& \frac{1}{L^d}\sum_{r,r'} C(r-r') \nonumber \\
&=& \frac{1}{L^d}\langle \sum_{i=1}^{N_{\mathcal R}} m_i^2 \rangle \nonumber \\
&=& \frac{1}{L^d} \sum_m m^2 N_m \; ,
\label{chidefn}
\end{eqnarray}
where $d$ is the spatial dimensionality.

Finally, with an eye towards the possibility of a percolation transition associated with the random geometry of ${\mathcal R}$-type regions, we keep track of the wrapping probability $P$ which is defined in both two and three dimensions as the probability that a sample has at least one ${\mathcal R}$-type region that wraps around in at least one direction around the torus. In addition, in the three-dimensional case, we also count, for each sample, the number of ${\mathcal R}$-type regions that wrap around the $3$-torus simultaneously in three independent directions, as well as the number of odd ${\mathcal R}$-type regions that wrap around the $3$-torus simultaneously in three independent directions. These measurements are all made using efficient techniques borrowed from the percolation theory literature~\cite{Machta_etal,Newman_Ziff,Pruessner_Moloney}.

  \section{Random geometry of the Dulmage-Mendelsohn decomposition: Basic picture}
  \label{Geometry:Basic}
  
We now present the basic picture that emerges from our computational study of the random geometry of the Dulmage-Mendelsohn decomposition of site-diluted square, honeycomb and cubic lattices. Our initial focus is on thermodynamic densities, which converge rapidly to a nonzero thermodynamic limit at any nonzero dilution $n_{\rm vac}$. This is followed by a characterization of the important length-scales that control the statistical properties of this decomposition.

\subsection{Thermodynamic densities}
\label{subsec:thermodynamicdensities}

How do $M_{\rm tot}$, $N_{\mathcal R}$, $N_{\mathcal P}$, and $W$ scale with system size $L$ and
vacancy concentration $n_{\rm vac}$ in the small-$n_{\rm vac}$ limit? Our answers to these questions are displayed in Fig.~\ref{NRNPmw_combined}.

 From the behaviour of $n_{\mathcal R}$ 
and $n_{\mathcal P}$ for the
honeycomb and square lattices in two dimensions and the cubic lattice in three dimensions, we see that $n_{\mathcal R}$ and $n_{\mathcal P}$ both tend rapidly to a nonzero value in the thermodynamic limit, with finite-size corrections that are not readily visible at the sizes studied here. The same also holds (although we have not displayed it) for $n^{\rm odd}_{{\mathcal R}}$, the corresponding number density of odd ${\mathcal R}$-type regions. From the $n_{\rm vac}$ dependence of these quantities, we see that these densities decrease rapidly to zero as $n_{\rm vac} \to 0$ (Fig.~\ref{NRNPmw_combined}).

From Fig.~\ref{NRNPmw_combined}, we also see that the density of zero modes $w$ on the honeycomb and square lattices in two dimensions and the cubic lattice in three dimensions  saturates to a nonzero value in the thermodynamic limit, with finite-size corrections that are not readily visible in the range of sizes studied here. As expected, we also see that $w$ tends to zero as $n_{\rm vac} \to 0$. From Fig.~\ref{NRNPmw_combined}, we see that $m_{\rm tot}$ saturates rapidly to a nonzero value in the thermodynamic limit in both two and three dimensions, with finite-size corrections that are not visible in the size range studied. Moreover, it is apparent that $m_{\rm tot}$ tends towards the value $m_{\rm tot}  = 1-n_{\rm vac}$ as $n_{\rm vac}$ goes to zero in both two and three dimensions. In conjunction with the behaviour of $n_{\mathcal R}$, this implies that the entire sample is taken over by a few ${\mathcal R}$-type regions in this limit. 

This is, at first sight,  a surprising and counter-intuitive result, since $n_{\rm vac}=0$ corresponds to the pure lattice, which admits a perfect matching. In other words, {\em at} $n_{\rm vac} = 0$, we have $w=0$. In the language of the Dulmage-Mendelsohn decomposition, there are no ${\mathcal R}$-type regions, and the undiluted system consists of a single ${\mathcal P}$-type region at $n_{\rm vac}=0$. How are we to reconcile this counter-intuitive result  for $m_{\rm tot}$ with the existence of perfect matchings at $n_{\rm vac}=0$? As we detail below, the answer has to do with the order in which the thermodynamic limit and the limit of zero dilution are taken.
     \begin{figure*}
  \begin{tabular}{cc}
 		a) \includegraphics[width=\columnwidth]{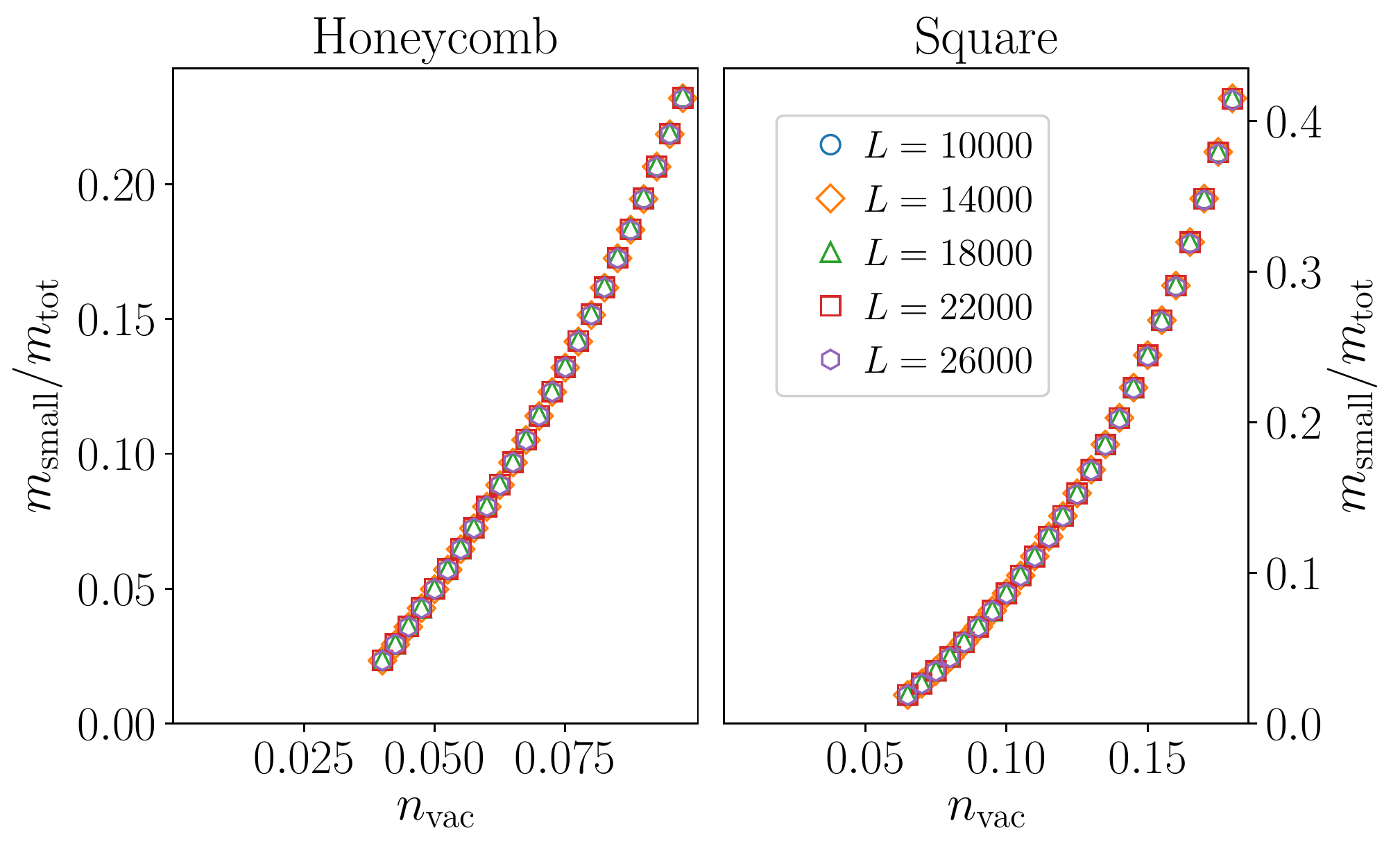} & 
		b) \includegraphics[width=\columnwidth]{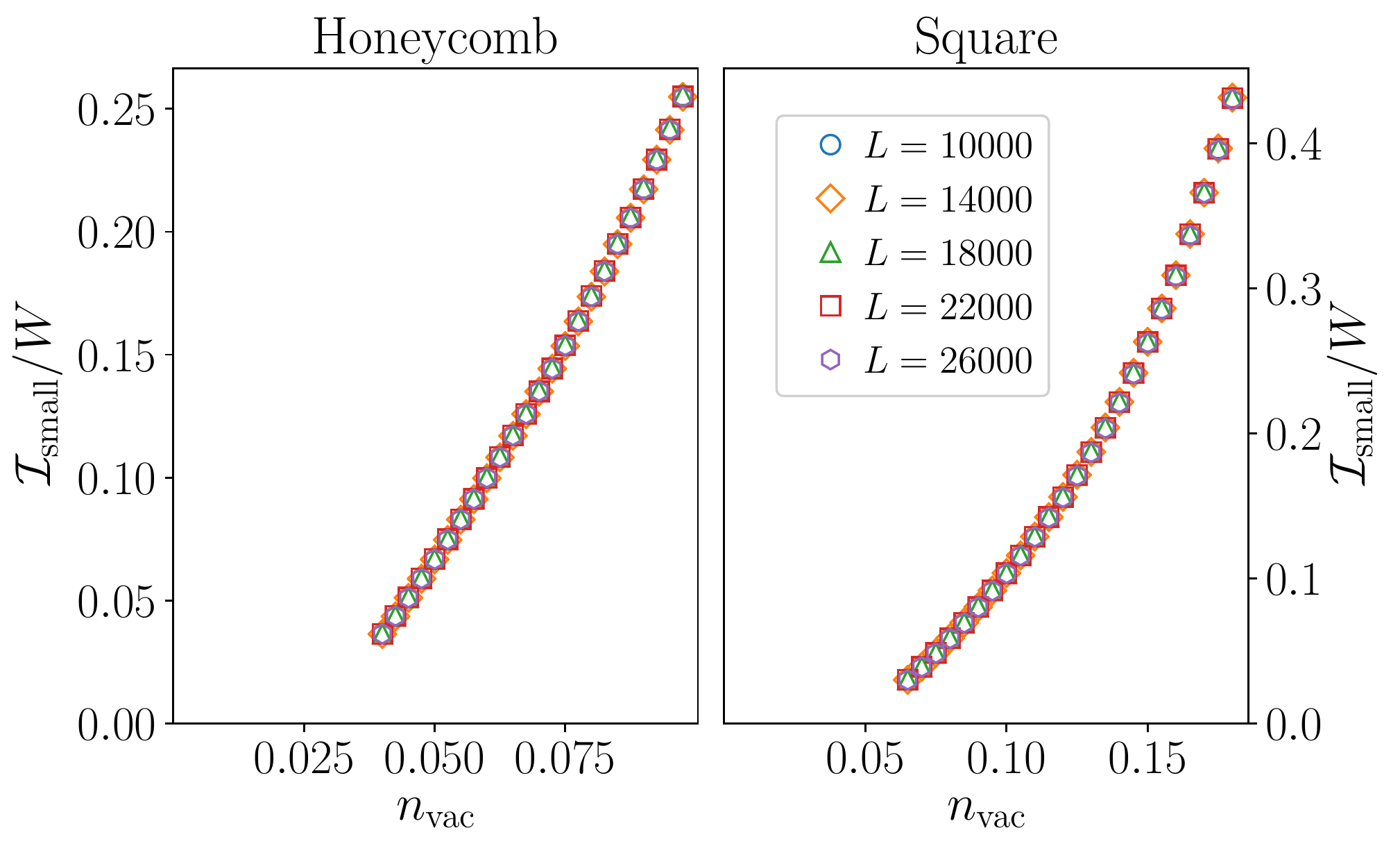}
  \end{tabular}
\caption{a) The sample averaged ratio $m_{\rm small}/m_{\rm tot}$, where $m_{\rm small}$ is the contribution of ``small'' ${\mathcal R}$-type regions (with less than  $10000$ vacancies associated with them) decreases rapidly with $n_{\rm vac}$. b)  The same holds true for the corresponding ratio ${\mathcal I}_{\rm small}/W$ of zero modes contributed by ``small'' ${\mathcal R}$-type regions. See Sec.~{\protect{\ref{Geometry:Basic}}} for details.}  
\label{smallfraction_m}
\end{figure*}

\subsection{Diverging length-scales}
\label{subsec:Length-scales}

As noted earlier, a nonzero density of vacancies is naturally associated with a length scale $l_{\rm vac}$ that corresponds to the typical distance between vacancies, while a nonzero $w$ defines a second length scale $l_w$ which is a measure of the ``typical distance'' between zero modes if one thinks of them as localized objects. If we take the thermodynamic limit at fixed nonzero $n_{\rm vac}$ and then take the limit $n_{\rm vac} \to 0$, we are studying the limit $L/l_{\rm vac} \to \infty$,
$l_{\rm vac} \to \infty$. In this case, we see from our computations that $m_{\rm tot} \to 1$ in this
 $n_{\rm vac} \to 0$ limit. On the other hand, if we first send $n_{\rm vac}$ to zero while keeping $L$ fixed, and then take the thermodynamic limit, we are studying the limit $l_{\rm vac} /L\to \infty$, $L \to \infty$. This corresponds to the limiting case of the pure system for which $m_{\rm tot} \to 0$, since the entire sample has a perfect matching.

From the $n_{\rm vac}$-dependence of $w$ in the small-$n_{\rm vac}$ limit, we see that
$l_w/l_{\rm vac} \to \infty$ in this limit in both two and three dimensions, since $w$ goes to zero very rapidly in the limit of small dilution. Thus,  these two length-scales have a parametrically large separation $l_w \gg l_{\rm vac}$ at small $n_{\rm vac}$. In other words, zero modes in this limit are a cooperative effect of a very large number of vacancies, both in two and in three dimensions.

%Nevertheless, we see that the typical size of ${\mathcal R}$-type regions of a finite-size sample in the small-$n_{\rm vac}$ limit depends {\em a priori} on these two length scales $l_{\rm vac}$ and $l_w$, in addition to the sample size $L$.

 We now come to the key observation that motivates much of our subsequent analysis. In Fig.~\ref{length-scalebasic}, we plot $R_{\rm max}$, $\xi$, $l_w$ and $l_{\rm vac}$ as a function of $n_{\rm vac}$ for the largest sizes studied both in two and three dimensions, while Fig.~\ref{Rgandxi} shows $R_{\rm max}$ and $\xi$ for different sizes as a function of $n_{\rm vac}$. It is clear from the displayed results
 that $R_{\rm max}$ and $\xi$ more or less track each other, with
 both these length scales dominating over $l_{\rm vac}$ and over $l_w$ in the small-$n_{\rm vac}$ limit in the two-dimensional case. For values of $L$ accessible to our computational method, we see that both these length scales appear to be limited only by the system size $L$ in the small-$n_{\rm vac}$ limit in two dimensions. The three-dimensional case also displays similar behaviours, which set in much ``earlier'', {\em i.e.}, below a much larger threshold in the vicinity of $n_{\rm vac} \approx 0.6$. 
 
 Thus, in the thermodynamic limit, we conclude that the typical size of a ${\mathcal R}$-type region grows without bound in the $n_{\rm vac} \to 0$ limit in two dimensions. In three dimensions, the corresponding quantity exhibits the same behaviour below $n_{\rm vac}  \approx 0.6$. The presence of this diverging length scale suggests that the random geometry of ${\mathcal R}$-type regions may be independent of lattice-scale details in the corresponding regimes both in two and in three dimensions. Additionally, as is clear from the insets in Fig.~\ref{Rgandxi}, we see that curves of $R_{\rm max}/L$ and  $\xi/L$ corresponding to cubic lattices of different linear sizes $L$ appear to cross near this threshold when plotted as a function of $n_{\rm vac}$. This suggests that this threshold is associated with a critical point in the cubic-lattice case.

\subsection{Dominance of large ${\mathcal R}$-type regions}
\label{subsec:dominance}
 
 In two dimensions, additional support for this scenario comes from a study of the $n_{\rm vac}$ dependence of $m_{\rm small}/m_{\rm tot}$, where $m_{\rm small}$ is the contribution to $m_{\rm tot}$ from ``small'' ${\mathcal R}$-type regions of absolute mass $m < m^*(n_{\rm vac})$. For the cutoff value $m^*$, we choose $m^*(n_{\rm vac}) = V_{\rm small} /n_{\rm vac}$ to ensure that clusters of mass $m^*$ can be expected on average
 to be associated with some fixed ``small'' number of vacancies, $V_{\rm small}$. Another characterization of $m^*(n_{\rm vac})$ is that it corresponds to the expected number of sites in a region of linear size $l_{\rm small} = \sqrt{V_{\rm small}} \times l_{\rm vac}$ in the pure system.
 
 The results of this study with $V_{\rm small}=10000$, {\em i.e.}, $l_{\rm small} = 100 \times l_{\rm vac}$, are shown in Fig.~\ref{smallfraction_m}. Analogous results for the $n_{\rm vac}$ dependence of ${\mathcal I}_{\rm small}/W$, where ${\mathcal I}_{\rm small}$ is the contribution to $W$ from these small ${\mathcal R}$-type regions of mass $m < m^*(n_{\rm vac})$, are also shown in 
 Fig.~\ref{smallfraction_m}. From the data displayed in this figure, it is clear that the small-$n_{\rm vac}$ limit is dominated by the physics of large ${\mathcal R}$-type regions whose size diverges {\em even when measured in units of $l_{\rm vac}$}. 
 
\begin{figure}
 		\includegraphics[width=\columnwidth]{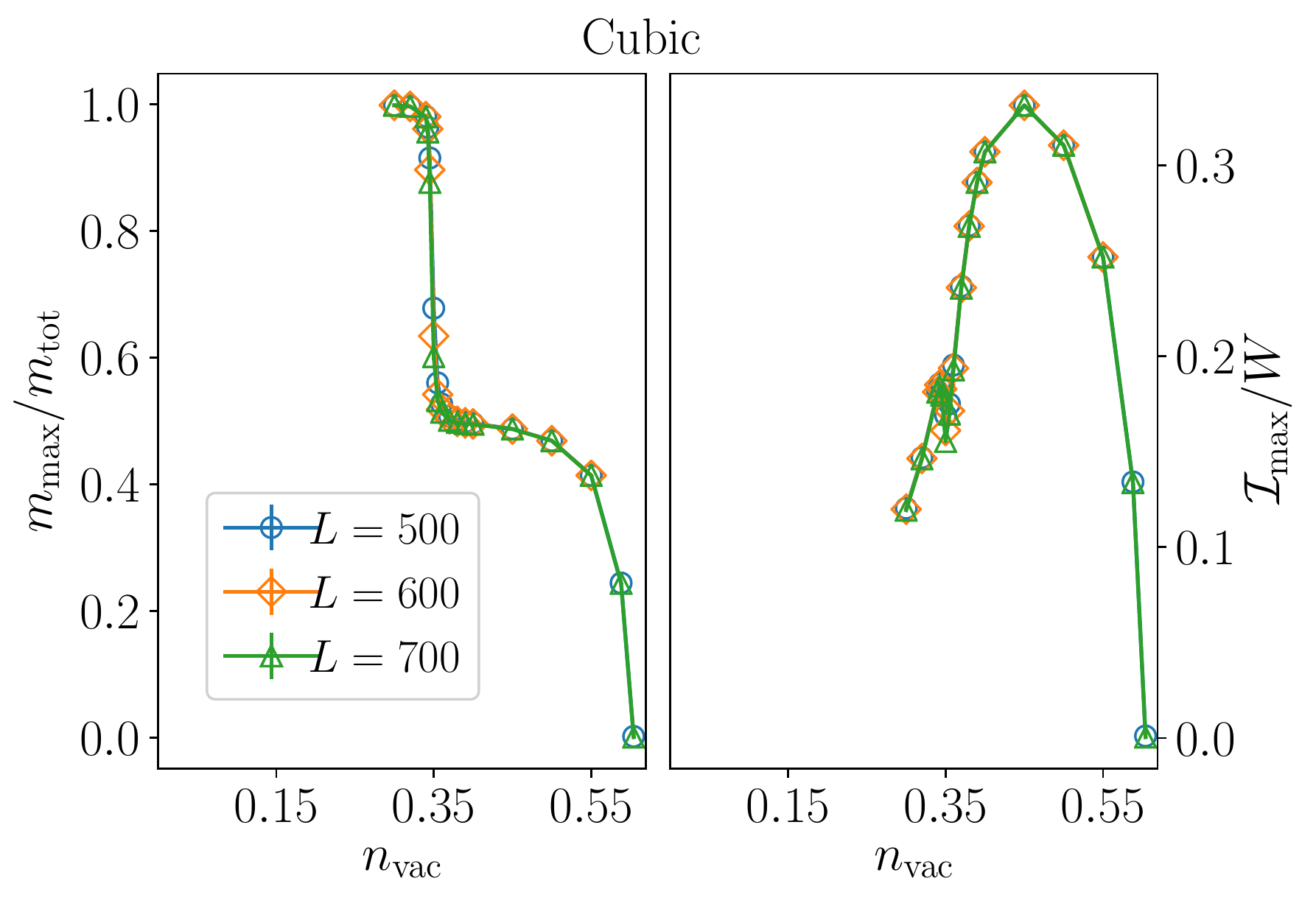}
\caption{On the site-diluted cubic lattice below $n_{\rm vac} \approx 0.6$, the largest ${\mathcal R}$-type region contributes a nonzero fraction of the total mass in ${\mathcal R}$-type regions in the thermodynamic limit, as well as a nonzero fraction of the total number of zero modes in this limit, as is clear from the corresponding sample-averaged ratios. See Sec.~{\protect{\ref{Geometry:Basic}}} for details.}
\label{maxfraction_mwcubic}
 \end{figure}

In three dimensions, a similar picture emerges in a somewhat more direct way from a study of the $n_{\rm vac}$ dependence of $\langle m_{\rm max}/M_{\rm tot} \rangle$, where $M_{\rm tot}$ is the total mass in all ${\mathcal R}$-type regions of a  sample,
the angular brackets denote an ensemble average, and $m_{\rm max}$ is the mass of the largest ${\mathcal R}$-type region in a sample, {\em i.e.}, the larger of $m_{\rm max}^{A}$ and $m_{\rm max}^{B}$.  As is clear from Fig.~\ref{maxfraction_mwcubic}, the largest ${\mathcal R}$-type region contains a nonzero fraction of the total mass as well as a nonzero fraction of the total number of zero modes once $n_{\rm vac}$ is reduced below a threshold in the vicinity of $n_{\rm vac} \approx 0.6$. 

When the dilution falls below a lower threshold in the vicinity of $n_{\rm vac} \approx 0.35$, we see that this mass fraction climbs rapidly to $1$, suggesting that a single large ${\mathcal R}$-type region dominates in the low-dilution regime below this lower threshold. This lower threshold is also associated with a cusp-like feature in $\langle {\mathcal I}_{\rm max}/W \rangle$, suggesting an abrupt qualitative change in the morphology of the largest ${\mathcal R}$-type regions below this threshold. 

In the context of the quantum percolation problem or in the context of diluted quantum antiferromagnets and closely-related particle-hole symmetric Hubbard models on such diluted lattices, the behaviour of ${\mathcal I}_{\rm small}/W$ ($\langle {\mathcal I}_{\rm max}/W \rangle$) in two (three) dimensions implies that the low-energy physics of interest to us is completely dominated by the fact that the typical size of ${\mathcal R}$-type regions diverges in the limit of low dilution. 

In the Appendix, we establish that odd ${\mathcal R}$-type regions behave similarly to their even cousins, {\em i.e.}, they also have a divergent typical size in the low-dilution limit both in two and in three dimensions. Before proceeding further, it is however useful to note the following distinction: In contrast to the above, the implication of this divergent typical size of odd ${\mathcal R}$-type regions for the physics of Majorana networks is somewhat different.  This is because each odd ${\mathcal R}$-type region only hosts a single stable zero-energy Majorana excitation of the network. As a result, although long-distance properties are completely controlled by the large odd ${\mathcal R}$-type regions, the {\em density} of
collective zero-energy Majorana fermion excitations is {\em not} dominated by this large-scale physics, since small odd ${\mathcal R}$-type regions also contribute significantly to this density.

\begin{figure}
		 \includegraphics[width=\columnwidth]{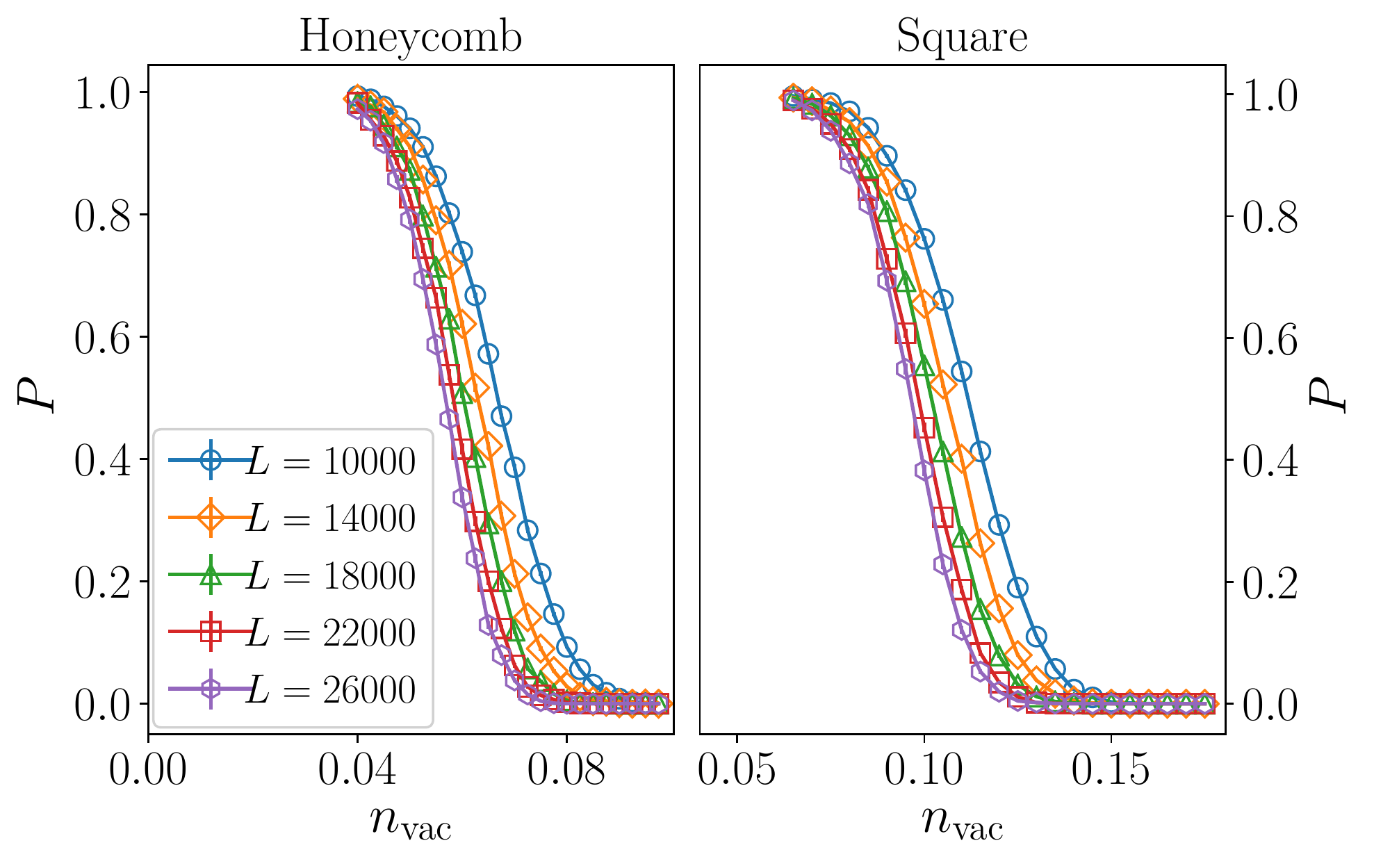}
		 \includegraphics[width=\columnwidth]{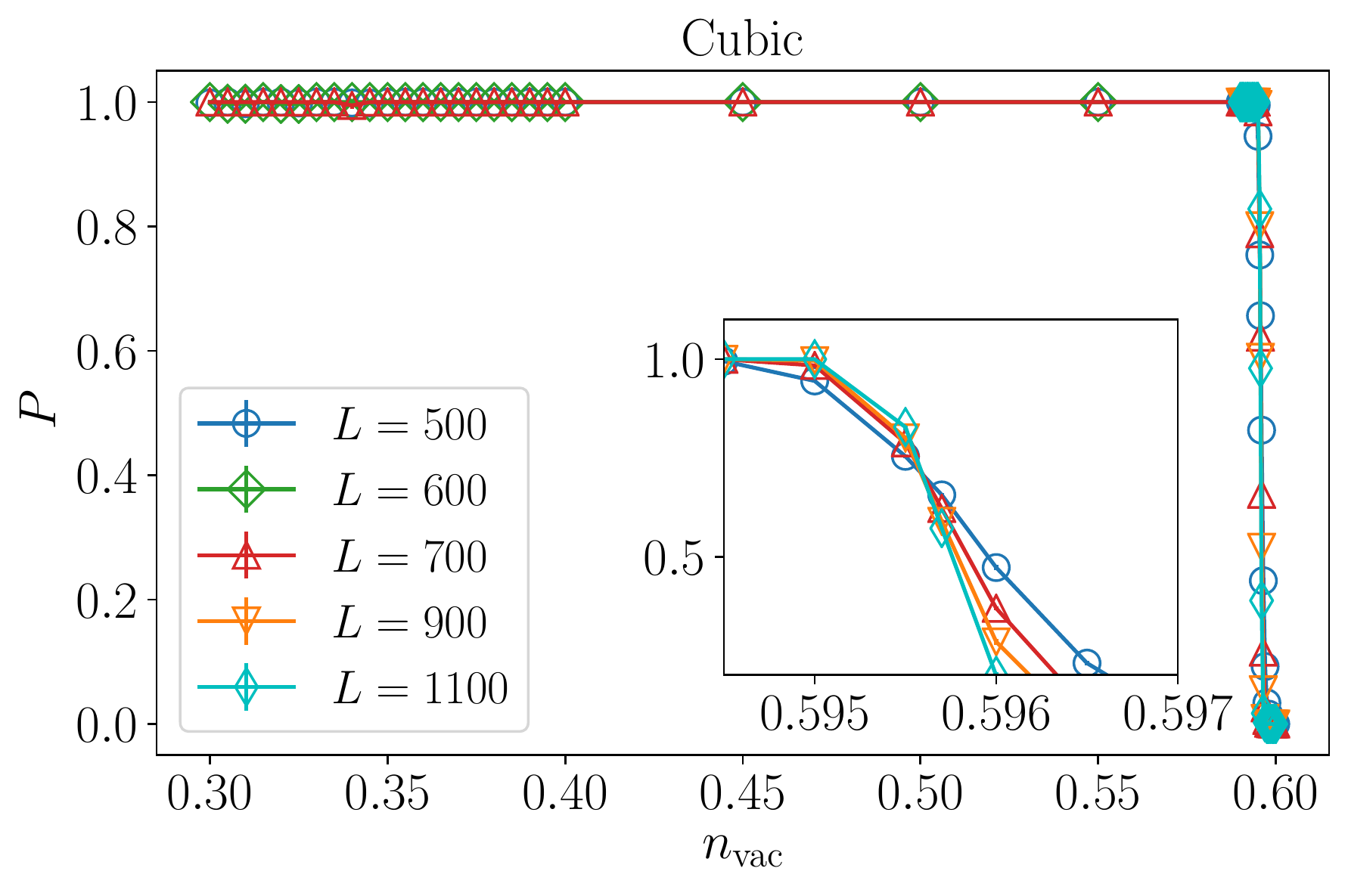}
		\caption{Figures display the probability $P$ that a sample with periodic boundary conditions has at least one ${\mathcal R}$-type region that wraps around the torus in at least one direction. In two dimensions, this tends towards $P=1$ in the small-$n_{\rm vac}$ limit, but curves for different sample sizes $L$ never cross each other at any nonzero $n_{\rm vac}$ accessible to our numerical work. In contrast, for the cubic lattice, this rises sharply to $P \approx 1$ around $n_{\rm vac} \approx 0.6$, with curves of different sizes crossing one another at a sharply-defined threshold value. Sec.~{\protect{\ref{Geometry:Basic}}} for details.}
		\label{Pw}
 \end{figure}

 Finally, in both two and three dimensions, we consider the probability $P$ that a sample with periodic boundary conditions has at least one ${\mathcal R}$-type region that wraps around the torus in at least one direction (Fig.~\ref{Pw}). In classical percolation theory, the analogous quantity provides a simple diagnostic of the percolation transition~\cite{Cardy_conformalinvariancepercolation,Langlands_etal}: If one plots this wrapping probability as a function of concentration for various $L$, these curves cross each other at the critical concentration. In the percolated phase, this probability is closer to $1$ for larger $L$, while the unpercolated phase is characterized by the opposite behaviour. In our two-dimensional case, it is clear that the corresponding curves for different $L$ all tend to $1$ as $n_{\rm vac} \to 0$, but there is no crossing point at any nonzero
 $n_{\rm vac}$ accessible to our numerics. However, on the cubic lattice, we see that there is a clear crossing point in the vicinity of the previously identified threshold near $n_{\rm vac} \approx 0.6$, signalling the presence of a percolation transition at a nonzero critical dilution in this three-dimensional case.

Thus, our two-dimensional results strongly suggest that the $n_{\rm vac} \to 0$ limit in two dimensions may be fruitfully thought of in terms of universality and critical scaling associated with an incipient percolation phenomenon corresponding to a critical point at $n_{\rm vac} = 0$. By the same token, our three-dimensional results motivate a scaling description in the vicinity of the threshold $n_{\rm vac} \approx 0.6$, associated with a transition to a percolated phase. In addition, the results at lower dilution point towards the possibility of a second transition near $n_{\rm vac} \approx 0.35$ within the percolated phase.

\section{Random geometry of ${\mathcal R}$-type regions}
\label{percolation&incipientpercolation}
With this motivation, we now present  finite-size scaling analyses of incipient percolation of ${\mathcal R}$-type regions in the $n_{\rm vac} \to 0$ limit in two dimensions, and  of the corresponding percolation transition in the vicinity of $n_{\rm vac} \approx 0.6$ in the three-dimensional cubic-lattice case. In the three-dimensional case, we also analyze the geometry of the largest ${\mathcal R}$-type regions in the vicinity of a second transition deep within the percolated phase, and show that this is a sublattice symmetry breaking transition. A study of the analogous scaling behaviour of just the odd ${\mathcal R}$-type regions, {\em i.e.}, with odd imbalance ${\mathcal I}$, is also interesting, since each such region hosts a robust zero-energy Majorana fermion excitation of the corresponding bipartite Majorana network described by Eq.~\ref{H_Majorana}. This is presented in the Appendix, which also describes our detailed study of the morphology of the largest ${\mathcal R}$-type regions as a function of the dilution. 
    \begin{figure}
		a)~\includegraphics[width=\columnwidth]{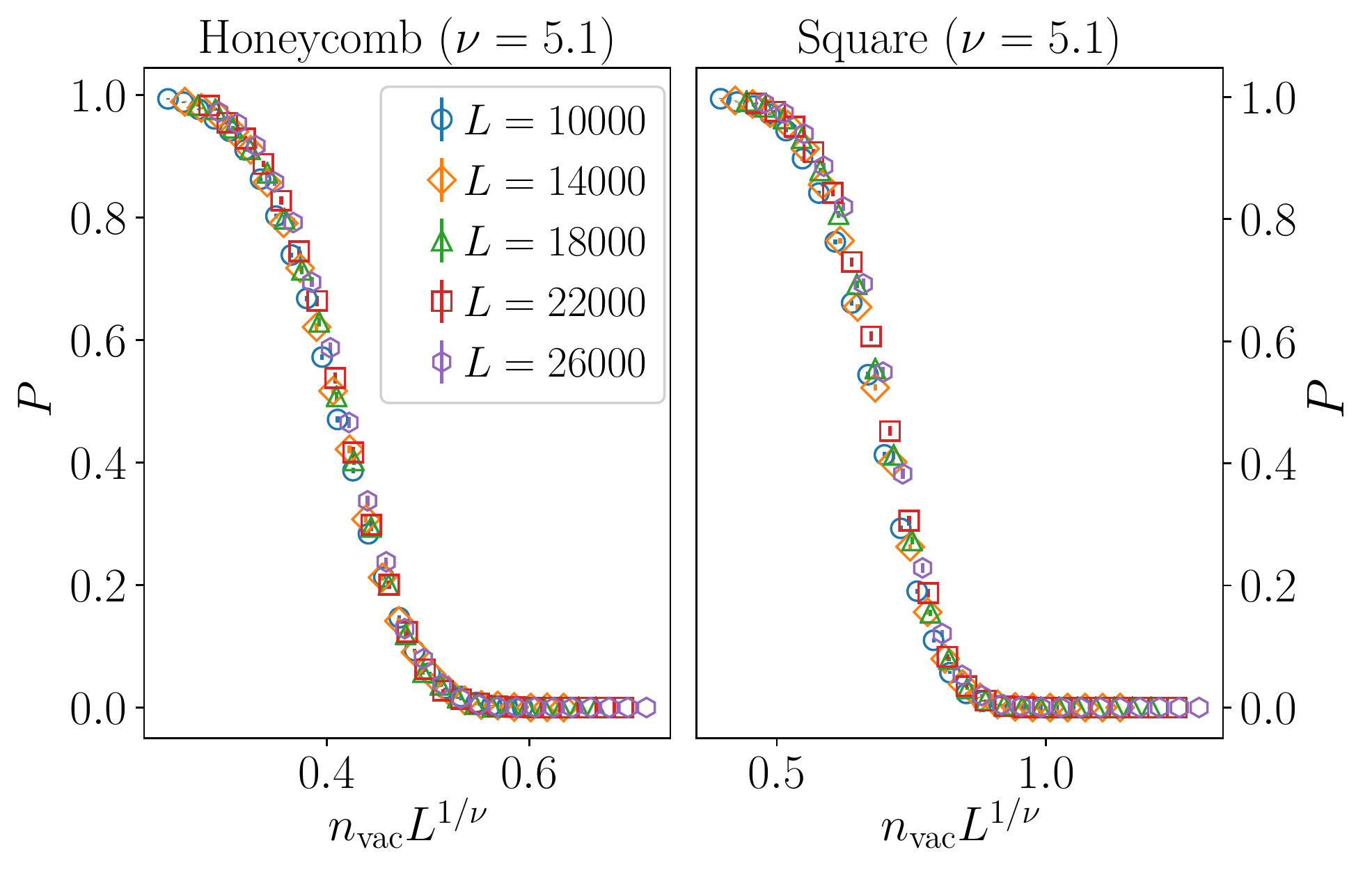}\\
		b)~\includegraphics[width=\columnwidth]{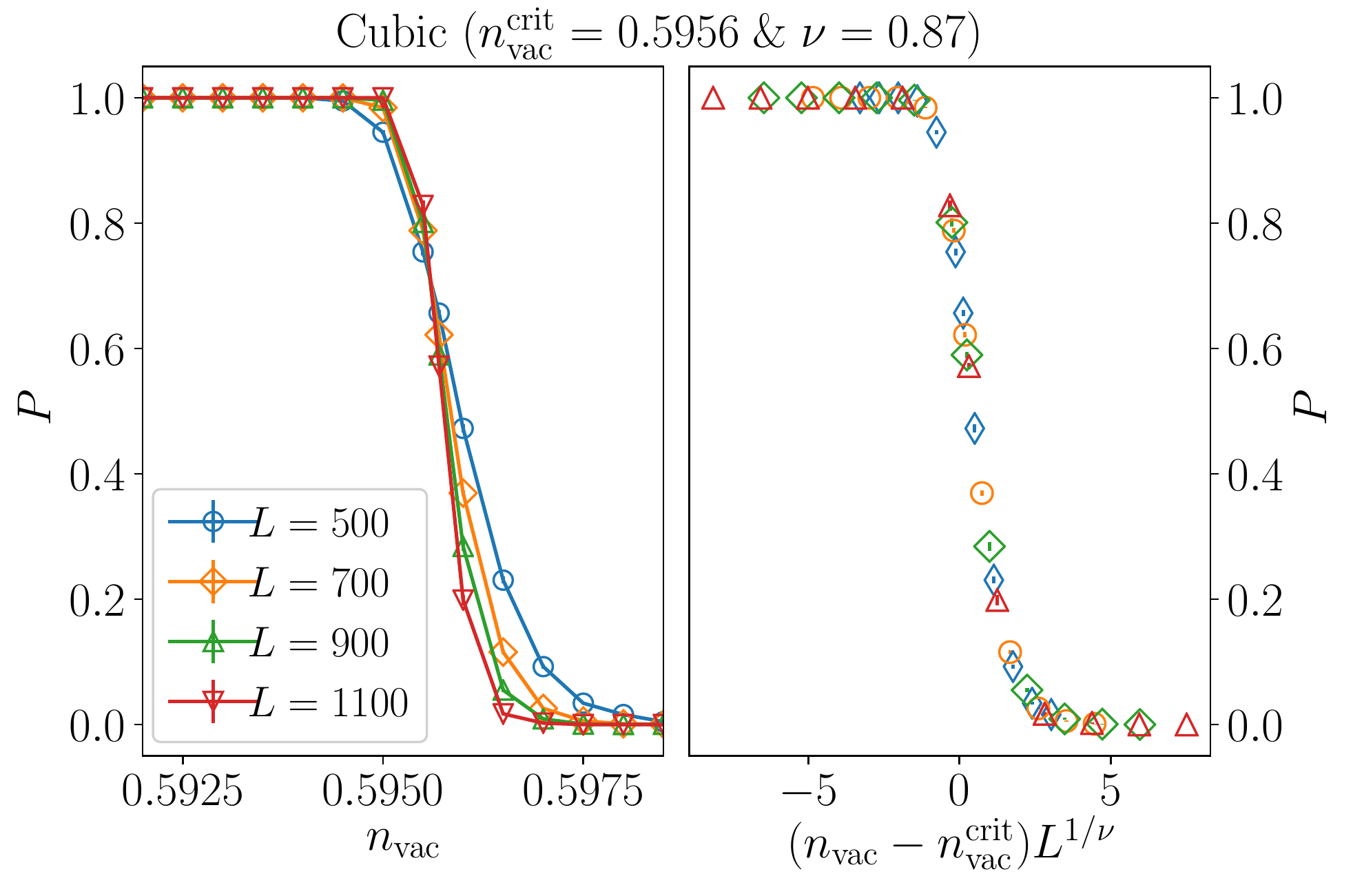}
		\caption{ a) The probability $P$ that a two-dimensional $L \times L$ sample with periodic boundary conditions has at least one ${\mathcal R}$-type region that wraps around the torus in at least one direction collapses onto a single scaling curve when data for various $L$ and $n_{\rm vac}$ are plotted against the scaling variable $n_{\rm vac}L^{1/{\nu}}$ for small values of $n_{\rm vac}$. b) (left panel): A plot against $n_{\rm vac}$ of the corresponding probability $P$ for a $L \times L \times L$ site-diluted cubic lattice. Note that curves corresponding to different $L$ cross at a threshold $n_{\rm vac}^{\rm crit}$ in the vicinity of $n_{\rm vac} = 0.6$. b) (right panel): $P$ in this three-dimensional case for various values of $n_{\rm vac}$ and $L$ collapses on to a single scaling curve when plotted  against the scaling variable $(n_{\rm vac} - n_{\rm vac}^{\rm crit})L^{1/\nu}$ for $n_{\rm vac}$ close to $n_{\rm vac}^{\rm crit}$. See Sec.~{\protect{\ref{percolation&incipientpercolation}}} for details.}
		\label{scalingP}
\end{figure}

\subsection{Overview of analysis}
\label{subsec:analysisoverview}
Our computational resources do not allow  a study of large-enough samples in the small-dilution regime with $n_{\rm vac} < 0.04$ ($n_{\rm vac} < 0.06$) on the honeycomb (square) lattice at present. As a result, we cannot definitively rule out the possibility that there is a percolation transition at a small {\em nonzero} $n_{\rm vac}^{h} \ll 0.04$ ($n_{\rm vac}^{s} \ll 0.06$) on the honeycomb (square) lattice, rather than a $n_{\rm vac}=0$ 
critical point on both lattices.
Nevertheless, Occam's razor dictates that we first explore the simpler description in terms of a $n_{\rm vac}=0$ critical point in two dimensions. On the cubic lattice, we are limited to $n_{\rm vac} \gtrsim 0.2$. Fortunately, both transitions we study on the cubic lattice occur at much higher dilutions, and this limitation is therefore not a serious constraint. Thus, the main caveat attached to the various rather interesting implications of our two-dimensional results is the possibility of a percolation threshold at an extremely small but nonzero $n_{\rm vac}$, both on the square and on the honeycomb lattice. This possibility would leave all our conclusions unaffected for $n_{\rm vac}$ greater than this currently-inaccessible threshold value in two dimensions, while possibly introducing a new regime with volume-law entanglement entropy in the phase diagram of the quantum dimer models of Sec.~\ref{Executivesummary} and a delocalized phase of the particle-hole symmetric quantum percolation problem.

Guided by the usual finite-size scaling ideas~\cite{Cardy_book}, and by an analogy to the scaling picture of the standard geometric percolation transition~\cite{Stauffer_Aharony_book,Christensen_Moloney_book,Cardy_book,Cardy_conformalinvariancepercolation,Langlands_etal}, we ask if various observables in the vicinity of $n_{\rm vac} = 0$ ($n_{\rm vac}^{\rm crit}$) in two dimensions (on the three-dimensional cubic lattice), when rescaled by a suitable power of the system size $L$, collapse onto scaling functions of the scaling variable $\bar{\delta}_s \equiv n_{\rm vac} L^{1/\nu}$
($\bar{\delta}_s \equiv (n_{\rm vac} - n_{\rm vac}^{\rm crit})L^{1/\nu}$)  for a suitable dimension-specific choice of $\nu$, and, in the three-dimensional cubic case, a lattice-specific choice of $n_{\rm vac}^{\rm crit}$.

As in any such finite-size scaling analysis, one can either identify a single best-fit set of exponents that simultaneously achieves a good scaling collapse of all critical properties, or one can find the best-fit exponents individually for each critical observable. In the former approach, the observed quality of the simultaneous scaling collapse serves to validate the underlying scaling hypothesis. In the latter approach, the spread in the values of best-fit exponents obtained from different observables contribute to the uncertainty in estimates of various exponents. We have explored both approaches. Here, we display scaling collapses corresponding to a single set of best-fit exponents to provide the reader with a direct visual confirmation of scaling behaviour, but quote conservative error bars that are dominated by the spread in their values within the second approach.
\begin{figure}
		a)~\includegraphics[width=\columnwidth]{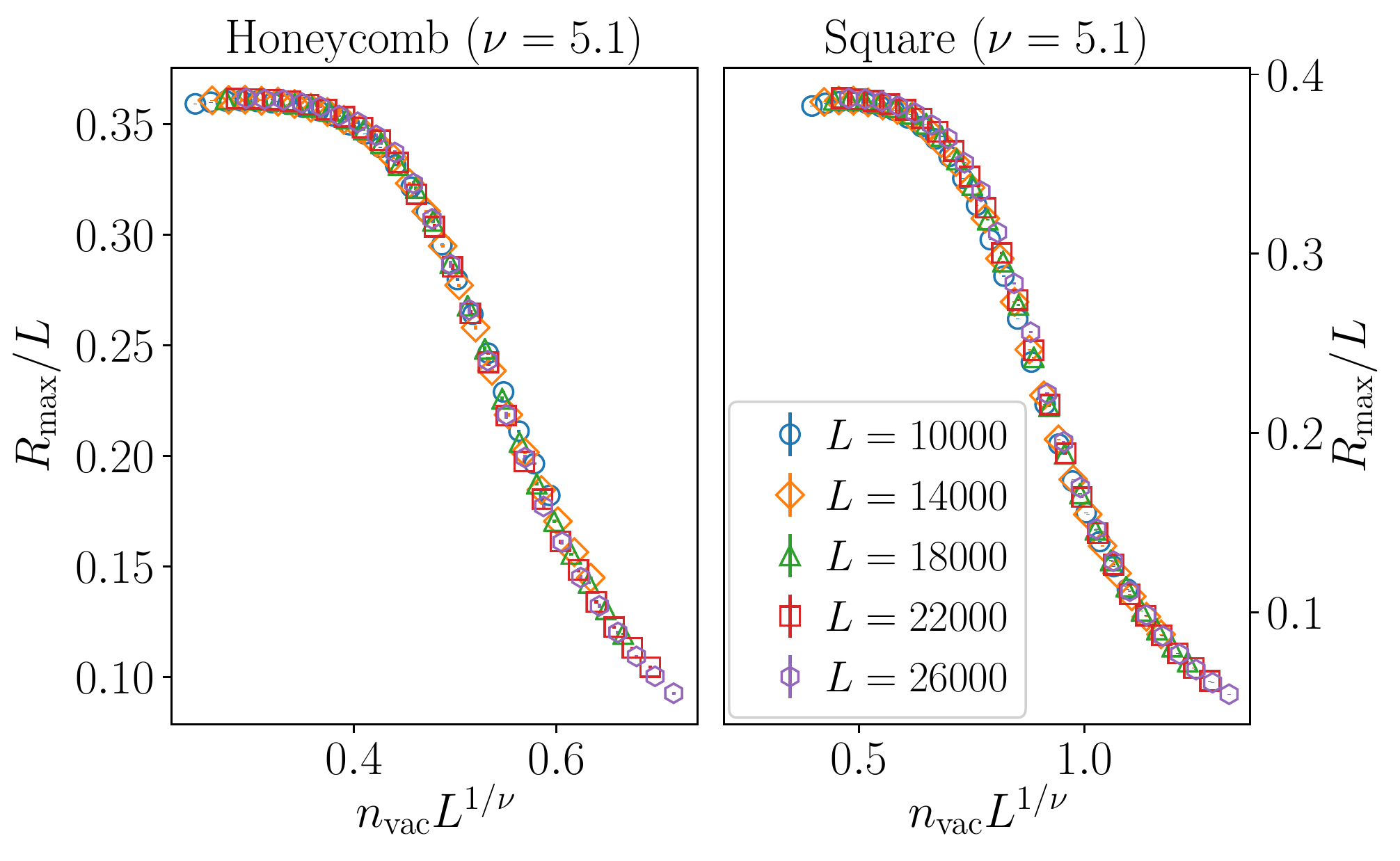}\\
        b)~\includegraphics[width=\columnwidth]{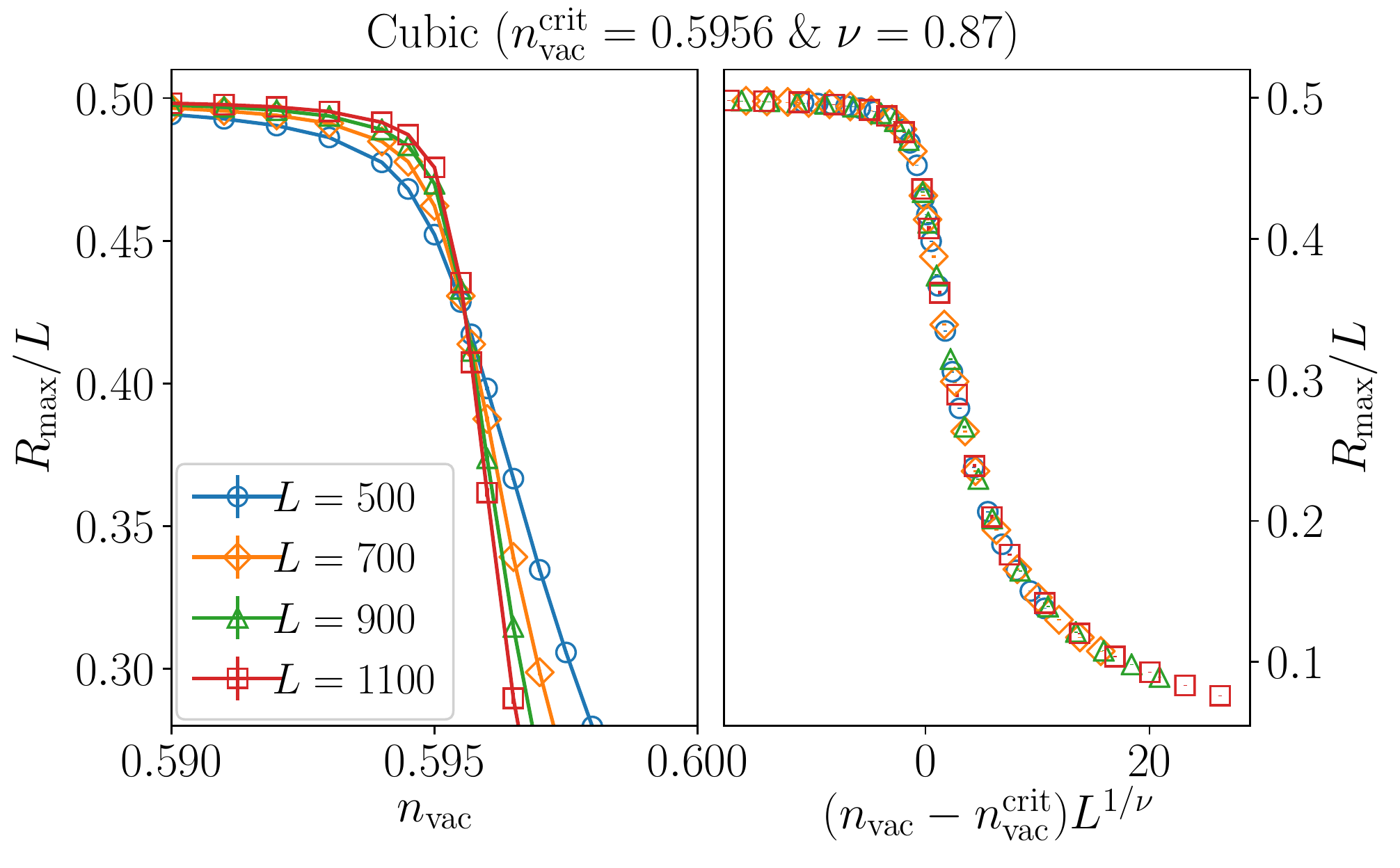}
		\caption{a): The sample-averaged radius of gyration $R_{\rm max}$ of the largest ${\mathcal R}$-type region in a two-dimensional $L \times L$ sample at various values of $n_{\rm vac}$ collapses onto a single scaling curve when plotted as a function of $n_{\rm vac}L^{1/{\nu}}$ for small values of $n_{\rm vac}$. b) (left panel): The same quantity on three-dimensional $L \times L \times L$ cubic lattices, plotted as a function of $n_{\rm vac}$.  Note that curves corresponding to different $L$ display a sharp crossing at a threshold $n_{\rm vac}^{\rm crit}$ near $n_{\rm vac} \approx 0.6$. b) (right panel): Data for $R_{\rm max}$ from $L \times L \times L$ cubic lattices in the vicinity of this threshold collapses onto a single scaling curve when plotted as a function of the scaling variable $(n_{\rm vac} - n_{\rm vac}^{\rm crit})L^{1/\nu}$. See Sec.~{\protect{\ref{percolation&incipientpercolation}}} for details.}
		\label{scalingRmax}
\end{figure}

\subsection{Universal scaling at the percolation threshold}
\label{scalingcollapse}

\begin{figure}
			a)~\includegraphics[width=\columnwidth]{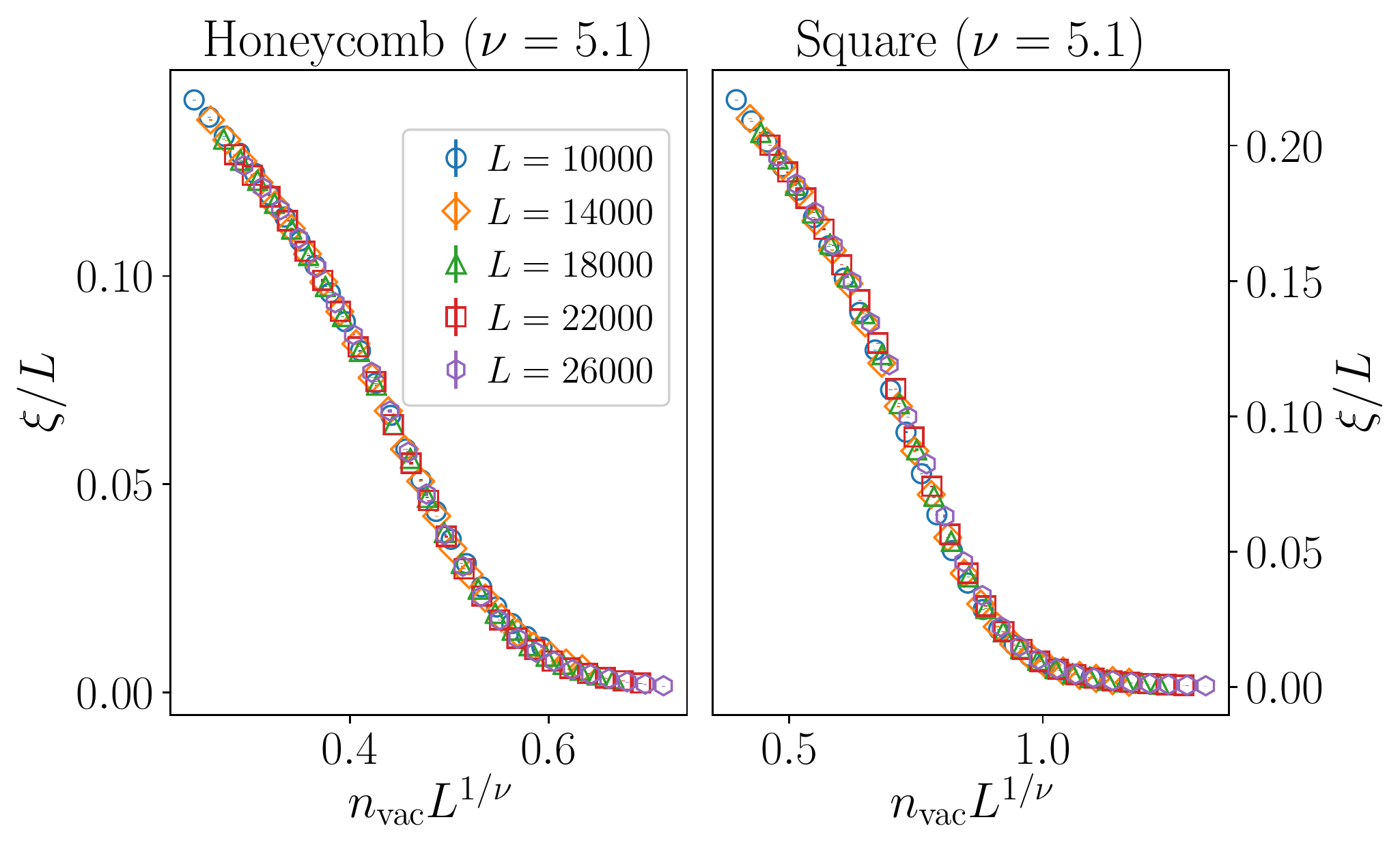} \\
            b)~\includegraphics[width=\columnwidth]{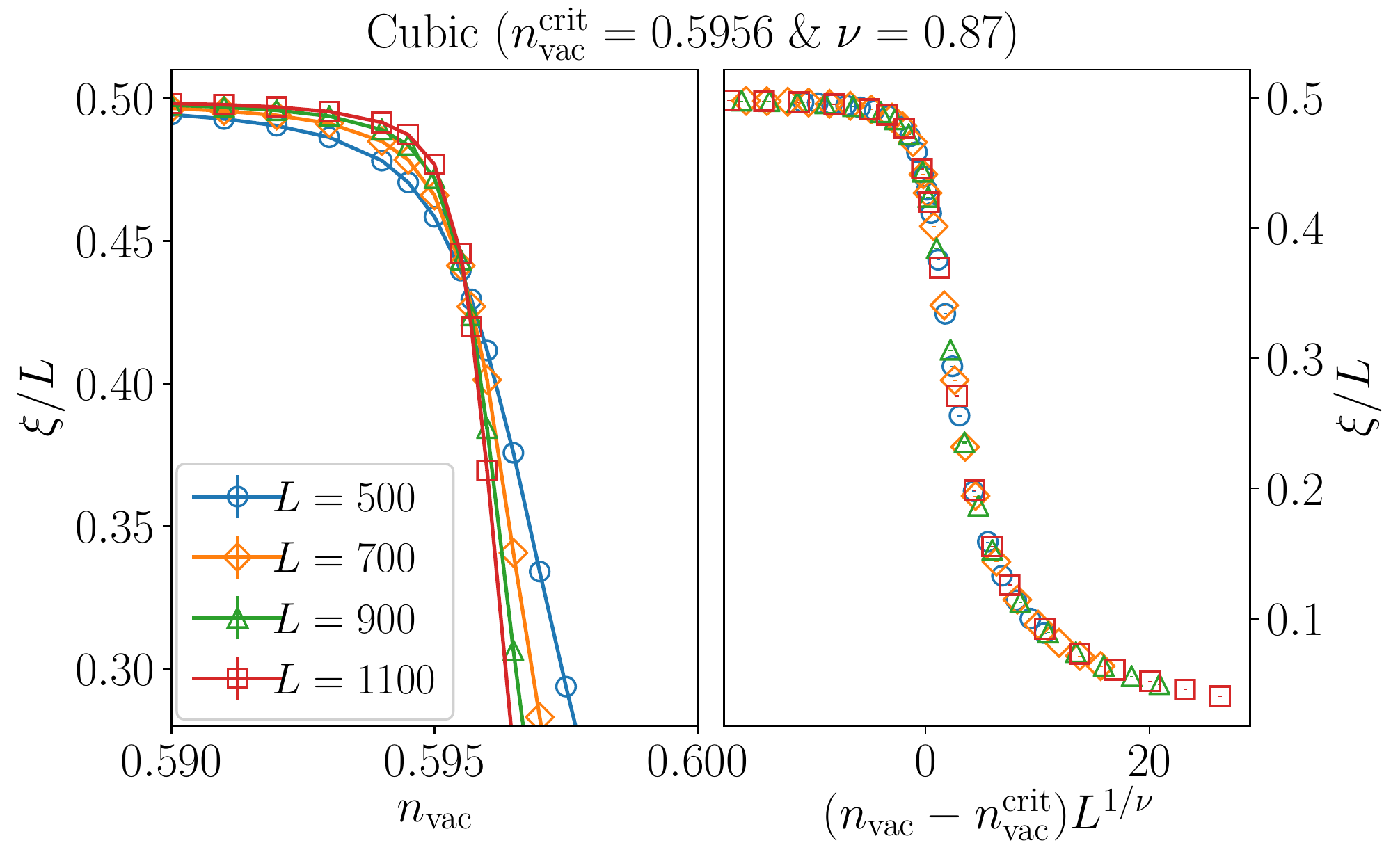} 
		\caption{a) The correlation length $\xi$ corresponding to the sample-averaged geometric correlation function $C(r-r')$ of a two-dimensional $L \times L$ sample at various values of $n_{\rm vac}$ collapses onto a single scaling curve when plotted as a function of $n_{\rm vac}L^{1/{\nu}}$ for small values of $n_{\rm vac}$. Left panel of b): The same quantities on three-dimensional $L \times L \times L$ cubic lattices, plotted as a function of $n_{\rm vac}$.  Note that curves corresponding to different $L$ display a sharp crossing at a threshold $n_{\rm vac}^{\rm crit}$ near $n_{\rm vac} \approx 0.6$. Right panel of b): Data for $\xi$ from $L \times L \times L$ cubic lattices in the vicinity of this threshold collapses onto a single scaling curve when plotted as a function of the scaling variable $(n_{\rm vac} - n_{\rm vac}^{\rm crit})L^{1/\nu}$. See Sec.~{\protect{\ref{percolation&incipientpercolation}}} for details.}
		\label{scalingxi}
 \end{figure}

We begin by studying the scaling behaviour of the wrapping probability $P$. In Fig.~\ref{scalingP}, we see that this scaling hypothesis gives a rather good account of our data for these wrapping probabilities.
Note that these wrapping probabilities, being dimensionless variables, require no rescaling by any power of $L$.
Next, we consider another dimensionless ratio $R_{\rm max}/L$. In Fig.~\ref{scalingRmax}, we again see that this scaling hypothesis gives an extremely good account of the data for $R_{\rm max}/L$. Since $\xi$ represents the correlation length associated with the sample-averaged geometric correlation function, we expect $\xi/L$ should also exhibit scaling behaviour analogous to $R_{\rm max}/L$. In Fig.~\ref{scalingxi}, we see that this is also the case.

\begin{figure}
		a)~\includegraphics[width=\columnwidth]{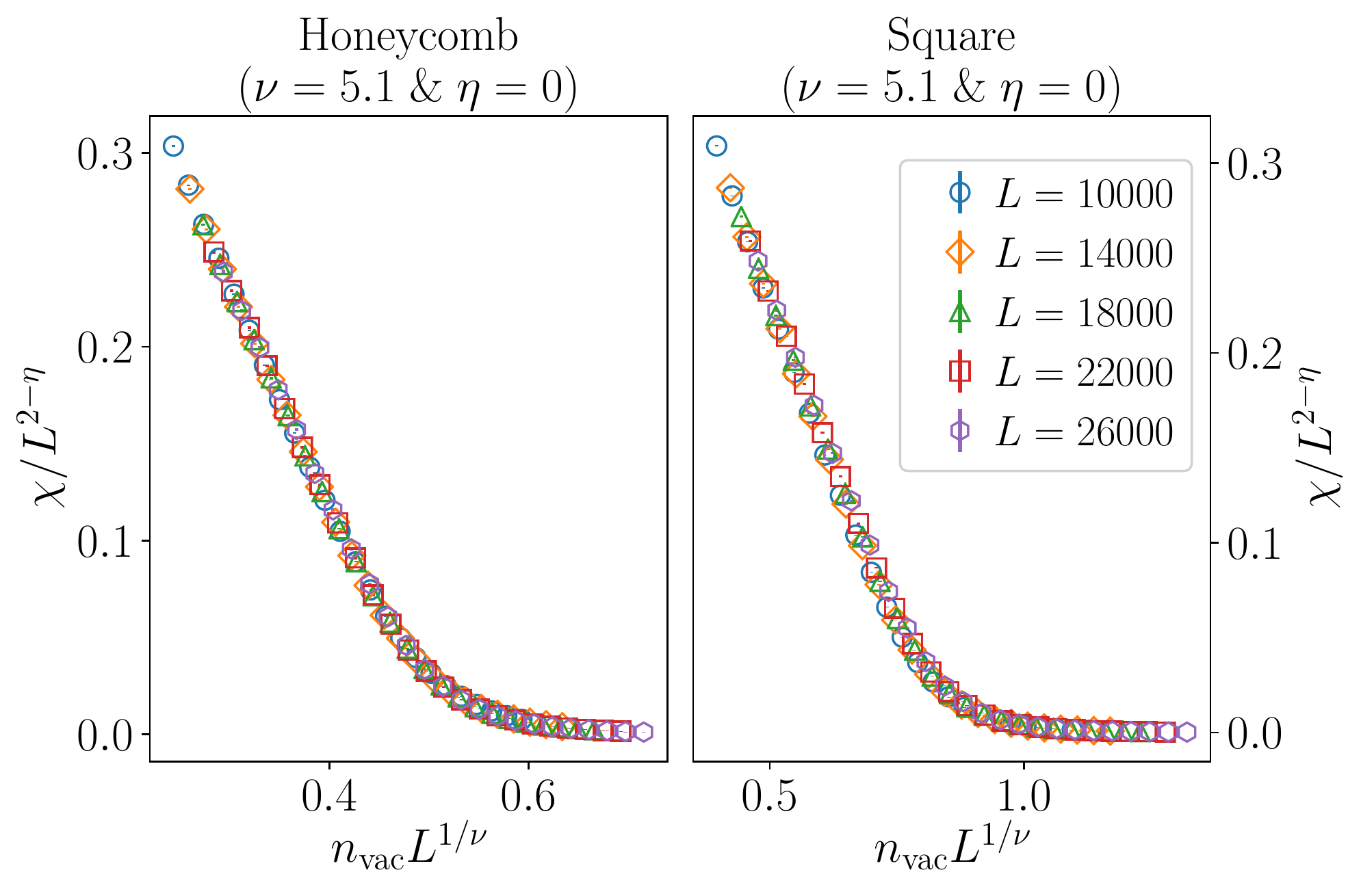}\\
		b)~\includegraphics[width=\columnwidth]{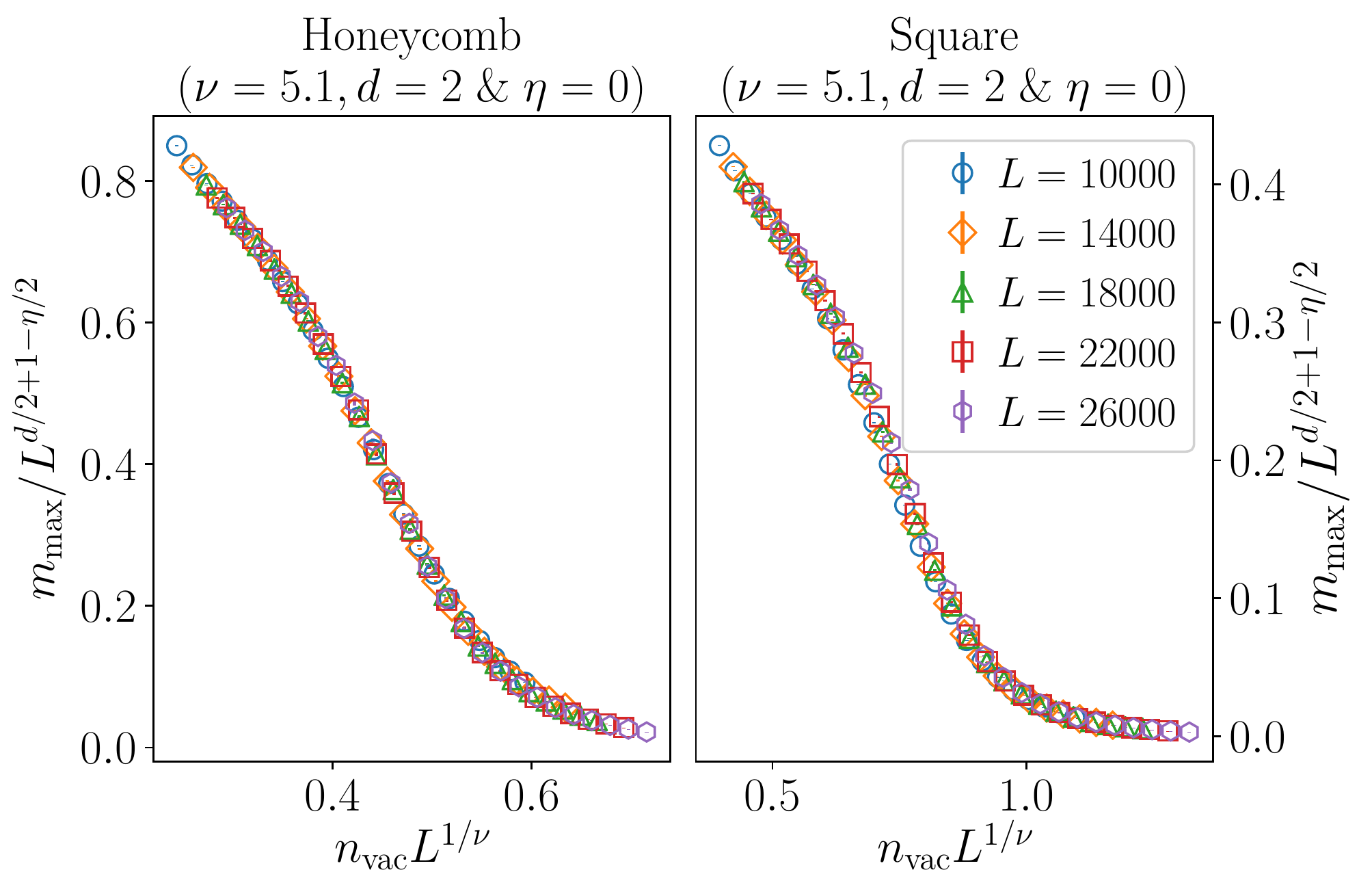}
		\\
		c)~\includegraphics[width=\columnwidth]{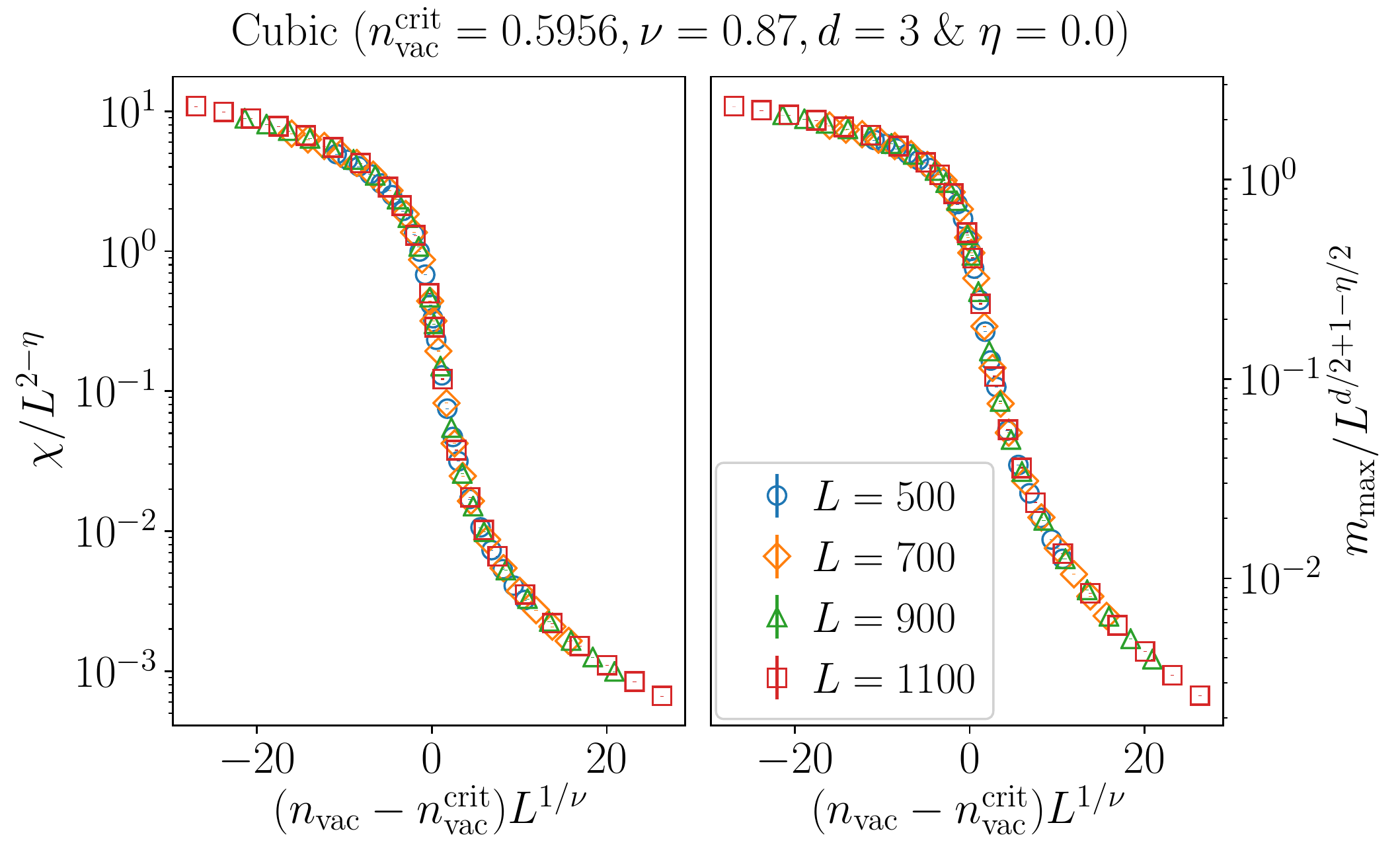}
		\caption{a) and left panel of c): For appropriate dimension-dependent choices of $\nu$ and $\eta$, the susceptibility $\chi$ associated with the sample-averaged geometric correlation function $C(r-r')$ in two-dimensional $L\times L$ samples (three-dimensional $L \times L \times L$ samples), when rescaled by $L^{2-\eta}$, collapse on to a single curve when plotted as a function of $n_{\rm vac}L^{1/\nu}$ ($(n_{\rm vac} - n_{\rm vac}^{\rm crit})L^{1/\nu}$)  for small $n_{\rm vac}$ (for $n_{\rm vac}$ close to $n_{\rm vac}^{\rm crit}$). b) and right panel of c): $m_{\rm max}$, the sample-averaged mass of the largest ${\mathcal R}$-type region in two-dimensional $L \times L$ samples (three-dimensional $L \times L \times L$ samples), when rescaled by $L^{d/2 +1 - \eta/2}$ with $d=2$ (with $d=3$) show analogous scaling behaviour.  See Sec.~{\protect{\ref{percolation&incipientpercolation}}} for details.}
		\label{scalingchimmax}
\end{figure}

Next, we consider two dimensionful quantities, the susceptibility $\chi$, and the mean mass of the largest ${\mathcal R}$-type region, defined as $m_{\rm max} \equiv \langle {\rm max}(m_{\rm max}^A, m_{\rm max}^B)\rangle$. For the susceptibility, we expect that our data for $\chi/L^{2-\eta}$ at various $L$ collapses onto a single curve when plotted versus the scaling variable $\bar{\delta}_s$ for $n_{\rm vac}$ in the vicinity of the putative critical point. Similarly, we expect that $m_{\rm max}/L^{d/2 +1 -\eta/2}$ exhibits an analogous scaling collapse. In Fig.~\ref{scalingchimmax}, we see that these expectations are both borne out by our data. 

We have also studied the scaling behaviour of the analogous observables constructed by considering only the `odd' ${\mathcal R}$-type regions, with odd imbalance ${\mathcal I}$. Our basic conclusion is that restricting attention to such odd ${\mathcal R}$-type regions does not change the scaling picture. Some illustrative examples of this scaling behaviour are presented in the Appendix. From this, we conclude that the spatial extent of topologically-robust zero-energy Majorana excitations also exhibits the same critical behaviour in the $n_{\rm vac} \to 0$ limit. However, we caution that their density $n_{\mathcal R}^{\rm odd}$ does {\em not} exhibit critical behaviour since it is dominated by small odd ${\mathcal R}$-type regions.

Although we have set $\eta=0$ in all the tests of scaling displayed here, we emphasize that scaling collapses of comparable quality can also be achieved for $\eta \lesssim 0.06$ in two dimensions, and $\eta \lesssim 0.03$ in the cubic case. Nonzero values of $\eta$ are generically accompanied by slightly larger best-fit values for the correlation length exponent $\nu$. Folding the results of this systematic analysis into our error estimates, we conclude that $\nu_{\rm 2d} = 5.1 \pm 0.9 $, 
$\nu_{\rm 3d} = 0.87 \pm 0.1$, $\eta_{\rm 2d} \lesssim 0.06$, $\eta_{\rm 3d} \lesssim 0.03$. Additionally, we obtain a rather accurate determination of the Dulmage-Mendelsohn percolation threshold of the cubic lattice: $n_{\rm vac}^{\rm crit} = 0.5956(5)$.
\begin{figure}
		\includegraphics[width=\columnwidth]{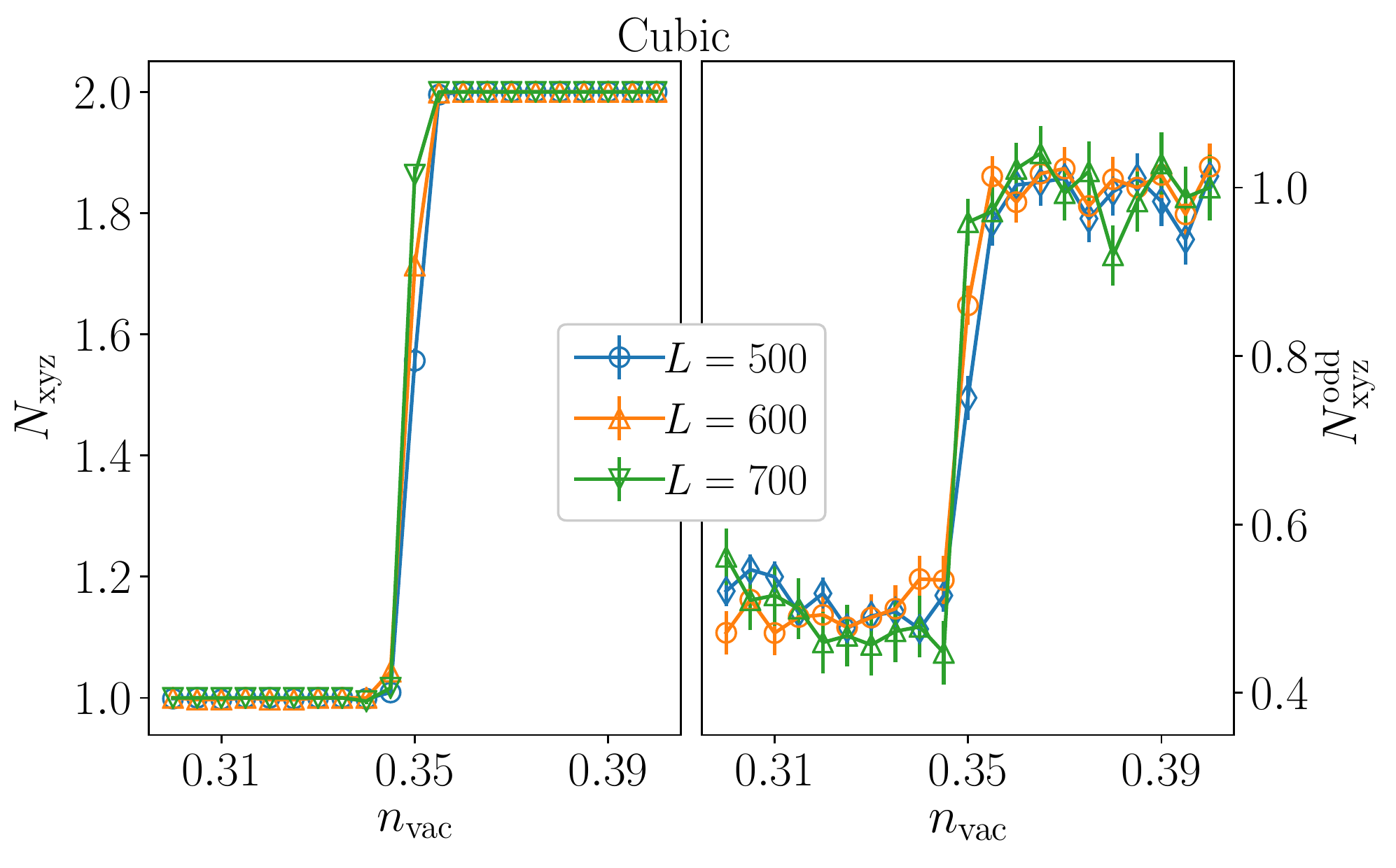}
		\caption{The number $N_{\rm xyz}$ of ${\mathcal R}$-type regions that wrap around the cubic lattice in three independent directions jumps at a sharply defined threshold $n_{\rm vac}^{\rm low} = 0.35(1)$ from $2$ to $1$ in the thermodynamic limit. Correspondingly, the number $N_{\rm xyz}^{\rm odd} $ of such odd ${\mathcal R}$-type regions jumps from a mean value of $1$ to a mean value close to $0.5$. See Sec.~\ref{chiralsymmbreak} for details.}
		\label{nwrapping}
 \end{figure}

\subsection{Sublattice symmetry breaking transition}
\label{chiralsymmbreak}
As noted earlier several properties dominated by the geometry of the largest ${\mathcal R}$-type regions display a distinctive feature in the vicinity of a dilution of $n_{\rm vac} \approx 0.35$ on the cubic lattice. Here we argue that this signals an interesting sublattice symmetry-breaking transition at a sharply defined threshold $n_{\rm vac}^{\rm low} = 0.35(1)$.
To this end, we first note that just below the percolation threshold of $n_{\rm vac}^{\rm crit}$ in the limit of large $L$, we find that there are always exactly two clusters that both wrap in three independent directions around the torus, and no clusters that wrap in just one or two independent directions. A closer inspection of our data reveals that one of these is always a ${\mathcal R}_A$-type region while the other is a ${\mathcal R}_B$-type region.
  \begin{figure}
		\includegraphics[width=\columnwidth]{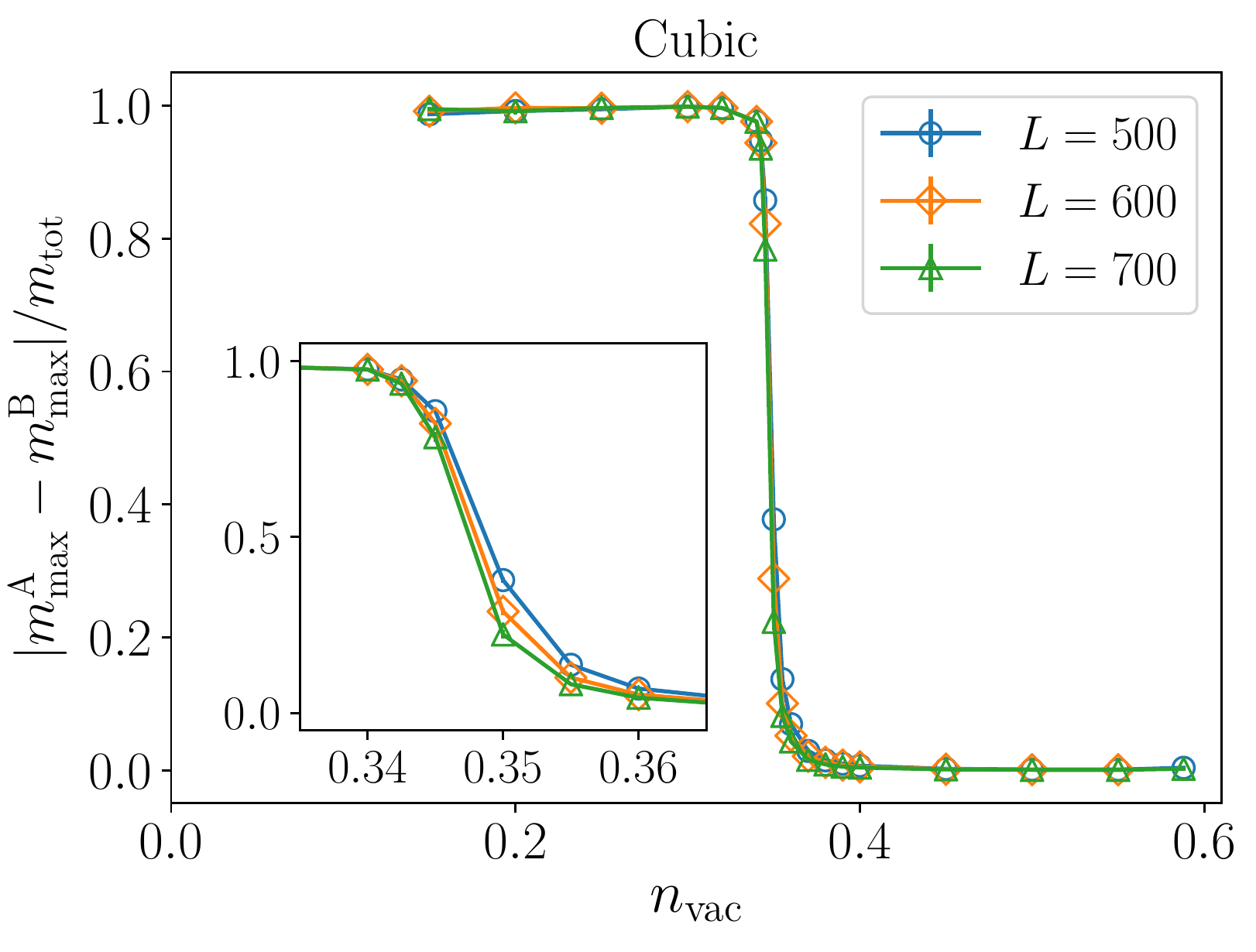}
		\caption{The sample averaged ratio $|m_{\rm max}^{A} - m_{\rm max}^{B}|/m_{\rm tot}$ displays a jump in the thermodynamic limit at a sharply-defined threshold value $n_{\rm vac}^{\rm low} = 0.35(1)$, corresponding to an abrupt onset of sublattice symmetry breaking. See Sec.~\ref{chiralsymmbreak} for details.}
		\label{sublatsymmbreak}
 \end{figure}

To explore this further, we keep track of the number of ${\mathcal R}$-type regions in a sample that wrap in three independent directions in the large-size limit and study the dilution dependence of the corresponding ensemble average $N_{\rm xyz} $ in the low and intermediate dilution regime. From Fig.~\ref{nwrapping}, we see that this number drops from $2$ to $1$ at the threshold value of $n_{\rm vac}^{\rm low} = 0.35(1)$, with this threshold becoming increasingly sharp as one approaches the thermodynamic limit. Unlike the critical scaling in the vicinity of the percolation transition at
$n_{\rm vac}^{\rm crit}$, the behaviour in the vicinity of this threshold has a distinct first-order character, with curves of $N_{\rm xyz}$ corresponding to different $L$ showing no crossing point. The corresponding number for odd ${\mathcal R}$-type regions also shows a sharp drop, from a mean value of $1$ to a mean value close to $0.5$. We also monitor the sample average of the ratio $|m_{\rm max}^{A} - m_{\rm max}^{B}|/m_{\rm tot}$ as a function of the dilution (Fig.~\ref{sublatsymmbreak}). In the thermodynamic limit, we find that this exhibits a sharply defined transition, becoming nonzero as soon as we cross into the low-dilution phase with $n_{\rm vac} < n_{\rm vac}^{\rm low}$. This nonzero value serves as the order parameter corresponding to the spontaneous breaking of sublattice symmetry in this low-dilution phase.

\section{Discussion}
\label{Discussion}

We have provided convincing evidence for a new universality class of percolation phenomena in site-diluted bipartite lattices in two and three dimensions. These phenomena are associated with the end-to-end connectivity of regions of the lattice with local sublattice imbalance, identified using the structure theory of Dulmage and Mendelsohn. This geometric criticality is expected to be associated with monomer percolation in the maximally-packed dimer model on these lattices, with a transition from  area law to volume law in the entanglement entropy of arbitrary many-body eigenstates of the corresponding quantum dimer models, with the quantum percolation transition of a zero-energy particle hopping on such a lattice, and with a Majorana percolation transition governing the spatial extent of collective zero-energy Majorana fermion excitations of the corresponding bipartite network of localized Majorana modes.  In addition, these results on the geometry of regions with local sublattice imbalance also shed light on the origin of low-energy triplet excitations in diluted quantum antiferromagnets.

Here, we first provide a heuristic reinterpretation of our geometric results, and then discuss in more detail their implications in various contexts. In addition, we describe how our results lead to a unified perspective on two apparently disparate strands of recent work, one of which deals with dimer models on the quasi-periodic Penrose tiling, while the other studies tight-binding models on the same lattice.

 \subsection{Heuristic reinterpretation of results}
  \label{Coulomb}
  At the risk of oversimplification, we present here a heuristic interpretation of our results on the random geometry of the Dulmage-Mendelsohn decomposition of diluted bipartite lattices. This is based on the coarse-grained Coulomb phenomenology of fully-packed dimers on the parent bipartite lattices~\cite{Youngblood_Axe_McCoy,Youngblood_Axe,Huse_Krauth_Moessner_Sondhi,Henley_review}. As is well-known, in this way of thinking, a fully-packed dimer configuration corresponds to a divergence-free configuration of a polarization field ${\mathbf P}$. An unmatched site on the $A$ ($B$) sublattice corresponds to the location of a unit positive (negative) electric charge in this description.

  Consider now the undiluted square or honeycomb lattice with boundary conditions that allow a perfect matching. If we remove exactly one $A$ site (say the site $A_0$ at the origin) and look for a maximum matching, it is clear that any such maximum matching leaves one $B$ site of the diluted lattice unmatched. The coarse-grained picture for this ensemble of maximum matchings then has a static unit charge $+1$ at the origin $A_0$, and a mobile unit negative charge that is attracted to the origin by an entropically-generated logarithmic ``Coulomb potential'' $ V(r) = \eta_m \ln(r)$ of strength $\eta_m$; $\eta_m$ is expected to be equal to $1/2$ for the square and honeycomb lattice dimer models~\cite{papanikolaou_luijten_fradkin_prb2007,alet_etal_pre2006,Rakala_Damle_Dhar}. At this coarse-grained level of description, one would say that this mobile charge can diffuse anywhere on the lattice, with the probability for being at distance $r$ from the origin falling off in equilibrium as $\exp(-V(r))$. In the three-dimensional case, the corresponding potential is the usual three-dimensional Coulomb potential of an isolated charge, $V(r) \sim -1/r$ at large $r$.
  In other words, one would heuristically expect that there is a single Dulmage-Mendelsohn region ${\mathcal R}_{A_0}$ that spans the whole lattice. For the square and honeycomb lattices, this can presumably be verified by a direct free-fermion (Pfaffian) calculation of the lattice level partition function with exactly two monomers fixed at two given locations.

  What about a nonzero density of vacancies? One may again think of each vacancy as being a static charge with a sign given by the sublattice. But it is no longer clear if the ``screening cloud'' provided by a mobile monomer on the other sublattice extends over all space. Indeed, in this heuristic language of fluctuating electrostatics, our results on the nontrivial geometry of ${\mathcal R}$-type regions of slightly-diluted samples corresponds to a phenomenon whereby groups of vacancies seed a ``screening-cloud'' of mobile charges that are all of like sign, and confined to a finite-region of the lattice (corresponding to an individual ${\mathcal R}$-type region). 
  
The typical size of this screening cloud grows as $\xi \sim n_{\rm vac}^{-\nu_{\rm 2d}}$ in the $n_{\rm vac} \to 0$ limit in two dimensions; on the cubic lattice, it undergoes a percolation transition at a nonzero dilution threshold $n_{\rm vac}^{\rm crit} = 0.5956(5)$. A curious aspect of our computational results is that such a group of vacancies does not correspond to static charges which all have the same sign, although vacancies with positive (negative) charge do outnumber those with negative (positive) charge whenever the screening cloud is made up exclusively of monomers of negative (positive) charge. The three-dimensional cubic lattice throws up another interesting phenomenon: Below a lower threshold of $n_{\rm vac}^{\rm low} = 0.35(1)$, only one spontaneously chosen kind of polarization cloud percolates in each sample, either with positive screening charge, or with negative screening charge. Screening clouds of the opposite charge remain finite in extent. This spontaneously breaks the sublattice symmetry of the underlying statistical ensemble from which diluted samples are drawn. In contrast, for $n_{\rm vac}^{\rm low} < n_{\rm vac} < n_{\rm vac}^{\rm crit}$, screening clouds of both signs of charge percolate through the sample. 

%It would be interesting if this reinterpretation of our results could be used as the basis for an effective field theory approach to the Dulmage-Mendelsohn percolation phenomena studied here.
It is natural to ask if this restatement of our results in Coulomb language suggests a scaling argument for the scaling behavior at low density. However, any attempt at a low-density scaling argument must first deal with the following difficulty: A key ingredient of any such argument would have to be some kind of large deviation estimate or Griffiths argument for the likelihood of having regions of a given size with extensive local sublattice imbalance {\em and a site boundary made up entirely of sites belonging to the sublattice that is locally in the minority}. It is this constraint on the nature of the boundary that complicates any such argument. We have not been able to overcome this difficulty, which represents an interesting question for future work.

\subsection{Implications for quantum percolation and Majorana percolation}
\label{QuantumPercolation&MajoranaPercolation}

As we have already remarked earlier, ${\mathcal R}$-type regions remain localized in two dimensions at any nonzero dilution $n_{\rm vac}$, no matter how small, but are characterized by a growing length scale $\xi \sim n_{\rm vac}^{-\nu_{\rm 2d}}$ in the limit of $n_{\rm vac} \to 0$. Our results thus imply that the corresponding quantum percolation problem does not have a delocalized phase at any nonzero $n_{\rm vac}$. 
Further our results suggest that the low-dilution limit is associated with interesting scaling behaviour of the localization length $\xi_{\rm loc}$ (defined in Sec.~\ref{GF}) which determines the conductivity tensor of the system at zero chemical potential. However, there is a subtlety associated with this, which we now highlight.

Strictly speaking, our results only tell us that $\xi_{\rm loc} < \xi$. However, since bipartite random hopping problems are characterized by a localization length that diverges as the band center $\epsilon=0$ is approached, we expect that $\xi_{\rm loc}$ will also grow without bound whenever $\xi$ diverges~\cite{Evers_Mirlin,Motrunich_Damle_Huse_GadeWegnerPRB}, although we emphasize that we have not studied this directly in our work here.

Likewise, in three-dimensions, on the cubic lattice at $\mu=0$, our identification of a Dulmage-Mendelsohn percolation threshold $n_{\rm vac}^{\rm crit}=0.5956(5)$ serves to establish the presence of a localized phase of the fermion system for $n_{\rm vac} \in (n_{\rm vac}^{\rm crit}, n_{\rm vac}^{\rm gp})$, where $n_{\rm vac}^{\rm gp} \approx 0.688$~\cite{Stauffer_Aharony_book,Christensen_Moloney_book} is the geometrical site percolation threshold of the cubic lattice. With the same caveat as above, the particle-hole symmetric fermionic system with $\mu = 0$ is thus expected to undergo a metal-insulator transition at $n_{\rm vac}^{\rm crit}$, allowing us to identify the Dulmage-Mendelsohn percolation transition with the $\mu = 0$ quantum percolation transition. What about the sublattice symmetry-breaking that sets in below $n_{\rm vac}^{\rm low}$ on the site-diluted cubic lattice? While the implications of this for transport in the free Fermi system are not entirely clear, it seems likely that a small Hubbard interaction $U$ in this regime will give rise to interesting magnetic properties associated with the formation of local moments whose spatial profile is controlled by $\Delta G(r, r')$. Some closely-related possibilities for follow-up work are discussed in Sec.~\ref{Outlook}.

Further, since ${\mathcal R}$-type regions with odd imbalance ${\mathcal I}$ exhibit the same signatures of Dulmage-Mendelsohn percolation as those with even ${\mathcal I}$, the spatial structure of the collective zero-energy Majorana fermion excitations of the corresponding Majorana networks is also expected to exhibit critical behaviour, with the same caveat as above.
All of this provides a strong motivation for a detailed numerical study of the character of the zero-energy wavefunctions inside the largest ${\mathcal R}$-type regions that dominate both the low-dilution limit in two dimensions, as well as the critical regime in the vicinity of Dulmage-Mendelsohn percolation transition in three dimensions.

\subsection{Area-law scaling of entanglement entropy in maximally-packed quantum dimer models}
\label{MBLprediction}

These computational results on the random geometry of ${\mathcal R}$-type and ${\mathcal P}$-type regions thus imply, via the argument presented in Sec.~\ref{QMD} for the tensor-product structure of arbitrary many-body eigenstates, the existence of a phase with area-law entanglement entropy~\cite{MBL_review}  of all many-body eigenstates of the corresponding maximally-packed quantum dimer models in two and three dimensions. 

In particular, we expect that any nonzero dilution leads to such behavior in such site-diluted quantum dimer models on the square and honeycomb lattices in two dimensions. The prefactor of this area-law entanglement entropy is expected to diverge in the $n_{\rm vac} \to 0$ limit.  Similarly, on the three-dimensional cubic lattice, the corresponding site-diluted quantum dimer models are expected to exhibit such area-law behavior for $n_{\rm vac} > n_{\rm vac}^{\rm crit}$, with the prefactor of the area law again expected to diverge as $n_{\rm vac} \to n_{\rm vac}^{\rm crit}$.  
Since these phases exist at dilutions well below the geometric percolation threshold both in two and in three dimensions, they represent genuinely interesting and nontrivial examples of area-law behavior. 

Likewise, the Dulmage-Mendelsohn percolation transition at $n_{\rm vac}^{\rm crit}$ is expected to correspond to a genuine entanglement transition to area-law behavior in such cubic-lattice quantum dimer models. Curiously enough, the only uncertainty in connection with this expectation has to do with the existence of a volume-law entanglement entropy in the presence of disordered couplings in such quantum dimer models when ${\mathcal R}$-type regions percolate for $n_{\rm vac} < n_{\rm vac}^{\rm crit}$; while our arguments conclusively establish area-law behavior for $n_{\rm vac} > n_{\rm vac}^{\rm crit}$, they do not say anything definite about the existence of a volume law for $n_{\rm vac} < n_{\rm vac}^{\rm crit}$ in the presence of quenched disorder in the coupling strengths.

As noted in Sec.~\ref{QMD}, the tensor-product structure of many-body eigenstates is not entirely fragile, and remains unaffected by additional ring-exchange and potential-energy terms defined on larger flippable loops, as well as next-nearest-neighbor interactions between monomers. Nevertheless, since the three-dimensional transition at $n_{\rm vac}^{\rm crit}$ on the cubic lattice is entirely driven by the underlying Dulmage-Mendelsohn percolation transition, a generic many-body-localization critical point in higher dimensions, if it exists, will most likely be in a different universality class.  Loosely speaking, the situation is akin to the ferromagnetic quantum phase transition of the two-dimensional or three-dimensional transverse field Ising model on a diluted lattice, which is also driven entirely by the underlying percolation transition of the diluted lattice. In that case, the percolation-driven transition~\cite{Senthil_Sachdev_diluted} does share some features with the strong-disorder fixed point~\cite{Mau_etal} controlling the ferromagnetic quantum phase transition of the random transverse field Ising model in higher dimensions. It is not clear if this will be the case for the geometrically-driven transition identified here, since area-law entanglement entropy is only one of several different ways in which many-body localization manifests itself~\cite{MBL_review}. In Sec.~\ref{Outlook}, we outline a specific suggestion for follow-up work that could address a related question.

\subsection{Implications for diluted quantum antiferromagnets}
\label{dilutedAFM}

Turning to implications for diluted quantum antiferromagnets, our work strongly suggests that it would be extremely interesting in follow-up work to repeat the earlier QMC calculations~\cite{Wang_Sandvik_PRL2006,Wang_Sandvik_PRB2010}, but at much smaller values of $n_{\rm vac}$, corresponding to the low-dilution limit of the site-diluted square lattice, rather than the vicinity of the geometric percolation threshold studied earlier~\cite{Wang_Sandvik_PRL2006,Wang_Sandvik_PRB2010}. 

The rationale for this suggestion is as follows:
In the valence bond picture, the antiferromagnetically-ordered ground state can be thought as a wavefunction with a broad distribution of valence bond lengths, so that two spins far away from each other have a sizeable amplitude for freezing into a singlet state. On the other hand, valence bond solid  and spin liquid ground states are expected to be described by a wavefunction in which the valence bonds are short ranged~\cite{Liang_Doucot_Anderson} 

In the extreme limit in which the valence bonds only form singlets between nearest neighbour pairs of spins, it becomes possible make a precise connection between monomers in the maximally-packed dimer model on the one hand and dangling free spins in the magnet; these dangling free spins are expected to lead to a Curie tail in the low-temperature susceptibility. Conversely, in an antiferromagnetically-ordered state, monomers of the maximally-packed dimer model are not expected to be related in any such a direct way to dangling free spins.

Nevertheless, the results of Ref.~\cite{Wang_Sandvik_PRB2010} provide clear evidence that anomalously low-energy triplet excitations in the antiferromagnetically-ordered phase of diluted square lattice antiferromagnets have a spatial distribution that is accounted for quite well by identifying regions of the lattice which can host monomers in the maximally-packed dimer model. These results of Ref.~\cite{Wang_Sandvik_PRB2010} are in a regime with fairly high levels of dilution, close to the geometric percolation threshold. 

Our heuristic explanation for this close correspondence between low-energy triplet excitations and monomer-caryring regions is as follows: As noted above, if valence bonds were strictly restricted to nearest-neighbour links, monomers would correspond to dangling free spins. From this point of view, the main effect of the longer range valence bonds in the antiferromagnetically ordered phase is then to form singlets between two such would-be dangling spins on opposite sublattices of the square lattice. 

Since monomers in a ${\mathcal R}_A$-type (${\mathcal R}_B$-type) region all live on the  $A$ ($B$) sublattice, such a longer range singlet between two dangling spins would need to extend from one ${\mathcal R}_A$-type region to a neighbouring ${\mathcal R}_B$-type region. The longer the range over which such a singlet is formed, the lower is the corresponding binding energy or singlet-triplet gap.
Thus, our picture is that the low-energy triplet excitations found in the antiferromagnetically-ordered phase in Ref.~\cite{Wang_Sandvik_PRB2010} correspond to triplet excitations of these longer-range valence bonds that connect a $A$-sublattice site of a ${\mathcal R}_A$ region with a $B$-sublattice site of an adjacent ${\mathcal R}_B$ region. 

This is plausible because our results show that ${\mathcal R}$-type regions tend to be rather small in extent in the high-dilution regime studied in Ref.~\cite{Wang_Sandvik_PRB2010}. As a result, the longer-range valence bonds characteristic of the antiferromagnetically ordered ground state can readily form singlets between two spins in two different ${\mathcal R}$-type regions, giving rise to low but nonzero energy triplet excitations.

However, in the $n_{\rm vac} \to 0$ limit, our results show that ${\mathcal R}$-type regions are very large in size. In this regime, valence bonds that straddle two adjacent ${\mathcal R}$-type regions would then need to be extremely long. Based on the picture developed above, the corresponding triplet excitation energy would be extremely small. In other words, slightly-diluted samples with very small $n_{\rm vac}$ may be reasonably expected to carry a signature of the diverging length scale $\xi \sim n_{\rm vac}^{-\nu_{\rm 2d}}$ in their triplet excitation spectrum.

The analogous system in three dimensions, say a diluted quantum antiferromagnet on the cubic lattice, presents interesting questions of its own: Does the Dulmage-Mendelsohn percolation transition identified here correspond to a qualitative change in the nature of these triplet excitations? Does the sublattice symmetry breaking found deep in the low-dilution phase give rise to observable effects in the triplet excitation spectrum of the antiferromagnet?  It would be interesting to explore these questions via QMC simulations, although the large length scales involved pose a significant challenge.
Other closely related suggestions for follow-up work are described in Sec.~\ref{Outlook}.

%\bibitem{Sandvik_PRB2002} A.~W.~Sandvik, ``Classical percolation transition in the diluted two-dimensional $S=\frac{1}{2}$ Heisenberg antiferromagnet'', Phys. Rev. B {\bf 66}, 024418 (2002).

%\bibitem{Wang_Sandvik_PRL2006} L.~Wang and A.~W.~Sandvik,``Low-Energy Dynamics of the Two-Dimensional $S=1/2$ Heisenberg Antiferromagnet on Percolating Clusters'',  Phys. Rev. Lett. {\bf 97}, 117204 (2006).

%\bibitem{Wang_Sandvik_PRB2010} L.~Wang and A.~W.~Sandvik, ``Low-energy excitations of two-dimensional diluted Heisenberg quantum antiferromagnets'', Phys. Rev. B {\bf 81}, 054417 (2010).

\iffalse
From our scaling analysis of the $n_{\rm vac} = 0$ critical point in two dimensions, we have obtained the estimate ${\nu_{\rm 2d}} = 5.1 \pm 0.9$ for
the correlation length exponent, and the bound $\eta_{\rm 2d} \lesssim 0.06$ for the corresponding anomalous dimension. From the corresponding scaling analysis in the vicinity of the Dulmage-Mendelsohn percolation transition at $n_{\rm vac}^{\rm crit} = 0.5956(5)$ on the cubic lattice, we have obtained the corresponding estimates $\nu_{\rm 3d} = 0.87 \pm 0.1$ and $\eta_{\rm 3d} \lesssim 0.03$.

These results leave open the intriguing possibility that the associated monomer percolation, wavefunction percolation, and Majorana percolation phenomena (that ride on top of the geometric criticality studied here) require an additional set of independent exponents for their full characterization. Further computational work would be needed to explore this possibility.

\fi

\subsection{An aside: Tight-binding and dimer models on the Penrose tiling}
\label{Aside:Penrose}
There is a fairly large body of work going back nearly four decades, whose focus has been the spectrum of tight-binding models
defined on quasiperiodic lattices, most prominently the Penrose tiling made up of rhombi. For instance, Kohmoto and Sutherland~\cite{ Kohmoto_Sutherland} noted that the hopping Hamiltonian for a quantum mechanical particle hopping along links of the Penrose tiling had extensively degenerate zero-energy states with localized wavefunctions. In subsequent work, Arai and collaborators provided a partial characterization of these localized wavefunctions by identifying certain geometric motifs that supported such states~\cite{Arai_etal}. 
Based on this, they also arrived at a conjecture for the density of such localized zero-energy states in the thermodynamic limit. 

In an insightful analysis~\cite{Koga_Tsunetsugu}, Koga and Tsunetsugu exploited the self-similar nature of the Penrose tiling to arrive at an essentially complete characterization of the geometric motifs that support such localized zero modes, and proved the conjecture of Arai and co-authors.
Recognizing that local sublattice imbalance was an essential feature of these geometric motifs, Koga and Tsunetsugu also obtained a detailed characterization of the antiferromagnetic order that develops for infinitesimal onsite repulsion in the Hubbard model
on this lattice. This local imbalance associated with these geometric motifs was also emphasized in very recent work~\cite{Day-Roberts_Fernandes_Kamenev}.

On the other hand, in the recent work of Flicker and collaborators~\cite{Flicker_Simon_Parameswaran}, essentially the same geometric motifs seem to arise as a by-product of their analysis of the density of monomers in any maximum matching of the Penrose tiling. Their result for the monomer density also corresponds exactly to the previously obtained density of zero modes of the hopping problem. Moreover, their characterization of regions accessible to monomers bears an uncanny resemblance to the earlier characterizations of the localized zero mode wavefunctions of the hopping problem.

Using the perspective developed here, we see that this is no coincidence: Indeed, it becomes apparent that the results of Koga and Tsunetsugu amount to an essentially complete analytic characterization of the Dulmage-Mendelsohn decomposition of the Penrose tiling, which was independently rediscovered in the context of maximum matchings by Flicker and collaborators. Moreover, this suggests other natural questions that appear worth studying. These are discussed in Sec.~\ref{Outlook}.

\section{Outlook}
  \label{Outlook}

  The foregoing results and their interpretation lead us to identify several natural lines of enquiry. We conclude by listing some of these as suggestions for potentially fruitful follow-up studies.

  We begin with a question about the $n_{\rm vac} \to 0$ limit in two dimensions. Our results imply that there is a hierarchy of growing length scales in this limit: $l_{\rm vac} \sim 1/\sqrt{n_{\rm vac}}$, $l_w \sim 1/\sqrt{w}$, and $\xi \sim 1/n_{\rm vac}^{\nu}$, with ${\nu_{\rm 2d}} = 5.1 \pm 0.9$ ensuring that $l_{\rm vac} \ll l_w \ll \xi$.
Thus, ${\mathcal R}$-type regions in finite-size systems with $l_w \ll L \ll \xi$ ``look'' critical, in the sense that that standard finite-size scaling ideas strongly suggest that their  properties would be controlled by the incipient critical point at $n_{\rm vac} = 0$. Given that the critical point of classical percolation in two dimensions has conformal invariance~\cite{Cardy_conformalinvariancepercolation,Langlands_etal}, is there some sense in which a similar enlarged symmetry governs the behaviour of such ${\mathcal R}$-type regions? 

The next two questions concern both this critical regime in the vicinity of $n_{\rm vac} =0$ in $d=2$, and the corresponding critical regime in the vicinity of $n_{\rm vac}^{\rm crit}  = 0.5956(5)$ on the cubic lattice: The localization length $\xi_{\rm loc}$ that controls the conductivity tensor of the free Fermi gas (Eq.~\ref{H_fermion}) obeys the bound $\xi_{\rm loc} < \xi$, but may be expected to also diverge whenever $\xi$ diverges (since the localization length in the closely-related bipartite random hopping problem is expected to diverge as one approaches the band center~\cite{Evers_Mirlin,Motrunich_Damle_Huse_GadeWegnerPRB}). Is the ratio
  $\xi_{\rm loc}/\xi$ in the critical regime characterised by an independent scaling exponent, and does this depend on the presence of hopping disorder? A similar question arises naturally for the monomer and dimer correlation lengths $\xi_M$ and $\xi_D$ in the associated maximally-packed dimer model (Eq.~\ref{Z_monomerdimer}): What is the behaviour of $\xi_M/\xi$ and $\xi_D/\xi$ in this critical regime, and how does it depend on the strength of the bond-disorder?

Fourth, it would clearly be interesting to ask if some of the universal aspects of our results extend to situations in which bonds are diluted randomly instead of sites, both in two and in three dimensions. Fifth, there is the natural question of generalizing to other diluted planar bipartite graphs, most notably hyperbolic graphs similar to those studied recently in the context of circuit quantum electrodynamics~\cite{Kollar_Fitzpatrick_Houck} and network theory~\cite{Krioukov_etal}. The percolation theory of such graphs is a well-developed subject in the mathematical literature~\cite{Benjamini_Schramm}, and it would be interesting to explore the possible critical behaviour of Dulmage-Mendelsohn clusters in this setting.
The next suggestion has to do with a natural generalization to random regular bipartite graphs. For such graphs, there is no notion of geometric distance between vertices, but the question of the distribution of sizes of the Dulmage-Mendelsohn clusters remains interesting. This is because recent work has already identified interesting algorithmic implications of the size of a maximum-matching for some problems in computer science~\cite{Frieze_Melsted}. Given our arguments about the factorization of the monomer-dimer partition function into factors associated with Dulmage-Mendelsohn clusters, it would also be interesting to study the size distribution of Dulmage-Mendelsohn clusters in this algorithmic context. Given that some results for such graphs can be obtained analytically, there is also the intriguing possibility of obtaining some exact results in this setting.

The  seventh question is a natural one regarding the statistics of overlap loops and paths. The dimer model on the undiluted square and honeycomb lattices has a useful coarse-grained description in terms of a compact scalar height field with a Gaussian action~\cite{Youngblood_Axe_McCoy,Youngblood_Axe,Henley_review,Kenyon_dimers}. Dimer correlations, and correlations of test-monomers are readily related to correlation functions of this Gaussian theory, which can be computed exactly. In a certain well-defined sense, long-distance behaviour of correlation functions in these dimer models exhibit conformal invariance~\cite{Kenyon_dimers}. The double dimer model~\cite{Kenyon_doubledimers} consists of two independent copies of the dimer model, with partition function given by the square of the dimer model partition function. In such a double-dimer model, the interesting observables are overlap loops built by tracing closed paths that alternately go along dimers in one copy and then the other. This defines an ensemble of loops, which is related to the contour lines of a Gaussian free field theory~\cite{Henley_Kondev_PRL,Desai_Pujari_Damle} and exhibits conformal invariance in the scaling limit~\cite{Kenyon_doubledimers}. In the diluted case, the analog of the double-dimer model involves two copies of the maximally-packed dimer model. This defines an ensemble of overlap loops and paths. The question then arises: What is the statistics of these loops and paths in the small-$n_{\rm vac}$ critical regime? The overlap loop ensemble defined by the fully-packed dimer model on the cubic lattice is also interesting~\cite{Nahum_thesis}. This motivates a study of the corresponding ensemble of loops and paths on the site-diluted cubic lattice, particularly in the vicinity of the Dulmage-Mendelsohn percolation transition.

 The next set of questions have to do with the physics of SU(N) antiferromagnets in a certain large-$N$ limit~\cite{Read_Sachdev} that has
 played a key role in the subsequent conceptual development of our understanding of quantum disordered phases of magnets. In these SU(N) models, one sublattice carries the fundamental representation and the other has SU(N) spins that transform under the complex-conjugate of the fundamental. Thinking in terms of the corresponding Hubbard model, each site has $N$ different fermion orbitals with the constraint that the total fermion number of $A$-sublattice sites is $1$, while that of $B$-sublattice sites is $N-1$.  In the large-$N$ limit of this model, the physics is dominated by the subspace of singlet states spanned by any fully-packed configuration of nearest-neighbor SU(N) singlet bonds, and reduces to a quantum dimer model on the lattice at leading order in
 $1/N$. At large but finite $N$, longer-range valence bonds come into play. These models are amenable to explicit computational study, for example using Quantum Monte Carlo (QMC) methods that work in the basis of bipartite (but not necessarily nearest-neighbor) valence bonds~\cite{Sandvik_Evertz,Beach_Sandvik,Mambrini_dimeraspects}. Such computational approaches have been used to study the physics of these systems as a function of $N$, finding a transition from quantum antiferromagnetism at $N=2$ to a valence-bond solid state above a threshold value of $N$~\cite{Beach_etal}.

 Clearly,  the analogous large-$N$ limit of the diluted model will exhibit interesting effects associated with the presence of a nonzero density of monomers in the corresponding maximally-packed dimer model, since these monomers are expected to be associated with SU(N) spinon degrees of freedom that cannot be quenched by short-ranged singlet bonds; this would be particularly interesting when the typical size $\xi$ of ${\mathcal R}$-type regions becomes very large. It would thus be interesting in the site-diluted case to revisit this large-$N$ limit, and to study corresponding finite-$N$ behaviour using QMC simulations. It would also be interesting to study resonating nearest-neighbor valence-bond wavefunctions~\cite{Tang_Henley_Sandvik_nnRVB,Albuquerque_Alet_nnRVB} for such
 SU(N) antiferromagnets: In the pure case, these are singlet wavefunctions which map to interesting loop ensembles that interpolate between the classical dimer model and the double-dimer model~\cite{Damle_Dhar_Ramola_nnRVB,Patil_Dasgupta_Damle_nnRVB,Albuquerque_Alet_Moessner_3d}. In the site-diluted case, their generalizations will describe degenerate ground states with a nonzero spinon number~\cite{Banerjee_Damle,Banerjee_Damle_Alet,Banerjee_Damle_Alet_SU3,Sanyal_Banerjee_Damle,Tang_Sandvik,Sanyal_Banerjee_Damle_Sandvik}, and map on to an ensemble of loops and paths closely related to the double-dimer model on the diluted lattice. It would be interesting to study the spinon localization properties of these wavefunctions on such slightly-diluted lattices, especially given the divergent size of the  ${\mathcal R}$-regions studied here. It seems likely that these studies will add to our understanding of the physics of dangling spins and local moments studied earlier~\cite{Sandvik_PRB2002,Wang_Sandvik_PRL2006,Wang_Sandvik_PRB2010,Changlani_Ghosh_Pujari_Henley_PRL2013,Ghosh_Changlani_HenleyPRBB2015}.
 
 Direct analogs of these questions are also potentially interesting in the context of Penrose tilings, since the results of Ref.~\cite{Koga_Tsunetsugu} and Ref.~\cite{Flicker_Simon_Parameswaran} provide us with an essentially complete analytic determination of the Dulmage-Mendelsohn decomposition of the Penrose tiling. Although there is no geometric criticality at play in this case, it would clearly be of interest to i) use these analytical results to explore the physics of the Read-Sachdev large-$N$ limit of SU(N) antiferromagnets, ii) perform a QMC study of the related physics at large but finite $N$, iii) to understand the nature of the SU(N) nearest-neighbor RVB wavefunctions mentioned above, and iv) to study the closely-related ensemble of overlap loops and paths defined by the double-dimer model.

Next, we note that our results read in conjunction with those of Ref.~\cite{Koga_Tsunetsugu} suggest some interesting questions about local-moment formation in the particle-hole-symmetric Hubbard model on site-diluted bipartite lattices. Localized states tied to the Fermi energy $\mu =0$ are expected to be intimately connected with the physics of local moment formation~\cite{Milovanovic_Sachdev_Bhatt} in such situations. How is this physics affected by the large length scale $\xi$ associated with the typical size of ${\mathcal R}$-type regions at low dilution, and by the presence of a nonzero density of coexisting zero modes in each such region? Does the topologically-protected nature of the zero modes lead to these moments being relatively robust to perturbations that preserve the particle-hole symmetry? How is the spontaneous breaking of sublattice symmetry reflected in the magnetic properties of this Hubbard model for $n_{\rm vac} < n_{\rm vac}^{\rm low}$ on the cubic lattice? 
Given that recent scanning tunneling microscopy experiments~\cite{Zhang_etal} have detected direct signatures of $\pi$-electron magnetism associated with vacancies in undoped graphene, these questions may be of particular interest in the context of vacancy defects in graphene.

Our work also throws up interesting questions about the thermodynamic susceptibility of Kitaev's honeycomb model with nonmagnetic impurities. From the detailed analysis of vacancy effects in Ref.~\cite{Willans_Chalker_Moessner_PRB}, it is clear that a vacancy-induced pile-up of low-but-nonzero energy fermionic excitations is associated with a weak singularity in the low temperature susceptibility. By analogy with the results of Ref.~\cite{Sanyal_Damle_Chalker_Moessner} on a SU(2) symmetric version~\cite{Yao_Lee} of the Kitaev model, the topologically-protected zero-energy states studied here are expected to lead to a stronger Curie-like singularity $\chi(T) \sim {\mathcal C}/T$ in the linear susceptibility. How does the Curie coefficient ${\mathcal C}$ scale in the small-$n_{\rm vac}$ limit
of a weakly-diluted honeycomb lattice? And does the topologically-protected nature of the zero-energy states endow this Curie term with some degree of protection against
time-reversal invariant perturbations such as a Heisenberg exchange term in the spin Hamiltonian?

Finally, we mention what is perhaps the most computationally-challenging suggestion for follow-up work: As noted in Sec.~\ref{MBLprediction}, our identification of a phase with area-law entanglement entropy of arbitrary eigenstates in a class of maximally-packed quantum dimer models (Fig.~\ref{Quantumdimermodel} of Sec.~\ref{Executivesummary}) on diluted bipartite lattices relies on the existence of a particular tensor-product structure (derived in Sec.~\ref{QMD}) of arbitrary many-body eigenstates. Since this structure is disrupted by the presence of additional next-next-nearest-neighbor interactions between monomers (see Sec.~\ref{QMD}), the question arises: Can the computational methods of Ref.~\cite{Fabien_MBLinQDM_1} and Ref.~\cite{Fabien_MBLinQDM_2} be extended to study the effect of this additional interaction on the area-law phases identified here? Does this area-law behavior survive when this additional interaction is weak but nonzero?

As one goes through this list of suggestions for follow-up studies, it is clear that our work opens up a number of interesting and potentially fruitful lines of enquiry.  It also becomes obvious that the elephant in the room throughout has been the bipartite nature of the underlying lattice.  Are there natural generalizations to the nonbipartite case of any of the geometric questions studied here? Are the corresponding results in the small-dilution limit interesting? The answer to the first question turns out to be in the affirmative~\cite{Damle_unpublished}, motivating follow-up studies aimed at addressing the second question.
%~\cite{Bhola_etal}

  \section{Acknowledgements}
  \label{Acknowledgements}
  
  We are grateful to T.~Kavitha and A.~Mondal for introducing us to the literature on graph decompositions, D.~Sen for stimulating discussions on the robustness of Majorana modes in various contexts, S.~Roy for a useful discussion on many-body localization, D.~Dhar and Mahan Mj for pointers to the literature on percolation theory, and D.~Dhar for helpful comments on a previous version of this paper. We also acknowledge useful discussions with S.~Bera, S.~Bhattacharjee, R.~Dandekar, F.~Evers, F.~Flicker, S.~A.~Parameswaran, J.~Radhakrishnan, K.~Ramola, S.~Ramasesha, A.~W.~Sandvik, R.~Sensarma, V.~Tripathi, and H.~Tsunetsugu. 
  Some of RB's work on the two-dimensional case contributed to his pre-registration DP-II project submission (2020) to the Tata Institute of Fundamental Research (TIFR) Deemed University, while the work of SB on the two-dimensional case formed a part of his Ph.D thesis submission (2019) to the TIFR Deemed University. SB thanks S.~Ramasesha and A.~W.~Sandvik for useful comments in their referee reports on his Ph.D thesis submission. KD gratefully acknowledges earlier collaborations with J.~T.~Chalker, R.~Moessner, O.~Motrunich, and S.~Sanyal which provided part of the motivation for this work. The work of RB, SB, and MMI was supported by graduate fellowships of the TIFR. KD was supported at the TIFR by DAE, India and in part by a J.C. Bose Fellowship  (JCB/2020/000047) of SERB, DST India, and by the Infosys-Chandrasekharan Random Geometry Center (TIFR). All computations were performed using departmental computational resources of the Department of Theoretical Physics, TIFR. During the completion of this manuscript, SB was supported by the European Research Council Grant 804213-TMCS while at Oxford University. 

{\em Statement of author contributions}: RB and SB contributed equally to this
work. SB and MMI devised efficient implementations of the various maximum matching algorithms~\cite{Duff_Kaya_Ucar} used here.  SB devised an efficient implementation of a burning algorithm to obtain all ${\mathcal R}$-type and 
${\mathcal P}$-type regions from a single maximum matching. RB and SB (RB) performed a study of the two-dimensional 
(three-dimensional) case using these computational tools. RB, SB, and KD (RB and KD)
analyzed the two-dimensional (three-dimensional) data. KD conceived and
directed this project, and wrote the manuscript based on results from these data analyses.

\clearpage

\appendix*

\section{Scaling of odd ${\mathcal R}$-type regions and morphology of the largest ${\mathcal R}$-type regions}
\label{morphology}

As noted in Sec.~\ref{Geometry:Basic}, the density $n_{\mathcal R}^{\rm odd}$ of odd ${\mathcal R}$-type regions is nonzero in the thermodynamic limit, signalling the presence of a thermodynamic density of collective zero-energy Majorana fermion excitations of bipartite Majorana networks described by Eq.~\ref{H_Majorana}. Here, we provide evidence of critical behavior of such excitations associated with large odd ${\mathcal R}$-type regions. This is important from the perspective of perturbatively stable Majorana modes hosted by the odd ${\mathcal R}$-type regions. Indeed, this scaling analysis confirms that these collective zero-energy Majorana fermion excitations of bipartite Majorana networks defined on such diluted lattices also exhibit the Dulmage-Mendelsohn percolation phenomena that are the focus of our work.

\begin{figure}
		\includegraphics[width=\columnwidth]{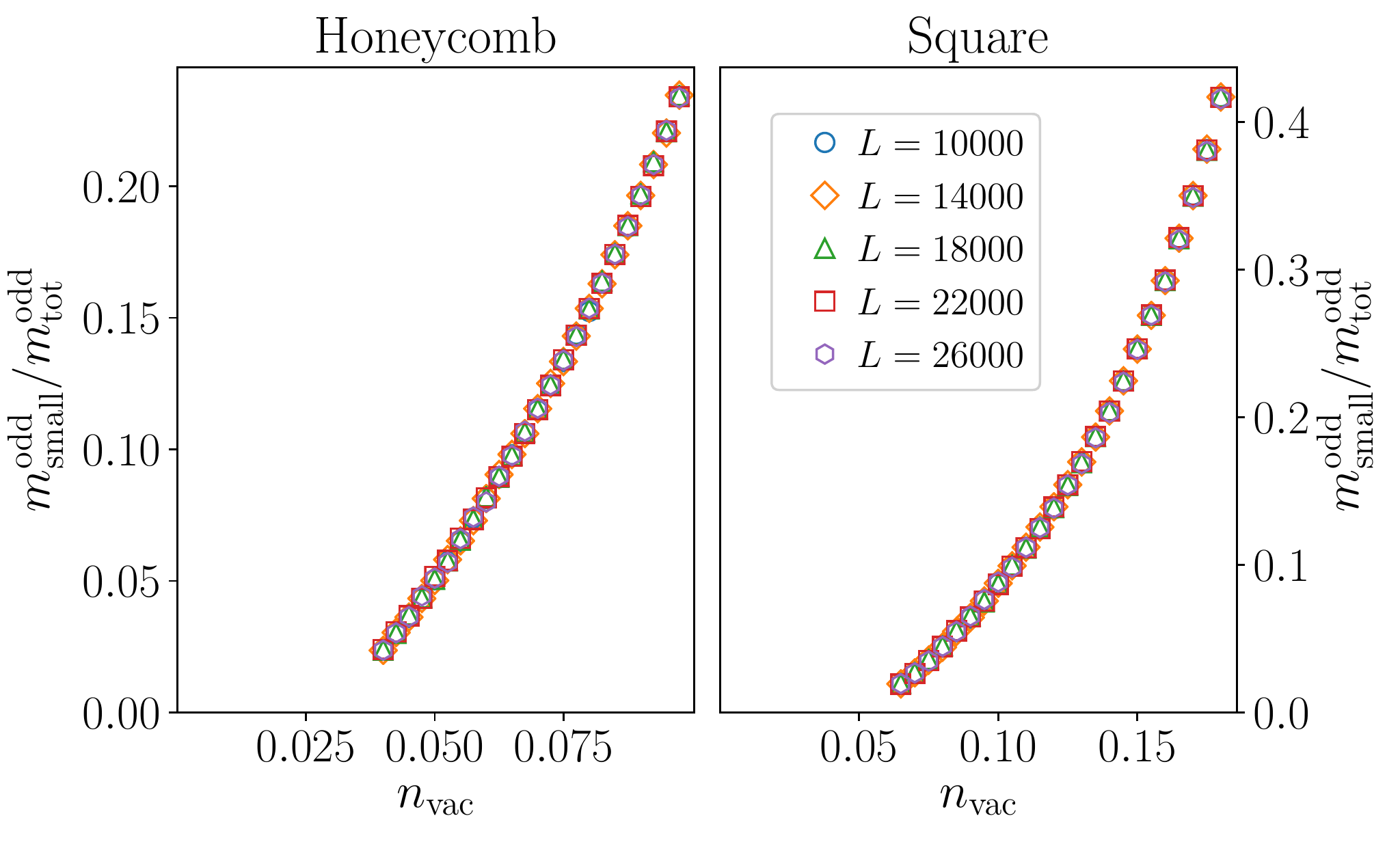}
		\caption{The sample averaged ratio $m_{\rm small}^{\rm odd}/m_{\rm tot}^{\rm odd}$ where $m_{\rm small}^{\rm odd}$ is the contribution of ``small'' odd ${\mathcal R}$-type regions (with odd imbalance and less than  $10000$ vacancies associated with them) decreases rapidly with $n_{\rm vac}$. This may be compared with results discussed in Sec.~{\protect{\ref{Geometry:Basic}}}}
		\label{oddsmallfraction_m}
\end{figure}

\begin{figure}
		\includegraphics[width=\columnwidth]{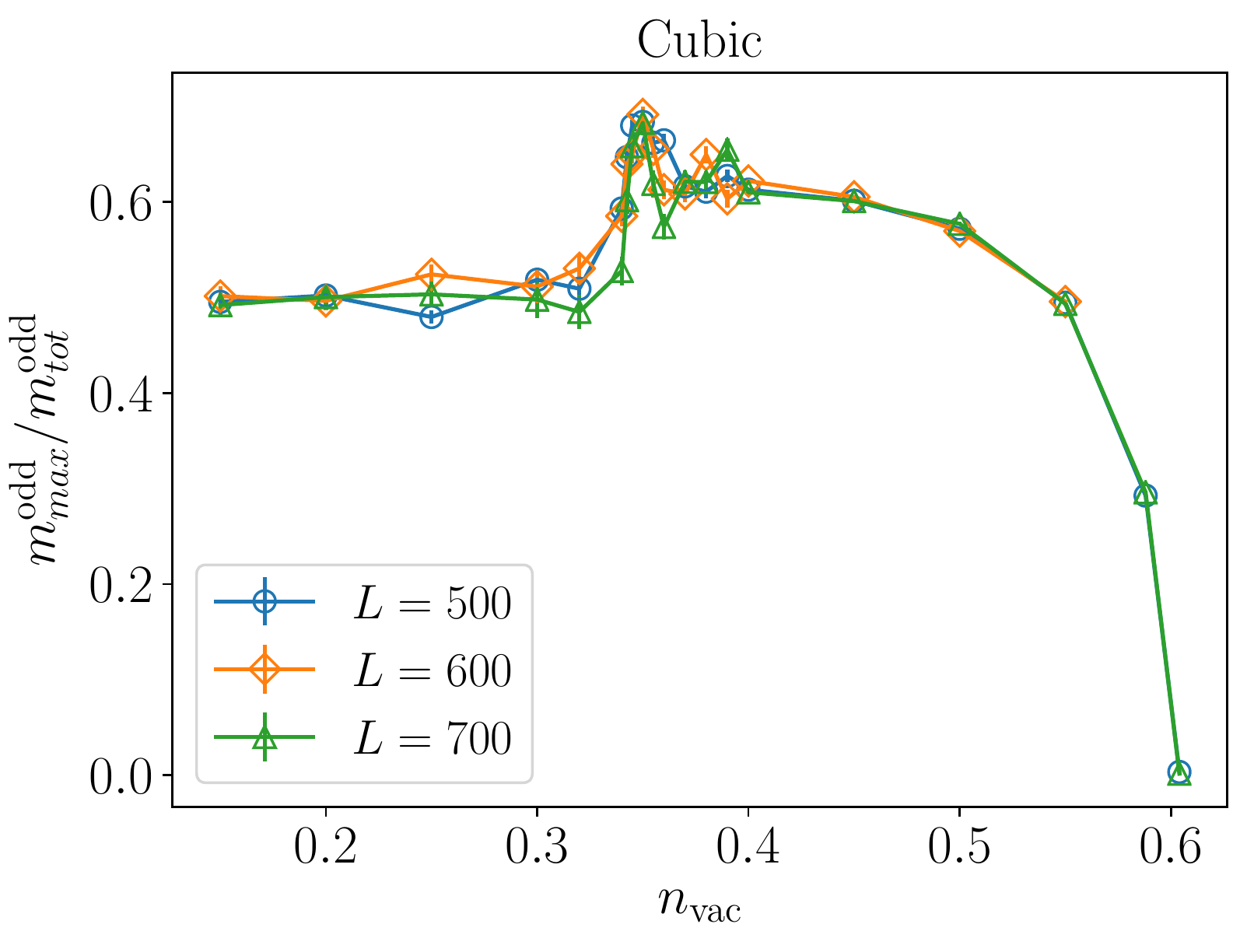}
		\caption{On the cubic lattice, the largest odd ${\mathcal R}$-type region contains a nonzero fraction of the total mass in such odd regions below a threshold value of $n_{\rm vac} \approx 0.6$ in the thermodynamic limit, as is clear from the sample average of this fraction shown here. This may be compared with results discussed in Sec.~{\protect{\ref{Geometry:Basic}}} }
		\label{oddmaxfraction_cubic}
\end{figure}

In Fig.~\ref{oddsmallfraction_m}, we display the fraction of the total mass of ``odd'' ${\mathcal R}$-type regions that is contained in small ``odd'' regions, with smallness defined as in the main text: small regions have absolute mass $m < V/n_{\rm vac}$ (with $V=10000$). Again, we see that most of the mass is in large odd regions. A similar conclusion follows more directly in the cubic-lattice case:
In Fig.~\ref{oddmaxfraction_cubic}, we display the fraction of mass of odd ${\mathcal R}$-type regions contained in the largest such region in the diluted cubic lattice. Again, we see that below $n_{\rm vac} \approx 0.6$, the largest odd region contains a nonzero fraction of the total mass of such regions in the thermodynamic limit, with this fraction displaying another kink-like feature at $n_{\rm vac} \approx 0.35$. As in the main text, we attribute this feature to the sublattice-symmetry breaking transition studied in Sec.~\ref{chiralsymmbreak}
\begin{figure}
			a)~\includegraphics[width=\columnwidth]{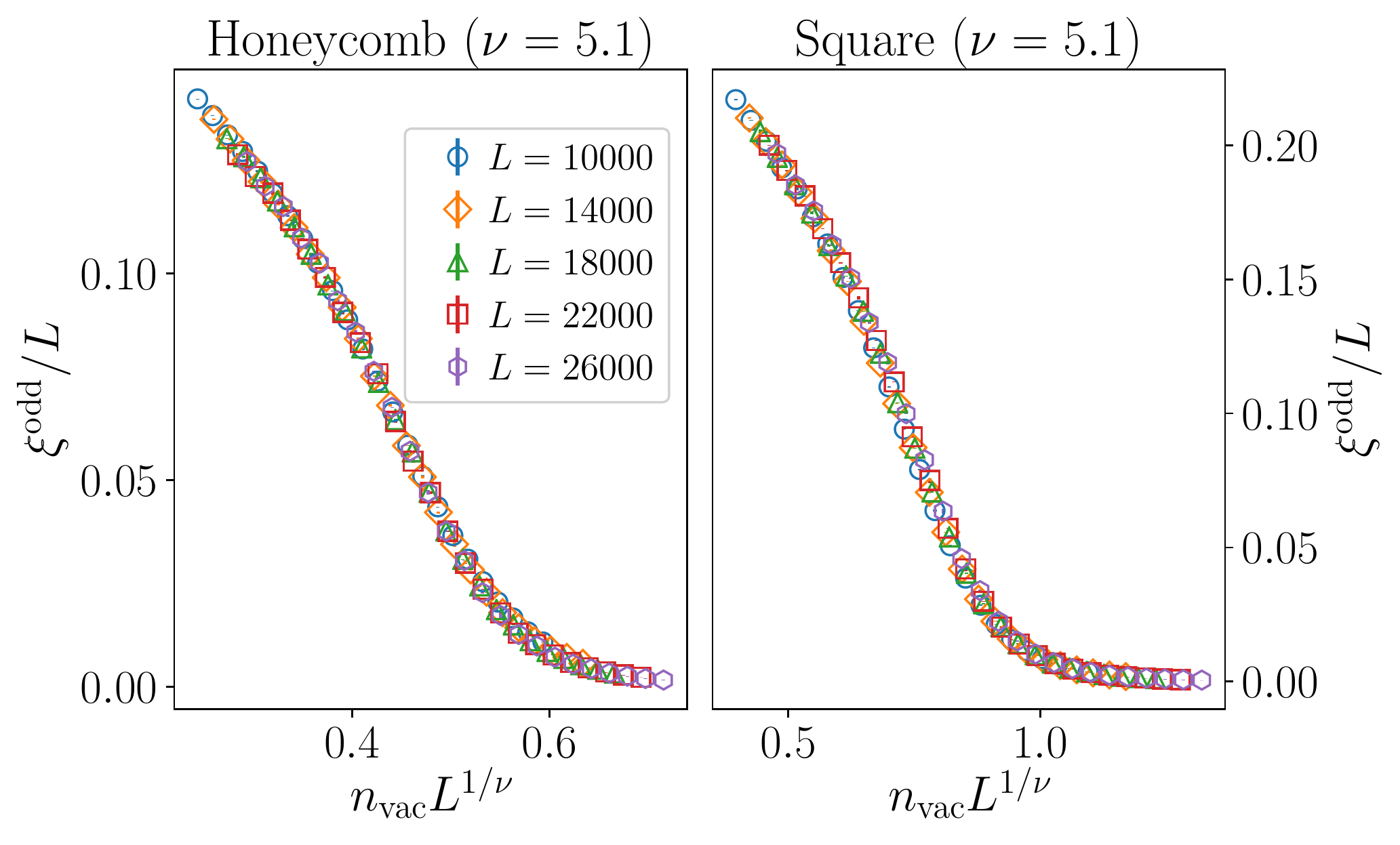}\\
		    b)~\includegraphics[width=\columnwidth]{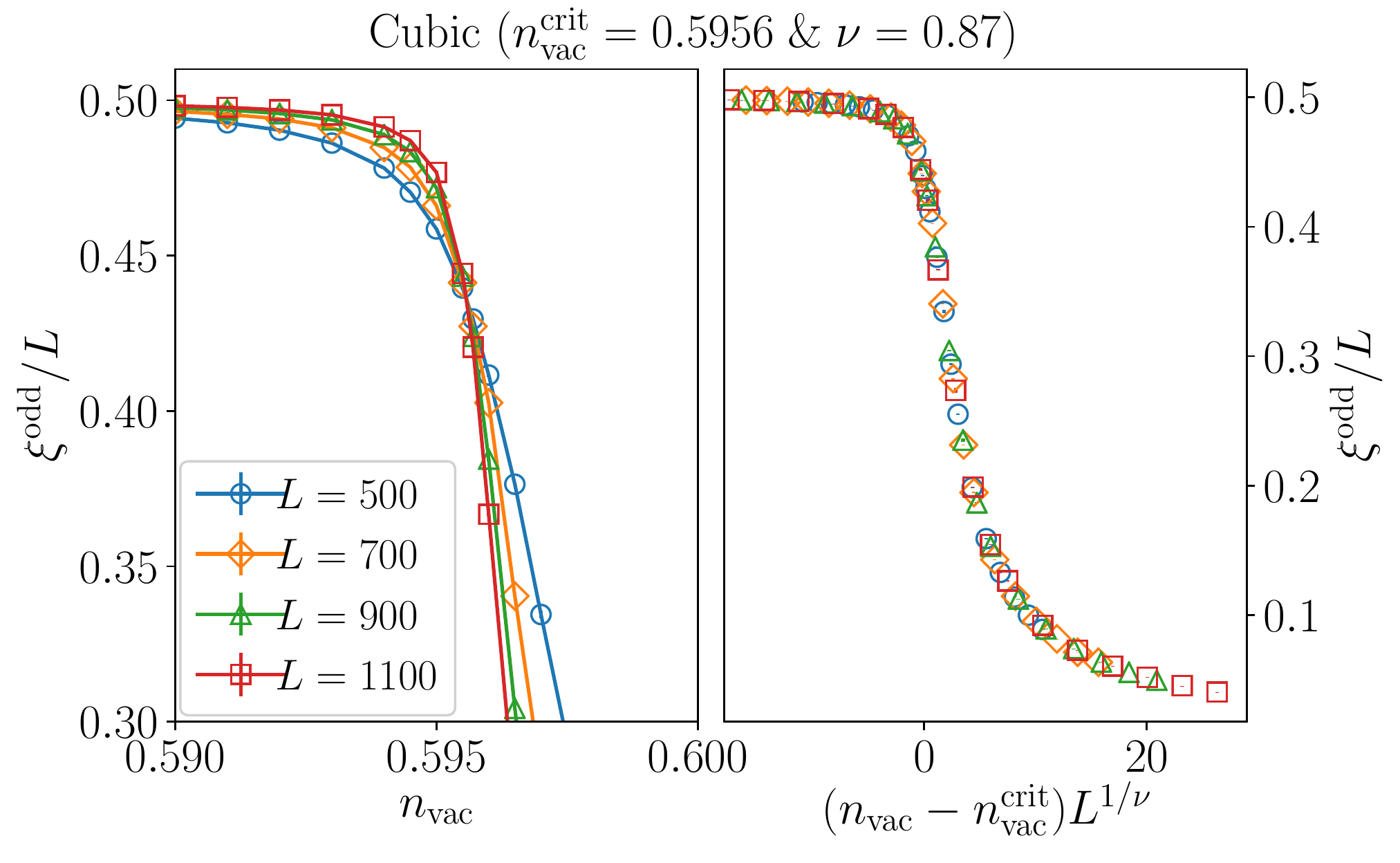}
		\caption{ a) The correlation length $\xi^{\rm odd}$, corresponding to the odd analogue $C^{\rm odd}(r-r')$ of the sample-averaged geometric correlation function in two-dimensional $L \times L$ samples at various values of $n_{\rm vac}$ collapses onto a single scaling curve when plotted as a function of $n_{\rm vac}L^{1/{\nu}}$ for small values of $n_{\rm vac}$. b) The same quantity on three-dimensional $L \times L \times L$ cubic lattices, plotted as a function of $n_{\rm vac}$.  Note that curves corresponding to different $L$ display a sharp crossing at a threshold $n_{\rm vac}^{\rm crit}$ near $n_{\rm vac} \approx 0.6$. Right panel: Data for $\xi^{\rm odd}$ from $L \times L \times L$ cubic lattices in the vicinity of this threshold collapses onto a single scaling curve when plotted as a function of the scaling variable $(n_{\rm vac} - n_{\rm vac}^{\rm crit})L^{1/\nu}$. This may be compared with results discussed in  Sec.~{\protect{\ref{percolation&incipientpercolation}}}.}
		\label{oddscalingxi}
 \end{figure}
 \begin{figure}
		a)~\includegraphics[width=\columnwidth]{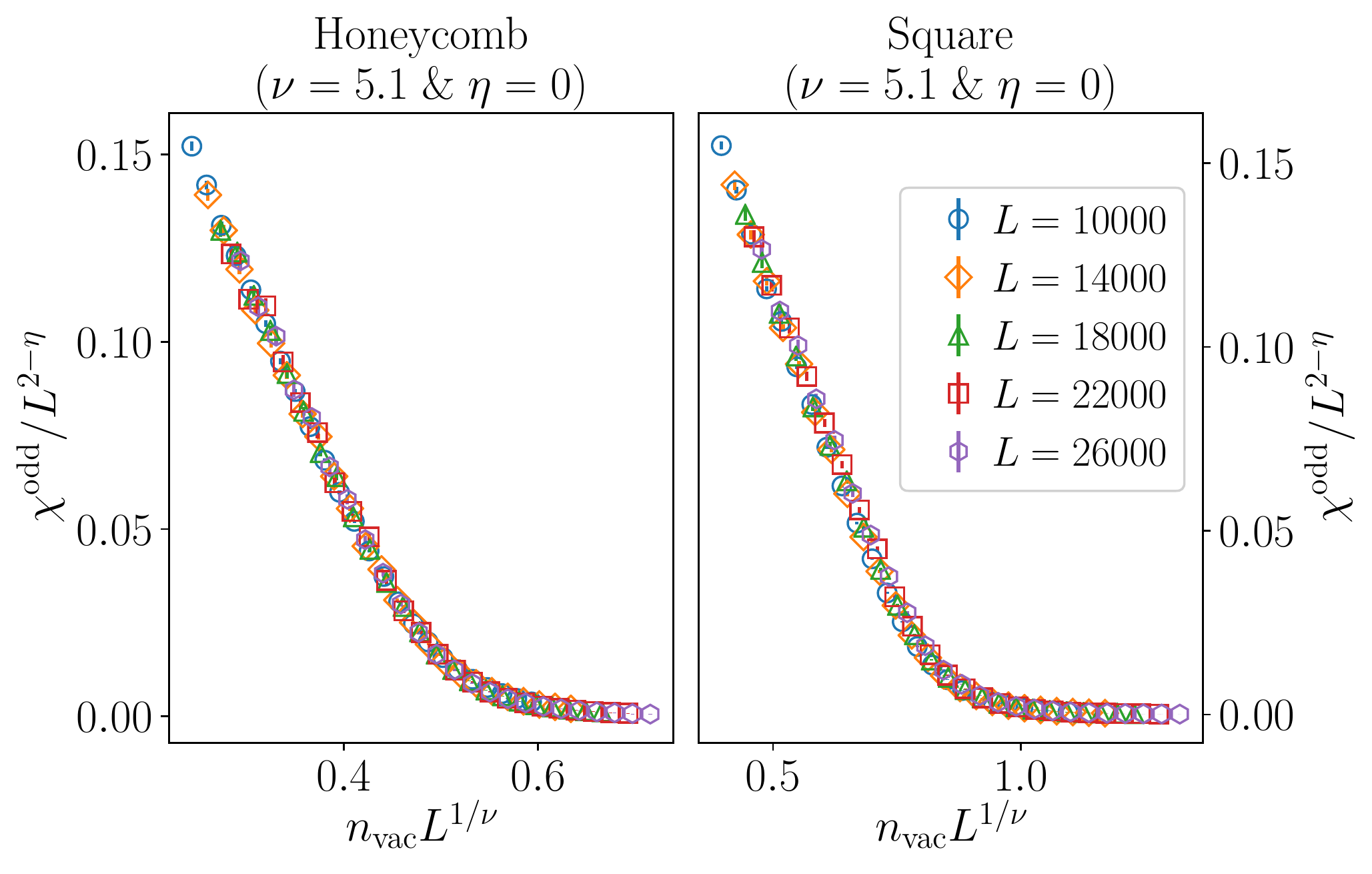}\\
		b)~\includegraphics[width=\columnwidth]{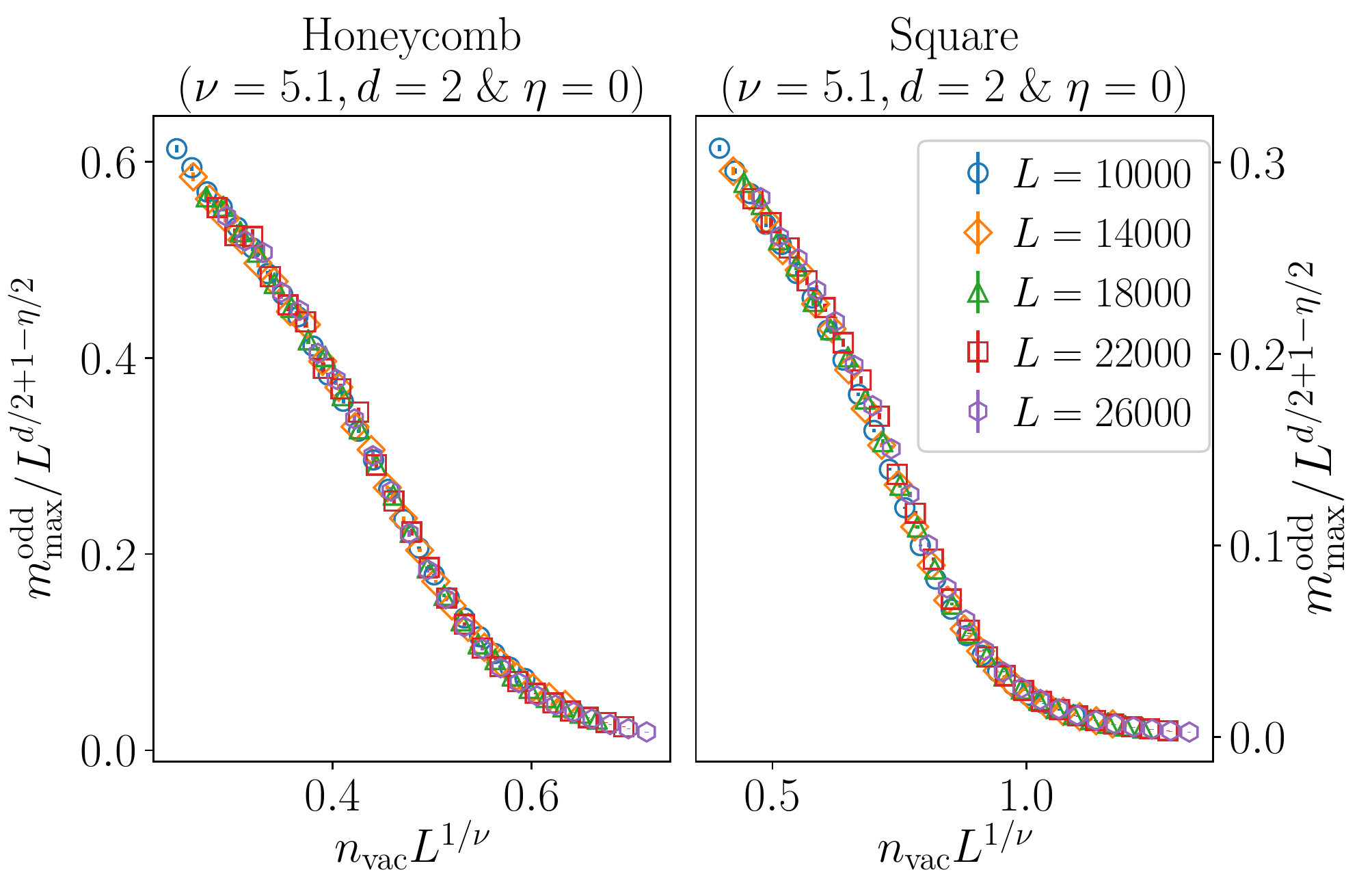}\\
		%c)~\includegraphics[width=\columnwidth]{./oddfinalplots_cubic/odd_mmax_chi_scaled.pdf}
		%\includegraphics[width=\columnwidth]{./finalplots_2d/chi_scaled1.pdf}\\
		%\includegraphics[width=\columnwidth]{./finalplots_2d/mmax_scaled1.pdf}\\
		%\includegraphics[width=\columnwidth]{./finalplots_cubic/mmax_chi_scaled.pdf}
		\caption{a) For appropriate choices of $\nu$ and $\eta$, the susceptibilities $\chi^{\rm odd}$ associated with the odd analog $C^{\rm odd}(r-r')$ of the sample-averaged geometric correlation function constructed from just odd ${\mathcal R}$-type regions of two-dimensional $L\times L$ samples, when rescaled by $L^{2-\eta}$, collapse on to a single curve when plotted as a function of $n_{\rm vac}L^{1/\nu}$ for small $n_{\rm vac}$. b) The masses $m_{\rm max}^{\rm odd}$ of the largest odd ${\mathcal R}$-type region in two-dimensional $L \times L$ samples, when rescaled by $L^{d/2 +1 - \eta/2}$ with $d=2$, show analogous scaling behaviour.  This may be compared with results discussed in Sec.~{\protect{\ref{percolation&incipientpercolation}}}.}
		\label{oddscalingchimmax_2d}
\end{figure}
\begin{figure}[b]
		\includegraphics[width=\columnwidth]{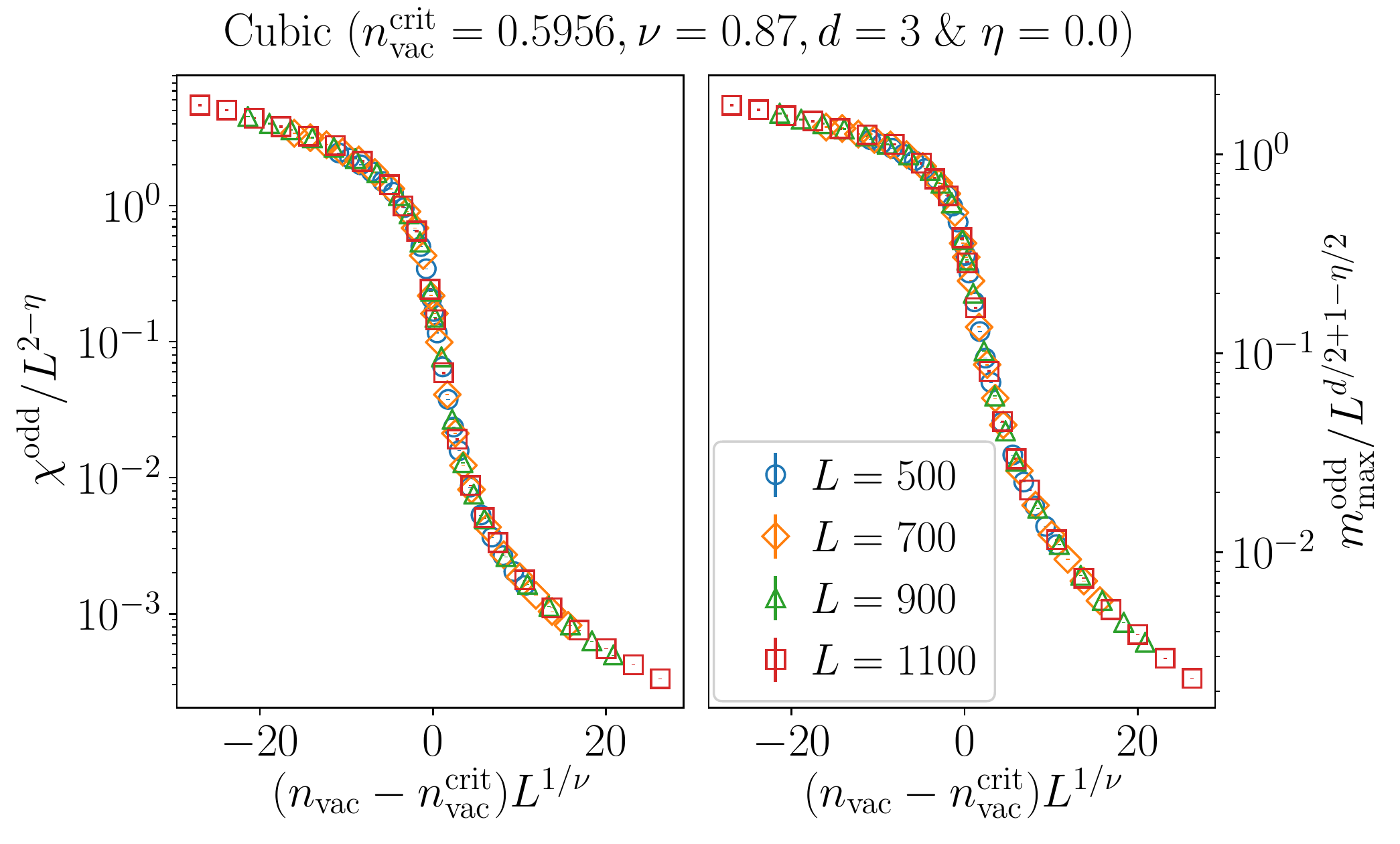}
		\caption{Left panel: For appropriate choice of $\nu$ and $\eta$, the susceptibilities $\chi^{\rm odd}$ associated with the odd analog $C^{\rm odd}(r-r')$ of the sample-averaged geometric correlation function constructed from just odd ${\mathcal R}$-type regions of three-dimensional $L \times L \times L$ samples, when rescaled by $L^{2-\eta}$, collapse on to a single curve when plotted as a function of $(n_{\rm vac} - n_{\rm vac}^{\rm crit})L^{1/\nu}$  for $n_{\rm vac}$ close to $n_{\rm vac}^{\rm crit}$. Right panel: The masses $m_{\rm max}^{\rm odd}$ of the largest odd ${\mathcal R}$-type regions in three-dimensional $L \times L \times L$ samples, when rescaled by $L^{d/2 +1 - \eta/2}$ with $d=3$, show analogous scaling behaviour.  This may be compared with results discussed in Sec.~{\protect{\ref{percolation&incipientpercolation}}}.}
		\label{oddscalingchimmax_cubic}
\end{figure}
\begin{figure}[t]
		a)~\includegraphics[width=\columnwidth]{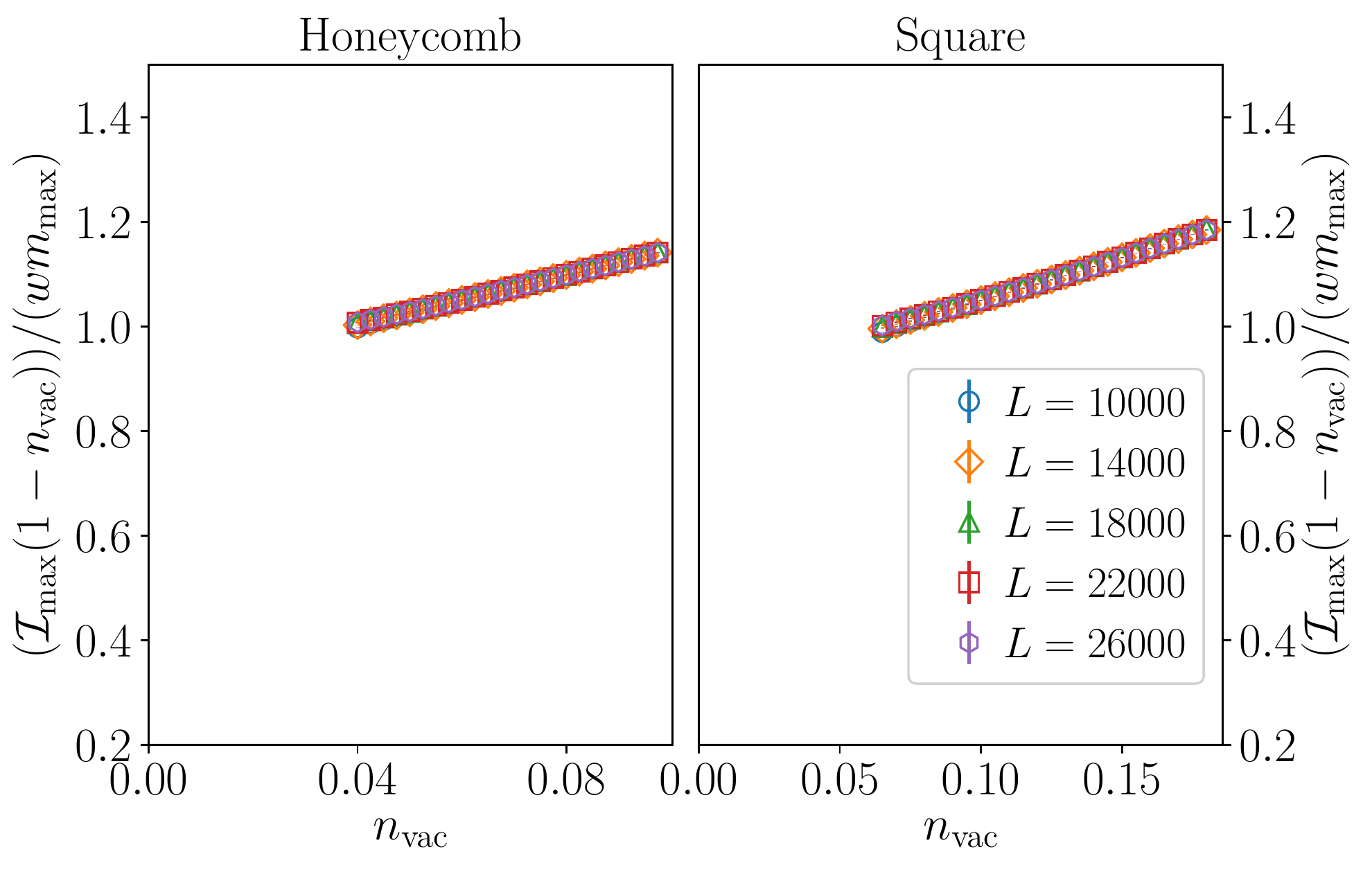}\\
	    b)~\includegraphics[width=\columnwidth]{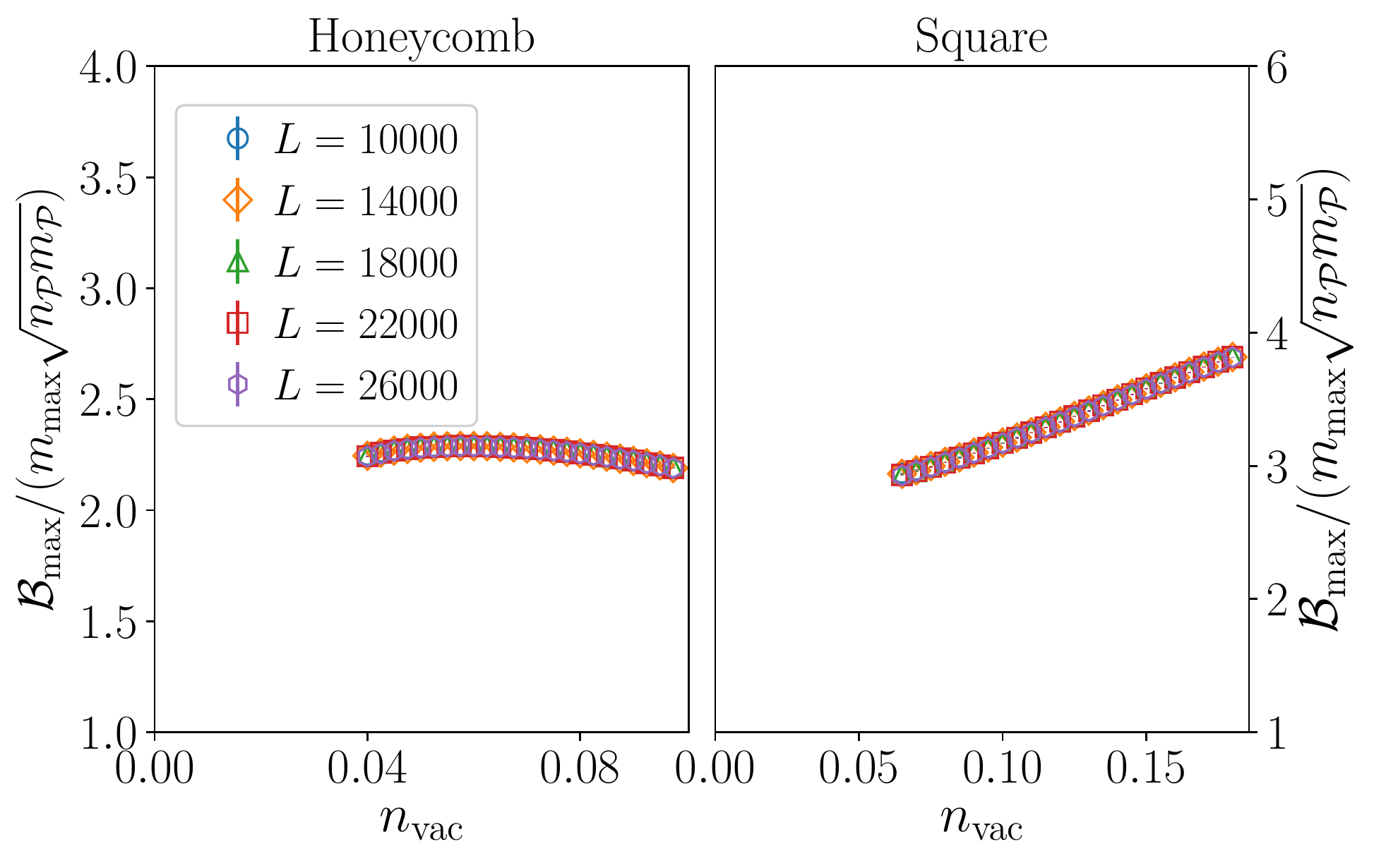}
		\caption{a), b): From the sample average of the corresponding ratio on the square and honeycomb lattices, we see that the size ${\mathcal B}_{\rm max}$ of the boundary of the largest ${\mathcal R}$-type region is proportional, with no discernible finite-size corrections, to $m_{\max} m_{\mathcal P}^{(d-1)/d}n_{\mathcal P}^{1/d}$, where $m_{\rm max}$ is the mass of this region and $n_{\mathcal P}$ ($m_{\mathcal P}$) is the total number (mass) density of ${\mathcal P}$-type regions in the thermodynamic limit in $d=2$ dimensions.
From the sample average of the corresponding ratio, ${\mathcal I}_{\rm max}$, the number of zero modes hosted by the largest ${\mathcal R}$-region scales with $w m_{\rm max}/(1-n_{\rm vac})$, with no discernible finite-size corrections.}
		\label{morphfigures1_2d}
\end{figure}

In Fig.~\ref{oddscalingxi}, we display the scaling of $\xi^{\rm odd}$, the correlation length that characterizes the sample-averaged odd correlation function $C^{\rm odd}(r-r')$ (to which each sample contributes $1$ if both $r$ and $r'$ belong to the same odd ${\mathcal R}$-type region in that sample, and zero otherwise). The scaling behaviour of the corresponding odd susceptibility $\chi^{\rm odd}$ is displayed in Fig.~\ref{oddscalingchimmax_2d} and Fig.~\ref{oddscalingchimmax_cubic}. The same figures also shows the scaling of the mass of the largest odd ${\mathcal R}$-type region. All these scaling analyses confirm that the odd ${\mathcal R}$-type regions share the same critical behaviour as the other ${\mathcal R}$-type regions.

\begin{figure}[H]
        \includegraphics[width=\columnwidth]{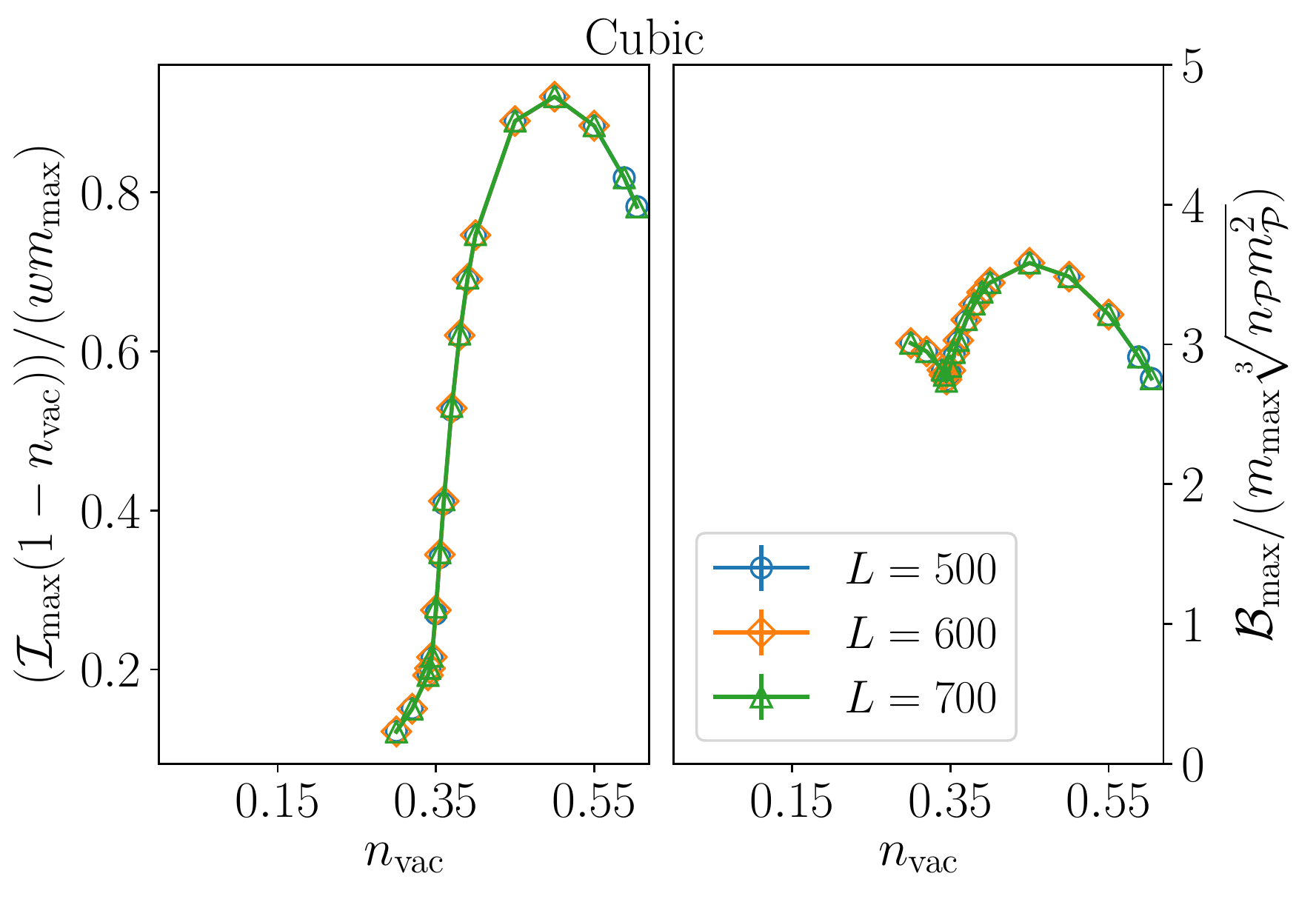}
		\caption{From the sample average of the corresponding ratio, we see that the size ${\mathcal B}_{\rm max}$ of the boundary of the largest ${\mathcal R}$-type region is proportional, with no discernible finite-size corrections, to $m_{\max} m_{\mathcal P}^{(d-1)/d}n_{\mathcal P}^{1/d}$, where $m_{\rm max}$ is the mass of this region and $n_{\mathcal P}$ ($m_{\mathcal P}$) is the total number (mass) density of ${\mathcal P}$-type regions in the thermodynamic limit in $d=3$ dimensions on the cubic lattice.
The cusp in the ratio near $n_{\rm vac} \approx 0.35$ is attributed to a sublattice symmetry breaking transition (Sec.~\ref{chiralsymmbreak}).  From the sample average of the corresponding ratio, ${\mathcal I}_{\rm max}$, the number of zero modes hosted by the largest ${\mathcal R}$-region scales with $w m_{\rm max}/(1-n_{\rm vac})$, with no discernible finite-size corrections. On the cubic lattice, the corresponding proportionality constant decreases rapidly in the low-dilution limit.}
		\label{morphfigures1_cubic}
\end{figure}

Next we turn to the morphology of the largest ${\mathcal R}$-type region in a finite-size sample. This is interesting especially since very large finite-size-limited ${\mathcal R}$-type regions dominate many observables at low dilution both in two and in three dimensions.
A plausible estimate for the size ${\mathcal B}_{\rm max}$ of its boundary is as follows: In the classical theory of percolation, the analog of ${\mathcal B}_{\rm max}$ is expected to scale as $m_{\rm max}$. This is due to the fact that
${\mathcal B}_{\rm max}$ counts both the size of the internal boundary (due to voids or holes) and the actual size of the external boundary. As a result, the boundary ${\mathcal B}_{\rm max} $ of the critical cluster in classical percolation is dominated by the contribution of these voids. In our case, it is reasonable to assume that the analogue of voids are ${\mathcal P}$-type regions since it is clear from our results that the size of the typical ${\mathcal P}$-type regions remain very small over the entire range of $n_{\rm vac}$ studied, both in two and three dimensions. We estimate that there are of order $m_{\rm max} n_{\mathcal P}$ such voids, with each void contributing $(m_{\mathcal P}/n_{\mathcal P})^{(d-1)/d}$ on average to
${\mathcal B}_{\rm max}$. Assembling this information, we arrive at the estimate ${\mathcal B}_{\rm max} \sim m_{\rm max} n_{\mathcal P} \times (m_{\mathcal P}/n_{\mathcal P})^{(d-1)/d}$ in $d$ spatial dimensions. Based on this argument, we expect
${\mathcal B}_{\rm max}/m_{\rm max} \sim  m_{\mathcal P}^{(d-1)/d}n_{\mathcal P}^{1/d}$. In Fig.~\ref{morphfigures1_2d} and Fig.~\ref{morphfigures1_cubic}, we see that this expectation is borne
out by the data, with the corresponding ratio converging very quickly to
its thermodynamic limit in two and three dimensions (indeed, at the sizes we study, no finite-size effects are readily discernible). In two dimensions, this ratio has a very mild and nonsingular dependence on $n_{\rm vac}$ in the limit of small dilution. On the cubic lattice, we see a noticeable 
cusp-like feature near $n_{\rm vac} \approx 0.35$, which we attribute to a second sublattice-symmmetry-breaking transition (see Sec.~\ref{chiralsymmbreak}).

In Fig.~\ref{morphfigures1_2d} and Fig.~\ref{morphfigures1_cubic}, we also see that the ratio ${\mathcal I}_{\rm max}(1-n_{\rm vac})/(wm_{\rm max})$ saturates quickly to the thermodynamic limit in both two and three dimensions, with no discernible finite-size corrections in the range of sizes studied here. In two dimensions, it has a ${\mathcal O}(1)$ value with a mild dependence on $n_{\rm vac}$. Thus, the largest ${\mathcal R}$-type region has a zero-mode density not very different from the globally averaged density $w$ of zero modes.
On the cubic lattice however, this ratio decreases noticeably with $n_{\rm vac}$ inside the percolated phase, suggesting that the percolated infinite cluster has a systematically smaller density of zero modes compared to the sample-wide average in the limit of low dilution.   
                      
\begin{figure}[H]
        \includegraphics[width=\columnwidth]{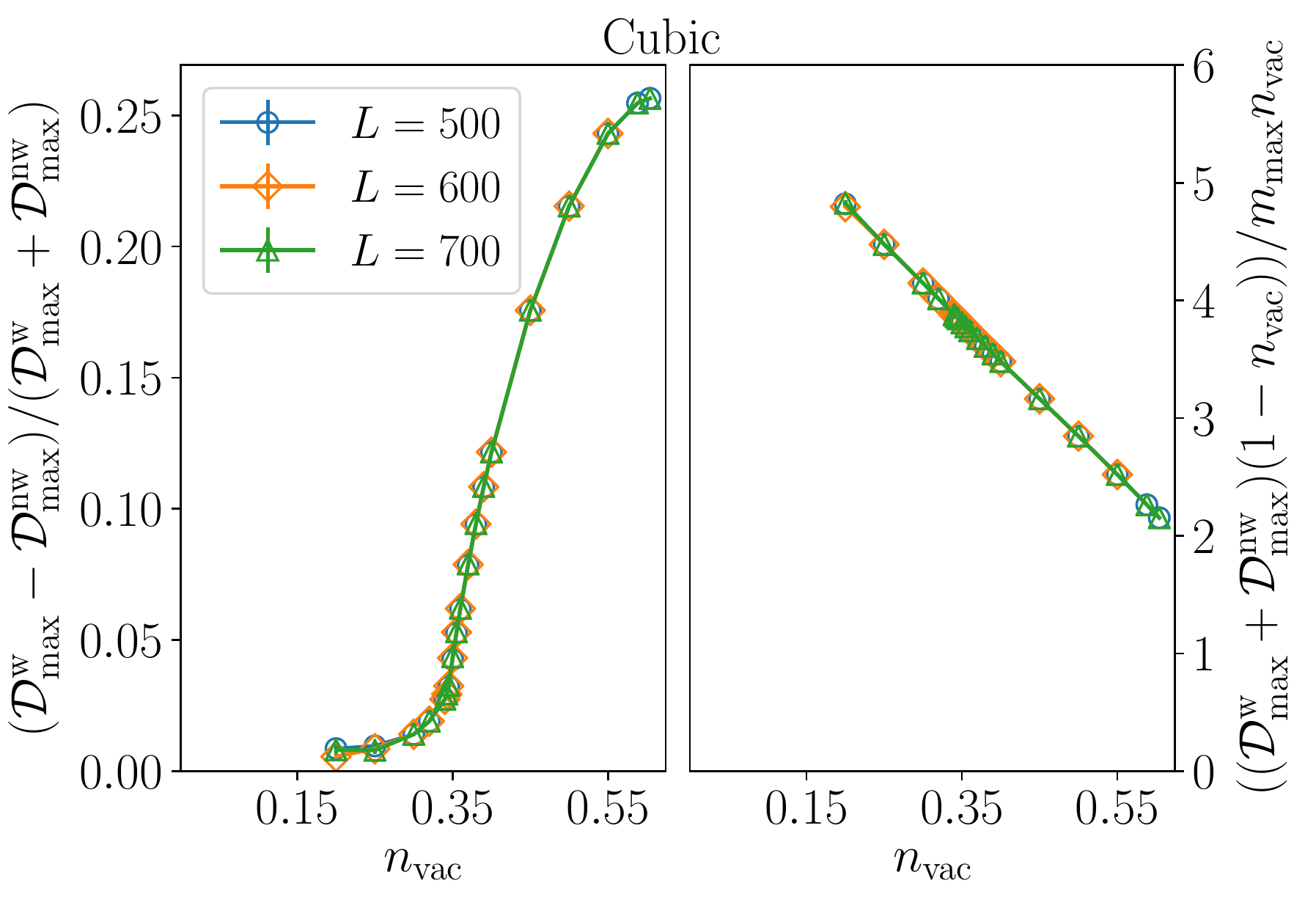}
		\caption{As is evident on the cubic latice from the corresponding sample-averaged ratios, ${\mathcal D}^{\rm nw}_{\rm max} + {\mathcal D}^{\rm w}_{\rm max}$ scales with $n_{\rm vac} m_{\rm max}/(1-n_{\rm vac})$, and ${\mathcal D}^{\rm nw}_{\rm max} - {\mathcal D}^{\rm w}_{\rm max}$ scales with ${\mathcal D}^{\rm nw}_{\rm max} + {\mathcal D}^{\rm w}_{\rm max}$, again with no discernible finite-size corrections even in the low-dilution limit. The $n_{\rm vac}$ dependence of $({\mathcal D}^{\rm nw}_{\rm max} - {\mathcal D}^{\rm w}_{\rm max})/({\mathcal D}^{\rm nw}_{\rm max} + {\mathcal D}^{\rm w}_{\rm max})$ broadly resembles that of ${\mathcal I}_{\rm max}(1-n_{\rm vac})/w m_{\rm max}$ in the low-dilution limit.}
		\label{morphfigures2_cubic}
\end{figure}

\begin{figure}[H]
	    a)~\includegraphics[width=\columnwidth]{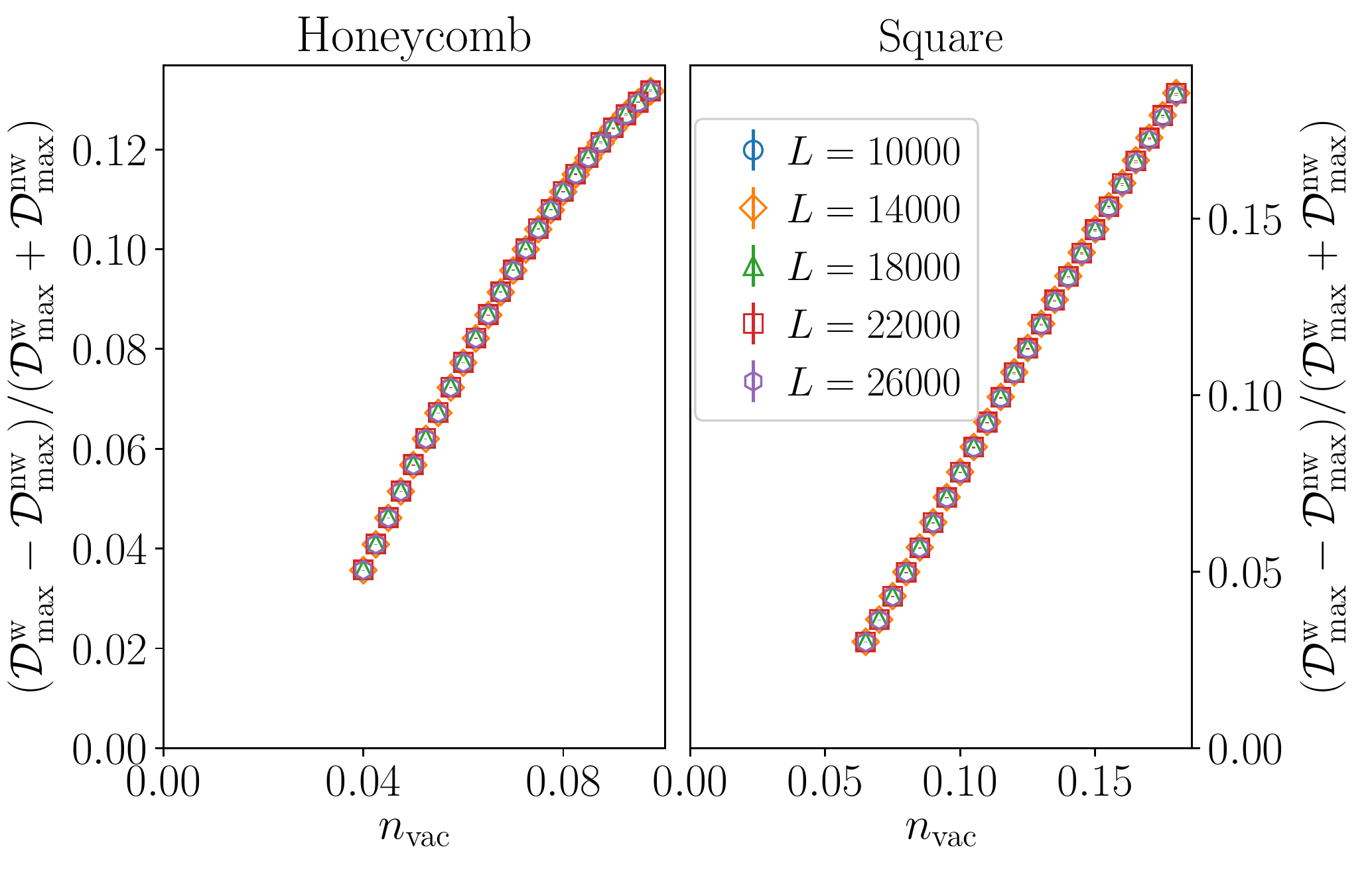}\\
	    b)~\includegraphics[width=\columnwidth]{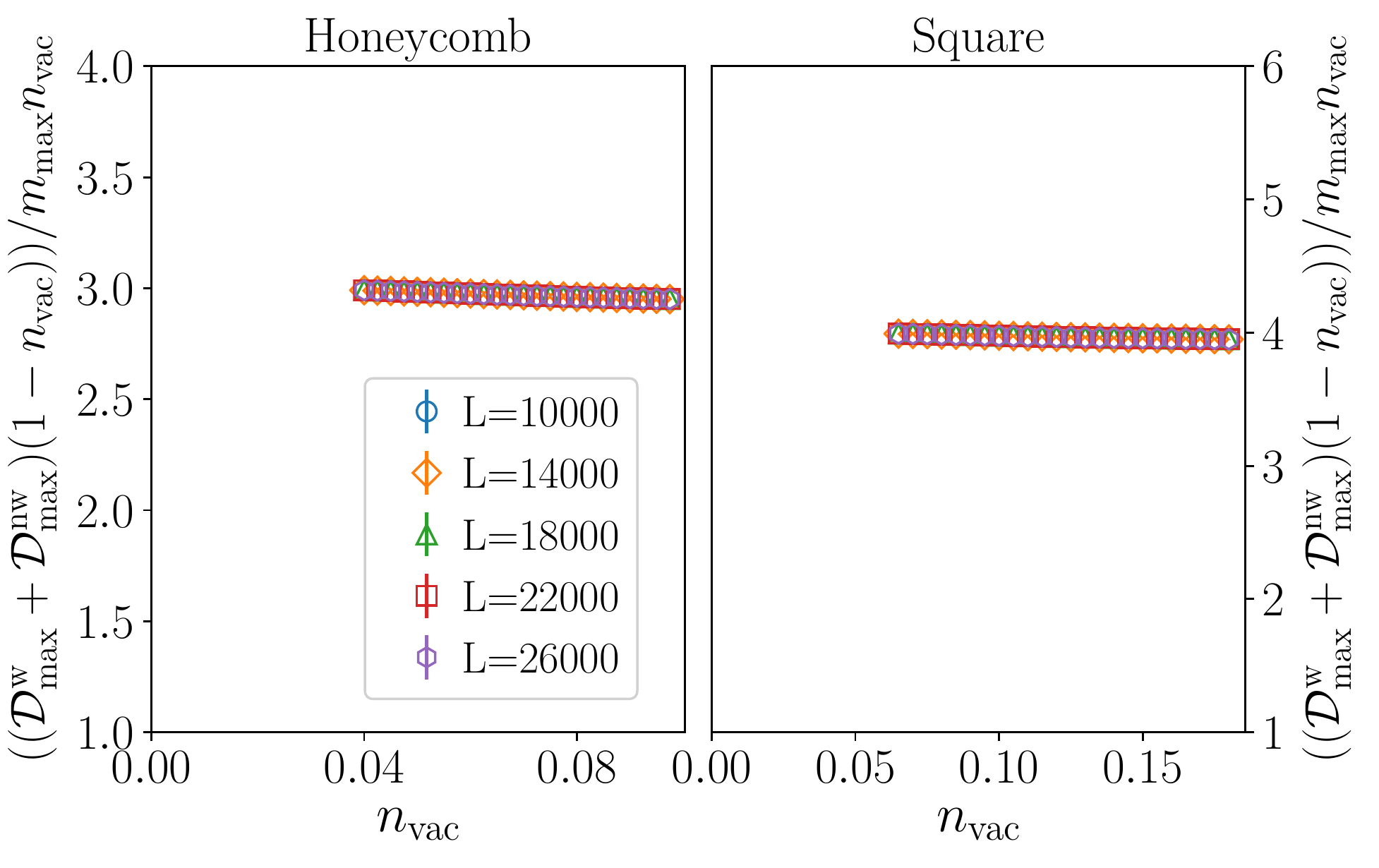}
		\caption{a), b): As is evident on the square and honeycomb lattices from the corresponding sample-averaged ratios, ${\mathcal D}^{\rm nw}_{\rm max} + {\mathcal D}^{\rm w}_{\rm max}$ scales with $n_{\rm vac} m_{\rm max}/(1-n_{\rm vac})$, and ${\mathcal D}^{\rm nw}_{\rm max} - {\mathcal D}^{\rm w}_{\rm max}$ scales with ${\mathcal D}^{\rm nw}_{\rm max} + {\mathcal D}^{\rm w}_{\rm max}$, again with no discernible finite-size corrections even in the low-dilution limit. The $n_{\rm vac}$ dependence of $({\mathcal D}^{\rm nw}_{\rm max} - {\mathcal D}^{\rm w}_{\rm max})/({\mathcal D}^{\rm nw}_{\rm max} + {\mathcal D}^{\rm w}_{\rm max})$ broadly resembles that of ${\mathcal I}_{\rm max}(1-n_{\rm vac})/w m_{\rm max}$ in the low-dilution limit.}
		\label{morphfigures2_2d}
\end{figure}

Next, we consider ${\mathcal D}^{\rm nw}_{\rm max} + {\mathcal D}^{\rm w}_{\rm max}$, which is a measure of the number of vacancies adjacent to sites that belong to the largest ${\mathcal R}$-type region (more accurately, it measures the number of links deleted due to these vacancies). If this region is not atypical in terms of the overall density of vacancies associated with it, one would expect $({\mathcal D}^{\rm nw}_{\rm max} + {\mathcal D}^{\rm w}_{\rm max} ) \sim n_{\rm vac} m_{\rm max}/(1-n_{\rm vac})$.
In Fig.~\ref{morphfigures2_2d} and Fig.~\ref{morphfigures2_cubic}, we see that the corresponding ratio saturates very quickly to the thermodynamic limit and has a very mild nonsingular dependence on $n_{\rm vac}$, thus confirming this basic picture in both two and three dimensions.

Turning to ${\mathcal D}^{\rm w}_{\rm max} - {\mathcal D}^{\rm nw}_{\rm max}$, we begin by making more explicit our intuitive picture for the zero modes and monomers hosted by ${\mathcal R}$-type regions: A local imbalance in the numbers of surviving sites on the two sublattices gives rise to these zero modes and mandates the existence of monomers in this region. 
Since the largest ${\mathcal R}$-type region hosts a nonzero monomer density, these regions must be atypical, in that ${\mathcal D}^{\rm w}_{\rm max} - {\mathcal D}^{\rm nw}_{\rm max}$ must scale in the same way as ${\mathcal D}^{\rm w}_{\rm max} + {\mathcal D}^{\rm nw}_{\rm max}$, {\em i.e.}, proportional to $m_{\rm max}$, rather than exhibiting the $\sqrt{m_{\rm max}}$ scaling that would characterize a truly random region of the diluted lattice. In Fig.~\ref{morphfigures2_cubic} and Fig.~\ref{morphfigures2_2d},  we see that this expectation is also borne out by the corresponding ratio, which converges rapidly to the thermodynamic limit, again with no discernible finite-size corrections for the range of sizes studied here. Moreover, the $n_{\rm vac}$ dependence of this ratio broadly resembles that of the ratio ${\mathcal I}_{\rm max}(1-n_{\rm vac})/(wm_{\rm max})$.

\end{document}